\documentclass[ejs]{imsart}

\doi{10.1214/154957804100000000}
\pubyear{0000}
\volume{0}
\firstpage{0}
\lastpage{0}

\usepackage{amssymb, amsmath, mathrsfs, graphicx, subfigure, psfrag, natbib}
\usepackage{adjustbox}
\RequirePackage[colorlinks, citecolor=blue, urlcolor=blue, breaklinks=true]{hyperref}

\numberwithin{equation}{section}
\newtheorem{lem}{Lemma}[section]
\newtheorem{them}{Theorem}[section]
\newtheorem{coro}{Corollary}[section]
\newtheorem{definition}{Definition}[section]
\def\T{{ \mathrm{\scriptscriptstyle T} }}

\newcommand{\be} {\mbox{\boldmath $e$}}
\newcommand{\bh} {\mbox{\boldmath $h$}}

\newcommand{\bv} {\mbox{\boldmath $v$}}

\newcommand{\bX} {\mathbf {X}}
\newcommand{\bY} {\mathbf {Y}}

\newcommand{\bS} {\mathbf {S}}
\newcommand{\bT} {\mathbf {T}}
\newcommand{\bW} {\mathbf {W}}
\newcommand{\bU} {\mathbf {U}}

\newcommand{\ISE}{\mathrm{ISE}}

\newcommand{\argmin}{\textrm{argmin}}

\begin{document}

\begin{frontmatter}

\title{Nonparametric modal regression in the presence of measurement error}
\runtitle{Modal regression in the presence of measurement error}


\author{\fnms{Haiming} \snm{Zhou} \ead[label=e1]{zhouh@niu.edu}}
\address{Division of Statistics, Northern Illinois University, DeKalb, IL 60115, USA \\ \printead{e1}}
\and
\author{\fnms{Xianzheng} \snm{Huang} \corref{} \ead[label=e2]{huang@stat.sc.edu}}
\address{Department of Statistics, University of South Carolina, Columbia, SC 29208, USA \\ \printead{e2}}

\runauthor{H. Zhou and X. Huang}

\begin{abstract}
In the context of regressing a response $Y$ on a predictor $X$, we consider estimating the local modes of the distribution of $Y$ given $X=x$ when $X$ is prone to measurement error. We propose two nonparametric estimation methods, with one based on estimating the joint density of $(X, Y)$ in the presence of measurement error, and the other built upon estimating the conditional density of $Y$ given $X=x$ using error-prone data. We study the asymptotic properties of each proposed mode estimator, and provide implementation details including the mean-shift algorithm for mode seeking and bandwidth selection. Numerical studies are presented to compare the proposed methods with an existing mode  estimation method developed for error-free data naively applied to error-prone data. 
\end{abstract}

\begin{keyword}[class=MSC]
\kwd[Primary ]{62G08}
\kwd[; secondary ]{62G20}
\end{keyword}

\begin{keyword}
\kwd{Bandwidth selection}
\kwd{deconvoluting kernel}
\kwd{Fourier transform}
\kwd{local mode}
\kwd{mean-shift algorithm}
\end{keyword}



\end{frontmatter}

\section{Introduction}
\label{s:intro}
The majority of statistical literature on regression analysis focuses on inferring the mean function of the response $Y$ given a predictor $X$. There are also a large body of work devoted to making inference for the quantiles of $Y$ given $X$ in the regression setting \citep[e.g.,][]{Koenker05}. Recently, researchers started to investigate methods to infer local modes of $Y$ given $X$ \citep[see][among others]{Yao2012, Yao2014, Chen2016}. These researchers pointed out valuable information about the association between the response and the predictor that conditional modes can provide yet the conditional mean/quantiles can miss. Advantages of modal regression compared to mean or quantile regression have been well appreciated in analyzing speed-flow data in traffic engineering \citep{Einbeck2006}, studying temperature patterns \citep{Hyndman96}, investigating galaxy properties conditioning on a given environment \citep{Rojas2008}, and in economics \citep{Huang2013} for instance. To address the practical issue of error-contaminated covariates, methods accounting for measurement error in mean regression \citep{LASbook} and methods for quantile regression in the presence of measurement error \citep{He2000, Wei2009, Ma2011, Wang2012} have been developed. In contrast, there is no existing research on modal regression when $X$ is prone to measurement error even though this issue often arises in the aforementioned applications, and ignoring measurement error in modal regression typically results in misleading inference. We tackle this important problem in our study. 

We propose two nonparametric methods for estimating conditional modes of a response variable $Y$ given an error-prone covariate $X$. The first method exploits a kernel density estimator of the joint density of $(X, Y)$ that accounts for measurement error in $X$. The second method is based on a local linear estimator of the conditional density of $Y$ given $X=x$. These methods are elaborated in Section~\ref{s:methods}. Asymptotic properties of the mode estimators resulting from these methods are presented in Section~\ref{s:asymptotics}. We provide a data-driven method for bandwidth selection to facilitate the implementation of the proposed methods in Section~\ref{s:implementation}. Numerical studies are presented in Section~\ref{s:empirical}, which include simulation experiments and an application to dietary data. Section~\ref{s:discussion} provides a discussion on follow-up open research questions.   

\section{Nonparametric estimation of local modes}
\label{s:methods}
\subsection{Data and measurement error model}
\label{s:data}
Suppose one wishes to collect data for a response $Y$ and a covariate $X$, $\{(X_j, Y_j),$\\$j=1, \ldots, n\}$, consisting of $n$ independent pairs from a bivariate distribution specified by the joint probability density function (pdf), $p(x,y)$. However, the reality is that $\bX=\{X_j\}_{j=1}^n$ cannot be observed directly due to error contamination, and $\bW=\{W_j\}_{j=1}^n$ are observed instead. More specifically, the observed covariate $W$ relates to the underlying true covariate $X$ via an additive measurement error model given by 
\begin{equation}
W_j=X_j+U_j, \textrm{ for $j=1, \ldots, n$}, 
\label{eq:wxu}
\end{equation}
where $\bU=\{U_j\}_{j=1}^n$ are nondifferential measurement errors \citep[][Section 2.5]{LASbook}, meaning that $\bU\perp \bX$ and $\bU\perp \bY=\{Y_j\}_{j=1}^n$. We assume a known distribution of $U$ specified by its pdf $f_{\hbox {\tiny $U$}}(u)$ in this study. Estimating the distribution of $U$ requires either replicate measures of each underlying $X_j$ or external validation data. For instance, if replicate measures are available and one assumes $f_{\hbox {\tiny $U$}}(u)$ known up to some parameters, such as the variance parameter, then one can follow equation (4.3) in \citet{LASbook} to estimate the measurement error variance consistently. One may also estimate the characteristic function of $U$ as proposed in \citet{Delaigle2008}, which is all our proposed inference methods need in terms of information regarding the measurement error distribution.  

The focal point of statistical inference presented in this study lies in local modes of the conditional pdf of $Y$ given $X=x$, $p(y|x)$. Denote by $\mathscr{X}$ the support of $X$ and by $\mathscr{Y}$ the support of $Y$. For a generic bivariate function $g(s,t)$, $g_s(s_1, t)$ refers to $(\partial/\partial s) g(s,t)$ evaluated at $s=s_1$, and notations for higher-order partial derivatives of $g(s,t)$ are similarly defined. Given $x\in \mathscr{X}$, a mode of $p(y|x)$, denoted by $y_{\hbox {\tiny $M$}}(x)$, is a value in $\mathscr{Y}$ such that $p_y\{y_{\hbox {\tiny $M$}}(x)|x\}=0$ and $p_{yy}\{y_{\hbox {\tiny $M$}}(x)|x\}<0$. Equivalently, $y_{\hbox {\tiny $M$}}(x)$ satisfies $p_y\{x,y_{\hbox {\tiny $M$}}(x)\}=0$ and $p_{yy}\{x,y_{\hbox {\tiny $M$}}(x)\}<0$. This latter viewpoint motivates our first proposed estimator of $y_{\hbox {\tiny $M$}}(x)$ described next. It is possible that $p(y|x)$ is multimodal at a given $x$, producing a mode set $M(x)=\{y\in \mathscr{Y}: \, p_y(x, y)=0 \textrm{ and } p_{yy}(x, y)<0\}$. Although multi-modality brings in certain challenges in the actual implementation of mode seeking, it adds little complication in asymptotics analyses of our proposed mode estimators. To avoid unnecessarily tedious notations, we assume a unimodal $p(y|x)$ for the methodology development and theoretical analysis in the main article. We repeat the key part of the preliminary theoretical development with the uni-modality assumption relaxed in Appendix I for illustration purposes. In addition, multimodal $p(y|x)$ is considered when illustrating the implementation of the proposed methods in Section~\ref{s:empirical}. 

Even though the conditional mode set $M(x)$ is formulated above in terms of the joint pdf $p(x, y)$, we shall point out that,  as $x$ moves in $\mathscr{X}$, the resultant conditional mode curve(s), $\{M(x),\, x\in \mathscr{X}\}$, characterize the mode structure of the conditional density of $Y$ given $X$. Hence, they typically differ from density ridges \citep{Genovese2014} and principal curves \citep{Hastie1989, Ozertem2011}, which focus on certain structures of the joint pdf. \citet[][Section 8]{Chen2016} provided detailed explanations on the distinction between conditional mode curves and density ridges, which are also helpful for one to see how they differ from principal curves. 

\subsection{Local constant estimator}
\label{s:localcon}
We first consider an estimator of $y_{\hbox {\tiny $M$}}(x)$, denoted by $\hat y_{\hbox {\tiny $M0$}}(x)$, as the solution to $\hat p_y(x,y)=0$, where $\hat p_y(x,y)$ is a kernel-based estimator of $p_y(x,y)$. The construction of $\hat p_y(x,y)$ traces back to the kernel density estimator of $p(x,y)$ in the absence of measurement error considered in \citet{Wand93},  
\begin{equation}
\tilde p(x, y)=\frac{1}{nh_1 h_2} \sum_{j=1}^n K_1\left(\frac{X_j-x}{h_1}\right)K_2\left(\frac{Y_j-y}{h_2}\right),
\label{eq:pdfest}
\end{equation} 
where $h_1$ and $h_2$ are bandwidths, and $K_1(t)$ and $K_2(t)$ are kernels. In the presence of measurement error, with $\bW$ observed in place of $\bX$, we follow the idea of deconvoluting kernel \citep{Carroll88, Stefanski90} and propose an estimator of $p(x,y)$ that accounts for measurement error as follows,   
\begin{equation}
\hat p(x, y)=\frac{1}{nh_1 h_2} \sum_{j=1}^n K_{\hbox {\tiny $U,0$}}\left(\frac{W_j-x}{h_1}\right)K_2\left(\frac{Y_j-y}{h_2}\right), 
\label{eq:jointpdfest}
\end{equation} 
where 
\begin{equation} 
K_{\hbox {\tiny $U,0$}}(t)=\frac{1}{2\pi}\int e^{-its} \frac{\phi_{\hbox {\tiny $K_1$}}(s)}{\phi_{\hbox {\tiny $U$}}(s/h_1)} ds 
\label{eq:KU0}
\end{equation}
is the deconvoluting kernel, in which $i$ is the imaginary unit, $\phi_{\hbox {\tiny $K_1$}}(\cdot)$ is the Fourier transform of $K_1(\cdot)$, and $\phi_{\hbox {\tiny $U$}}(\cdot)$ is the characteristic function of $U$, i.e., the Fourier transform of $f_{\hbox {\tiny $U$}}(\cdot)$. In this article, all integrations are over the entire real line unless otherwise specified. The estimator $\hat p(x,y)$ is motivated by the property of the deconvoluting kernel that $E[K_{\hbox {\tiny $U,0$}}\{(W-x)/h_1\}|X]=K_1\{(X-x)/h_1\}$. A direct implication of this property is that $E\{\hat p(x,y)\}=E\{\tilde p(x,y)\}$.

Differentiating (\ref{eq:jointpdfest}) with respect to (w.r.t.) $y$ yields an estimator of $p_y(x,y)$ based on $(\bW,\bY)$,   
\begin{equation}
\hat p_y(x, y)=\frac{-1}{nh_1 h^2_2} \sum_{j=1}^n K_{\hbox {\tiny $U,0$}}\left(\frac{W_j-x}{h_1}\right)K'_2\left(\frac{Y_j-y}{h_2}\right), \label{eq:py1}
\end{equation} 
where $K'_2(t)=(d/dt)K_2(t)$. To facilitate mode seeking, we choose $K_2(t)$ to be a radially symmetric kernel, that is, $K_2(t)=c_2k_2(t^2)$, in which $k_2(s)$ is a nonnegative-valued function, referred to as the profile of $K_2(t)$, and $c_2$ is a positive constant serving as a normalization constant so that $K_2(t)$ integrates to one. Furthermore, we choose $K_2(t)$ such that its profile $k_2(s)$ relates to the profile of another radially symmetric kernel $K_3(t)=c_3 k_3(t^2)$ via $k_3(s)=-k_2'(s)=-(d/ds)k_2(s)$. The Epanechnikov kernel and the normal kernel are examples where a kernel possesses the above desirable features for $K_2(t)$. Using the so-chosen $K_2(t)$, (\ref{eq:py1}) can be further elaborated as  
\begin{equation*}
\hat p_y(x, y)=\frac{2c_2}{nh_1 h_2^2 c_3} \sum_{j=1}^n K_{\hbox {\tiny $U,0$}}\left(\frac{W_j-x}{h_1}\right)K_3\left(\frac{Y_j-y}{h_2}\right)\left(\frac{Y_j-y}{h_2}\right).
\end{equation*} 
For illustration purposes, throughout the article, we set $K_2(t)=\exp(-t^2/2)/\sqrt{2\pi}$. Then, with $c_2=1/(2\sqrt{2\pi})$ and $k_2(s)=2\exp(-s/2)$, one has $K_3(t)=K_2(t)$, with $c_3=1/\sqrt{2\pi}$ and $k_3(s)=\exp(-s/2)$, and the above estimator becomes 
\begin{equation}
\hat p_y(x, y)=\frac{1}{nh_1 h_2^2} \sum_{j=1}^n K_{\hbox {\tiny $U,0$}}\left(\frac{W_j-x}{h_1}\right)K_2\left(\frac{Y_j-y}{h_2}\right)\left(\frac{Y_j-y}{h_2}\right).
\label{eq:py3}
\end{equation}

Based on $\hat p_y(x,y)$ in (\ref{eq:py3}), an estimator of $y_{\hbox {\tiny $M$}}(x)$ is the solution to the equation 
\begin{equation*}
\sum_{j=1}^n K_{\hbox {\tiny $U,0$}}\left(\frac{W_j-x}{h_1}\right)K_2\left(\frac{Y_j-y}{h_2}\right)(Y_j-y)=0.
\end{equation*} 
Rearranging terms in this equation yields a variant of the equation leading to the following updating formula that one evaluates iteratively until convergence in order to find a solution to it, 
\begin{equation}
y^{(k+1)}=\frac{\displaystyle{\sum_{j=1}^n K_{\hbox {\tiny $U,0$}} \left(\frac{W_j-x}{h_1}\right) K_2\left(\frac{Y_j-y^{(k)}}{h_2}\right)Y_j}}{\displaystyle{\sum_{j=1}^n K_{\hbox {\tiny $U,0$}} \left(\frac{W_j-x}{h_1}\right) K_2\left(\frac{Y_j-y^{(k)}}{h_2}\right)}},
\label{eq:iterate0}
\end{equation}
where $y^{(k+1)}$ is the value resulting from the $(k+1)$th iteration as an update of the value from the previous iteration, $y^{(k)}$, in search for $\hat y_{\hbox {\tiny $M0$}}(x)$, for $k=0, 1, \ldots$. One may view 
$$\omega_j^{(k)}=\frac{K_{\hbox {\tiny $U,0$}}\{(W_j-x)/h_1\}K_2\{(Y_j-y^{(k)})/h_2\}}{\sum_{j'=1}^n K_{\hbox {\tiny $U,0$}}\{(W_{j'}-x)/h_1\}K_2\{(Y_{j'}-y^{(k)})/h_2\}}$$
as a weight associated with the $j$th data point $(W_j, Y_j)$, for $j=1, \ldots, n$, of which the magnitude depends on the proximity of this data point to $(x, y^{(k)})$. Following this viewpoint, one can see that the right-hand side of (\ref{eq:iterate0}) is a weighted average of $\bY$, and one may interpret this updating formula as updating $y^{(k)}$ to $y^{(k+1)}$ using the weighted mean of the responses surrounding $(x, y^{(k)})$. In fact, this algorithm of finding an estimated mode is in line with the mean-shift algorithm used to search for modes of a distribution \citep{Cheng95, Comaniciu2002, Einbeck2006, Chen2016}. Compared to the existing mean-shift algorithm and its application, the complication caused by measurement error is that the weight $\omega_j^{(k)}$ can be negative because the deconvoluting kernel $K_{\hbox {\tiny $U,0$}}(\cdot)$ is not guaranteed to be nonnegative \citep{Stefanski90}. However, with careful choices of the bandwidths and starting values for the algorithm, as to be elaborated in Sections~\ref{s:implementation} and \ref{s:empirical}, our mean-shift algorithm can converge and produce a mode estimate $\hat y_{\hbox {\tiny $M0$}}(x)$. 

We call the so-obtained estimator $\hat y_{\hbox {\tiny $M0$}}(x)$ a local constant estimator of the mode because of the construction of $\hat p(x,y)$, which in nature is a local constant estimator of $p(x,y)$. This interpretation of $\hat p(x,y)$ is made clearer when compared to the way we estimate $p(y|x)$ in order to estimate the mode. 

\subsection{Local linear estimator}
\label{s:p1}
The second estimator of $y_{\hbox {\tiny $M$}}(x)$ we propose, denoted by $\hat y_{\hbox {\tiny $M1$}}(x)$, is a solution to $\hat p_y(y|x)=0$, where $\hat p_y(y|x)$ is an estimator of $p_y(y|x)$ obtained as follows. We start from evoking the local linear estimator of $p(y|x)$ in the absence of measurement error proposed by \citet{Fan96} given by 
\begin{equation}
\tilde p(y|x)=\be_1^\T \bS^{-1}_n(x) \bT_n(x,y),
\label{eq:cpdfnome}
\end{equation}
where $\be_1=(1,\, 0)^\T$, 
\begin{equation*}
\bS_n(x) = 
\begin{bmatrix} 
S_{n,0}(x) & S_{n,1}(x)\\
S_{n,1}(x) & S_{n,2}(x)
\end{bmatrix}, \, \,\,
\bT_n(x,y) = \left[T_{n,0}(x,y), \, T_{n,1}(x,y) \right]^\T,
\end{equation*}
in which 
\begin{align}
S_{n,\ell} (x) & = \frac{1}{nh_1}\sum_{j=1}^n \left(\frac{X_j-x}{h_1}\right)^\ell K_{1}\left(\frac{X_j-x}{h_1}\right), \textrm{ for $\ell=0, 1, 2$,} \label{eq:Snome}\\
T_{n, \ell}(x,y) & = \frac{1}{nh_1h_2}\sum_{j=1}^n \left(\frac{X_j-x}{h_1}\right)^\ell K_1\left(\frac{X_j-x}{h_1}\right)K_2\left(\frac{Y_j-y}{h_2}\right), \textrm{ for $\ell=0, 1$.} \label{eq:Tnome}
\end{align}
This estimator is motivated by the property that $E[K_2\{(Y-y)/h_2\}/h_2|X=x]\approx p(y|x)$ as $h_2\to 0$, and hence $p(y|x)$ can be approximately viewed as the mean function when regressing $K_2\{(Y-y)/h_2\}/h_2$ on $X$. Adopting this standpoint, one can employ the general strategy of estimating a mean function via local polynomial estimators \citep{Fanbook} to estimate $p(y|x)$, with $\{K_2\{(Y_j-y)/h_2\}/h_2\}_{j=1}^n$ being the response data and $\{X_j\}_{j=1}^n$ as the covariate data. In particular, the local linear estimator of $p(y|x)$ is as in (\ref{eq:cpdfnome}). 

In the presence of measurement error, we adjust $\tilde p(y|x)$ to account for measurement error in $X$ and propose the following local linear estimator of $p(y|x)$,  
\begin{equation}
\hat p(y|x)=\be_1^\T \hat \bS^{-1}_n(x) \hat\bT_n(x,y), \label{eq:cpdfme}
\end{equation}
where  
\begin{equation*}
\hat \bS_n(x) = 
\begin{bmatrix}
\hat S_{n,0}(x) & \hat S_{n,1}(x)\\
\hat S_{n,1}(x) & \hat S_{n,2}(x)
\end{bmatrix}, \,\,\,
\hat \bT_n(x,y) = \left[\hat T_{n,0}(x,y), \, \hat T_{n,1}(x,y) \right]^\T,
\end{equation*}
in which 
\begin{align}
\hat S_{n,\ell} (x) = & \frac{1}{n h_1}\sum_{j=1}^n K_{\hbox {\tiny $U$},\ell}\left(\frac{W_j-x}{h_1}\right), \textrm{ for $\ell=0, 1, 2$,} \label{eq:Sme}\\
\hat T_{n, \ell}(x,y) = & \frac{1}{n h_1h_2}\sum_{j=1}^n K_{\hbox {\tiny $U$},\ell}\left(\frac{W_j-x}{h_1}\right)K_2\left(\frac{Y_j-y}{h_2}\right), \textrm{ for $\ell=0, 1$,} \label{eq:Tme}
\end{align}
and, with $\phi^{(\ell)}_{\hbox {\tiny $K_1$}}(s)$ denoting the $\ell$-th derivative of $\phi_{\hbox {\tiny $K_1$}}(s)$,
\begin{equation}
K_{\hbox {\tiny $U$},\ell}(t)=i^{-\ell}\frac{1}{2\pi}\int e^{-its} \frac{\phi^{(\ell)}_{\hbox {\tiny $K_1$}}(s)}{\phi_{\hbox {\tiny $U$}}(s/h_1)}ds, \textrm{ for $\ell=0, 1, 2$.}\label{eq:KU1}
\end{equation}
The transform of $K_1(\cdot)$ in (\ref{eq:KU1}) is a generalization of the transform in (\ref{eq:KU0}) derived in \citet{Delaigle09}. This generalization leads to a generalized deconvoluting kernel $K_{\hbox {\tiny $U$}, \ell}(t)$ possessing the property that 
$E[K_{\hbox {\tiny $U$}, \ell}\{(W-x)/h_1\}|X]=\{(X-x)/h_1\}^\ell K_1\{(X-x)/h_1\}$, $\ell=0, 1, 2, \ldots$. Thanks to this property, given $(\bX,\bY)$, (\ref{eq:Sme}) and (\ref{eq:Tme}) are unbiased estimators of their counterparts in the absence of measurement error in (\ref{eq:Snome}) and (\ref{eq:Tnome}), respectively. Hence, $\hat p(y|x)$ is a sensible counterpart estimator of $\tilde p(y|x)$ in the presence of measurement error. 

Using (\ref{eq:cpdfme}), an estimator of $p_y(y|x)$ follows immediately by differentiating $\hat p(y|x)$ w.r.t. $y$. This gives  
\begin{equation}
\hat p_y(y|x)= \be_1^\T \hat \bS^{-1}_n(x) \hat\bT'_n(x,y), \label{eq:pyhat}
\end{equation}
where $\hat\bT'_n(x,y)=[\hat T'_{n,0}(x,y), \, \hat T'_{n,1}(x,y)]^\T$, in which, for $\ell=0, 1$,
$$
\hat T'_{n,\ell}(x,y) = \frac{\partial}{\partial y} \hat T_{n,\ell}(x,y) 
 = \frac{1}{n h_1 h_2^2} \sum_{j=1}^n K_{\hbox {\tiny $U$}, \ell}\left(\frac{W_j-x}{h_1}\right)K_2\left(\frac{Y_j-y}{h_2}\right)\left(\frac{Y_j-y}{h_2}\right).$$
Setting $\hat p_y(y|x)=0$ gives an equation to which the solution is the mode estimator $\hat y_{\hbox {\tiny $M1$}}(x)$. Elaborating (\ref{eq:pyhat}) reveals that $\hat y_{\hbox {\tiny $M1$}}(x)$ solves 
\begin{equation*}
\resizebox{1 \textwidth}{!}
{
$\sum_{j=1}^n \left\{K_{\hbox {\tiny $U,0$}}\left(\frac{W_j-x}{h_1}\right)\hat S_{n,2}(x)-K_{\hbox {\tiny $U,1$}}\left(\frac{W_j-x}{h_1}\right)\hat S_{n,1}(x)\right\} K_2\left(\frac{Y_j-y}{h_2}\right)(Y_j-y)=0$}.
\end{equation*}
This suggests the following updating formula one may use iteratively until convergence to find the solution to the equation, 
\begin{equation*}
\resizebox{.95 \textwidth}{!} 
{
$y^{(k+1)}=\frac{\displaystyle{\sum_{j=1}^n \left\{K_{\hbox {\tiny $U,0$}}\left(\frac{W_j-x}{h_1}\right)\hat S_{n,2}(x)-K_{\hbox {\tiny $U,1$}}\left(\frac{W_j-x}{h_1}\right)\hat S_{n,1}(x)\right\} K_2\left(\frac{Y_j-y^{(k)}}{h_2}\right)Y_j}}{\displaystyle{\sum_{j=1}^n \left\{K_{\hbox {\tiny $U,0$}}\left(\frac{W_j-x}{h_1}\right)\hat S_{n,2}(x)-K_{\hbox {\tiny $U,1$}}\left(\frac{W_j-x}{h_1}\right)\hat S_{n,1}(x)\right\} K_2\left(\frac{Y_j-y^{(k)}}{h_2}\right)}}$
}.
\end{equation*}
Like the previous updating formula, the right-hand side of this updating formula can also be viewed as a weighted average of $\bY$, and thus this algorithm of searching for an estimated mode is also in the spirit of the mean-shift algorithm. 

The common theme the above two proposed mode estimation methods follow is to solve an equation of the form $\hat g_y(x,y)=0$, where $\hat g_y(x,y)$ an estimator of $g_y(x,y)$, and $g(x,y)$ is a function such that $g_y\{x, y_{\hbox {\tiny $M$}}(x)\}=0$ and $g_{yy}\{x, y_{\hbox {\tiny $M$}}(x)\}<0$. When $g(x,y)=p(x, y)$, the solution is $\hat y_{\hbox {\tiny $M0$}}(x)$; and, when $g(x,y)=p(y|x)$, solving the equation yields $\hat y_{\hbox {\tiny $M1$}}(x)$. Furthermore, this common theme closely relates to the idea of corrected score \citep{Nakamura1990, Novick2002, LASbook}. More specifically, since $E\{\hat p(x,y)|(\bX, \bY)\}=\tilde p(x,y)$, one has $E\{\hat p_y(x,y)|(\bX, \bY)\}=\tilde p_y(x,y)$. Hence, the estimating equation one solves to obtain $\hat y_{\hbox {\tiny $M0$}}(x)$, i.e., $\hat p_y(x,y)=0$, is the corrected score estimating equation corresponding to $\tilde p_y(x,y)=0$, which is the equation one solves to estimate modes in the absence of measurement error. Although, given $(\bX, \bY)$, $\hat p_y(y|x)$ is not an unbiased estimator of $\tilde p_y(y|x)$, the building blocks in the former are unbiased scores of those in the latter, i.e., $E\{\hat \bS_n(x)|\bX\}=\bS_n(x)$ and $E\{\hat \bT_n(x, y)|(\bX, \bY)\}=\bT_n(x, y)$. When the solution to an equation associated with a method is not unique, the method leads to an estimated mode set, denoted by $\hat M_0(x)$ and $\hat M_1(x)$ for the first and the second method, respectively. In what follows, we present asymptotic properties of these mode estimators. For notational simplicity, we assume $\hat M_0(x)=\{\hat y_{\hbox {\tiny  $M0$}}(x)\}$ and $\hat M_1(x)=\{\hat y_{\hbox {\tiny  $M1$}}(x)\}$ in the next section. Also, $y_{\hbox {\tiny $M$}}$ is often used in place of $y_{\hbox {\tiny $M$}}(x)$ for brevity in the sequel, and $\hat y_{\hbox {\tiny $M$}}$ is used to refer to a mode estimator generically when we do not distinguish between $\hat y_{\hbox {\tiny $M0$}}(x)$ and $\hat y_{\hbox {\tiny $M1$}}(x)$. 

\section{Asymptotic properties}
\label{s:asymptotics}
\subsection{Preliminary}
\label{s:prelim}
We focus on convergence rates of three forms of error associated with $\hat y_{\hbox {\tiny $M$}}$ in this section, first, the pointwise error defined by $\Delta_n(x)=|\hat y_{\hbox {\tiny $M$}}(x)-y_{\hbox {\tiny $M$}}(x)|$; second, the mean integrated squared error (MISE), $\textrm{MISE}(\hat y_{\hbox {\tiny $M$}}) = E\{\int_{\mathscr{X}}\Delta^2_n(x) dx\}$; and third, the uniform error, $\Delta_n=\sup_{x\in \mathscr{X}}\Delta_n(x)$. We show next that the convergence rate of $\Delta_n(x)$ hinges on the bias and variance of $\hat g_y(x,y_{\hbox {\tiny $M$}})$. Given the pointwise error rate, the convergence rate of $\textrm{MISE}(\hat y_{\hbox {\tiny $M$}})$ follows straightforwardly. Under regularity conditions pointed out along the way, the uniform error rate can be established using existing results regarding the uniform consistency of kernel-based estimators \citep{Gine02, Einmahl05, Chen2015}.

Because $\hat g_y(x, \hat y_{\hbox {\tiny $M$}})=0$, by the mean-value theorem, one has
$\hat g_y(x, y_{\hbox {\tiny $M$}})= \hat g_y(x, y_{\hbox {\tiny $M$}})-\hat g_y(x, \hat y_{\hbox {\tiny $M$}})
= (y_{\hbox {\tiny $M$}}-\hat y_{\hbox {\tiny $M$}})\hat g_{yy}(x, y^*)$,
where $y^*$ lies between $y_{\hbox {\tiny $M$}}$ and $\hat y_{\hbox {\tiny $M$}}$. Thus, 
\begin{equation}
\hat y_{\hbox {\tiny $M$}}-y_{\hbox {\tiny $M$}} = -\frac{\hat g_y(x, y_{\hbox {\tiny $M$}})}{\hat g_{yy}(x, y^*)}. \label{eq:diff}
\end{equation}
Provided that $\hat g_{yy}(x, y_{\hbox {\tiny $M$}})$ and $g_{yy}(x, y_{\hbox {\tiny $M$}})$ are bounded away from zero, one has 
\begin{equation*}
|\{\hat g_{yy}(x, y^*)\}^{-1}-\{g_{yy}(x, y_{\hbox {\tiny $M$}})\}^{-1}|=O\left(\|\hat g_{yy} -g_{yy}  \|_\infty \right), 
\end{equation*}
where $\|\hat g_{yy} -g_{yy}  \|_\infty=\sup_{(x,y)\in \mathscr{X}\times \mathscr{Y}}|\hat g_{yy}(x,y)-g_{yy}(x,y)|$.
Then (\ref{eq:diff}) implies that 
\begin{equation}
\hat y_{\hbox {\tiny $M$}}-y_{\hbox {\tiny $M$}}  =  -\{g_{yy}(x, y_{\hbox {\tiny $M$}})\}^{-1}\hat g_y(x, y_{\hbox {\tiny $M$}})+O\left(\|\hat g_{yy} -g_{yy}  \|_\infty \right)\hat g_y(x, y_{\hbox {\tiny $M$}}). 
\label{eq:diffym}
\end{equation}
It follows that  
$\Delta_n(x)=|\{g_{yy}(x, y_{\hbox {\tiny $M$}})\}^{-1}||\hat g_y(x, y_{\hbox {\tiny $M$}})|+ O(\|\hat g_{yy} -g_{yy}  \|_\infty) |\hat g_y(x, y_{\hbox {\tiny $M$}})|$, 
or, equivalently, 
\begin{equation}
\frac{\Delta_n(x)}{|\{g_{yy}(x, y_{\hbox {\tiny $M$}})\}^{-1}||\hat g_y(x, y_{\hbox {\tiny $M$}})|}=1+ O(\|\hat g_{yy} -g_{yy}  \|_\infty)|g_{yy}(x, y_{\hbox {\tiny $M$}})|. 
\label{eq:Delta2}
\end{equation}
Under conditions (CK2), (CK5) and (CK9) in Appendix A, by Lemma 10 in \citet{Chen2015}, $\|\hat g_{yy} -g_{yy}\|_\infty$ converges to zero in probability. Therefore, under these conditions, (\ref{eq:Delta2}) suggests that $\Delta_n(x)$ can be approximated by $|\{g_{yy}(x, y_{\hbox {\tiny $M$}})\}^{-1}||\hat g_y(x, y_{\hbox {\tiny $M$}})|$, and thus the convergence rate of $\Delta_n(x)$ can be revealed through studying the convergence rate of $|\hat g_y(x, y_{\hbox {\tiny $M$}})|$. 

Once we turn to studying the convergence rate of $|\hat g_y(x, y_{\hbox {\tiny $M$}})|$, the bias and variance of $\hat g_y(x, y_{\hbox {\tiny $M$}})$ become highly relevant. This connection can be explained by first noting that $g_y(x, y_{\hbox {\tiny $M$}})=0$, and thus one has 
\begin{eqnarray}
|\hat g_y(x, y_{\hbox {\tiny $M$}})| 
& = & |\hat g_y(x, y_{\hbox {\tiny $M$}})-g_y(x, y_{\hbox {\tiny $M$}})| \nonumber \\
& \le & |\hat g_y(x, y_{\hbox {\tiny $M$}})-E\{\hat g_y(x, y_{\hbox {\tiny $M$}})\}|+|E\{\hat g_y(x, y_{\hbox {\tiny $M$}})\}-g_y(x, y_{\hbox {\tiny $M$}})|\nonumber \\
& = & O_{\hbox {\tiny $P$}}\left[\sqrt{\textrm{Var}\{ \hat g_y(x, y_{\hbox {\tiny $M$}}) \}}\right]+|\textrm{Bias}\{\hat g_y(x, y_{\hbox {\tiny $M$}})\}|. \label{eq:keyconnection}
\end{eqnarray}

The above preliminary asymptotic analysis leads to the road map we follow to study the error rates associated with $\hat y_{\hbox {\tiny $M$}}$, which is to, first, derive the asymptotic bias and variance of $\hat g_y(x, y_{\hbox {\tiny $M$}})$, which leads to the pointwise error rate by (\ref{eq:keyconnection}); second, establish the convergence rate of $\textrm{MISE}(\hat y_{\hbox {\tiny $M$}})$; third, provide the uniform error rate. This is also the order in which we present our theoretical findings for each of the proposed mode estimator in the next two subsections. Supporting materials of these findings are provided in the appendices. In particular, all coded conditions referenced henceforth are in Appendix A.  

\subsection{Convergence rates associated with $\hat y_{\hbox {\tiny $M0$}}(x)$}
\label{s:rate0}
With $g(x,y)=p(x,y)$, we establish the following asymptotic bias result for $\hat p_y(x, y_{\hbox {\tiny $M$}})$, with the proof given in Appendix C. 
\begin{lem}
\label{l:bias0}
Under conditions (CP1) and (CK1)--(CK4), 
\begin{equation}
\textrm{Bias}\{\hat p_y(x, y_{\hbox {\tiny $M$}})\}=0.5\left\{p_{xxy}(x, y_{\hbox {\tiny $M$}})\mu_2^{(1)}h_1^2+p_{yyy}(x, y_{\hbox {\tiny $M$}}) h_2^2\right\}+o(h_1^2+h_2^2),
\label{eq:bias0}
\end{equation}
where $p_{xxy}(x,y)=(\partial^3/\partial x^2 \partial y)p(x,y)$, $p_{yyy}(x,y)=(\partial^3/\partial y^3)p(x,y)$, and $\mu_2^{(1)}=\int t^2 K_1(t)dt$.  
\end{lem}
This bias result coincides with the result in \citet{Chacon2011}, where kernel-based estimators of derivatives of a multivariate joint pdf are considered in the absence of measurement error. This is expected since $E\{\hat p_y(x, y)\}=E\{\tilde p_y(x, y)\}$, as pointed out in Section~\ref{s:localcon}.

The variance of $\hat p_y(x, y_{\hbox {\tiny $M$}})$ depends on the smoothness of the measurement error distribution. There are two levels of smoothness of $f_{\hbox {\tiny $U$}}(u)$ considered in measurement error literature \citep{Fan91a, Fan91b, Fan91c}, ordinary smooth and super smooth. Their definitions are given below. 
\begin{definition}
\label{df:ordsm}
The distribution of $U$ is ordinary smooth of order $b$ if
$$\lim_{t\to +\infty} t^b \phi_{\hbox {\tiny $U$}}(t)=c \textrm{ and } \lim_{t\to +\infty} t^{b+1} \phi'_{\hbox {\tiny $U$}}(t)=-cb$$ 
for some positive constants $b$ and $c$. 
\end{definition}
\begin{definition}
\label{df:supsm}
The distribution of $U$ is super smooth of order $b$ if
$$d_0|t|^{b_0}\exp(-|t|^b/d_2)\le |\phi_{\hbox {\tiny $U$}}(t)| \le d_1|t|^{b_1}\exp(-|t|^b/d_2)  \textrm { as $|t|\to \infty$}$$ 
for some positive constants $d_0$, $d_1$, $d_2$, $b$, $b_0$ and $b_1$.
\end{definition}
In Appendix D, we derive the asymptotic variance of $\hat p_y(x, y_{\hbox {\tiny $M$}})$ and the results are given in the next lemma. 
\begin{lem}
\label{l:var0}
Assume conditions required for Lemma~\ref{l:bias0} hold. 
When $U$ is ordinary smooth of order $b$, under conditions (CU2) and (CK5)--(CK7), if $nh_1^{1+2b}h_2^3\to \infty$, then 
\begin{equation}
\textrm{Var}\{\hat p_y(x, y_{\hbox {\tiny $M$}})\}=\frac{\eta_0f_{\hbox {\tiny $W,Y$}}(x,y_{\hbox {\tiny $M$}})}{4\sqrt{\pi}c^2nh_1^{1+2b}h_2^3} +o\left(\frac{1}{nh_1^{1+2b}h_2^3}\right),
\label{eq:var0ord}
\end{equation}
where $\eta_0=\int |t|^{2b} |\phi_{\hbox {\tiny $K_1$}}(t)|^2dt/(2\pi)$, and $f_{\hbox {\tiny $W,Y$}}(\cdot, \cdot)$ is the joint pdf of $(W, Y)$. 

When $U$ is super smooth of order $b$, under conditions (CK5) and (CK8), if $nh_1^{1-2b_2}h_2^3 \exp(-2h_1^{-b}/{d_2})\to \infty$, then
\begin{equation}
\textrm{Var}\{\hat p_y(x,y_{\hbox {\tiny $M$}})\} \le \frac{\exp\left(2h_1^{-b}/{d_2}\right)}{nh_1^{1-2b_2}h_2^3}Cf_{\hbox {\tiny $W,Y$}}(x,y_{\hbox {\tiny $M$}})+o\left\{\frac{\exp\left(2h_1^{-b}/{d_2}\right)}{nh_1^{1-2b_2}h_2^3}\right\}, 
\label{eq:var0sm}
\end{equation}
where $C$ is some finite positive constant and $b_2=b_0 I(b_0<0.5)$. 
\end{lem}

Putting results in Lemmas~\ref{l:bias0} and \ref{l:var0} together, one has the convergence rate of $|\hat p_y(x, y_{\hbox {\tiny $M$}})|$, which leads to the pointwise error rates summarized in the following theorem. 
\begin{them}
\label{thm:deltanx}
Under the conditions in Lemma~\ref{l:bias0}, Lemma~\ref{l:var0}, and (CK9), 
when $U$ is ordinary smooth,
\begin{equation}
\Delta_n(x) = \frac{|p_{xxy}(x,y_{\hbox {\tiny $M$}}) \mu_2^{(1)}h_1^2+p_{yyy}(x, y_{\hbox {\tiny $M$}})h_2^2|}{-2p_{yy}(x, y_{\hbox {\tiny $M$}})}+O_{\hbox {\tiny $P$}}\left(\sqrt{\frac{1}{nh_1^{1+2b}h_2^3}}\right) +o(h_1^2+h_2^2);  \label{eq:pwe0ord}
\end{equation}
and, when $U$ is super smooth, 
\begin{equation}
\Delta_n(x) = \frac{|p_{xxy}(x,y_{\hbox {\tiny $M$}}) \mu_2^{(1)}h_1^2+p_{yyy}(x, y_{\hbox {\tiny $M$}})h_2^2|}{-2p_{yy}(x, y_{\hbox {\tiny $M$}})}+\displaystyle{O_{\hbox {\tiny $P$}}\left\{\frac{\exp(h_1^{-b}/d_2)}{\sqrt{nh_1^{1-2b_2}h_2^3}}\right\}}+o(h_1^2+h_2^2).
\label{eq:pwe0sup}
\end{equation}
\end{them}
\citet{Chen2016} estimated local modes based on $\tilde p(x,y)$ in (\ref{eq:pdfest}) with $h_1=h_2=h$, and they showed that the pointwise error rate of the resultant mode estimator is of order $O(h^2)+O_{\hbox {\tiny $P$}}\{\sqrt{1/(nh^4)}\}$. Comparing this with (\ref{eq:pwe0ord}) and (\ref{eq:pwe0sup}), one can see that the pointwise error tends to zero much slower with the added complication of measurement error, especially when it is super smooth. Hence, mode estimation in the presence of measurement error is substantially more challenging, as one would expect with noisier data. 

The convergence rate of $\textrm{MISE}(\hat y_{\hbox {\tiny $M0$}})$ can also be deduced from Lemmas~\ref{l:bias0} and \ref{l:var0}. To see this more clearly, first note that, assuming interchangeability of expectation and integration, one has $\textrm{MISE}(\hat y_{\hbox {\tiny $M$}}) = \int_{\mathscr{X}} \left\{ \textrm{Bias}^2(\hat y_{\hbox {\tiny $M$}}) +\textrm{Var}(\hat y_{\hbox {\tiny $M$}})\right\} dx$. As elaborated in Appendix E, the dominating terms in the integrated squared bias of $\hat y_{\hbox {\tiny $M0$}}$, $\int_{\mathscr{X}} \textrm{Bias}^2(\hat y_{\hbox {\tiny $M0$}})dx$, can be easily derived based on $\textrm{Bias}\{\hat p_y(x, y_{\hbox {\tiny $M$}})\}$; and the dominating terms in the integrated variance of $\hat y_{\hbox {\tiny $M0$}}$, $\int_{\mathscr{X}} \textrm{Var}(\hat y_{\hbox {\tiny $M0$}})dx$, can be deduced from $\textrm{Var}\{\hat p_y(x, y_{\hbox {\tiny $M$}}) \}$. Combining these dominating terms, we reach the following conclusion regarding $\textrm{MISE}(\hat y_{\hbox {\tiny $M0$}})$. 
\begin{them}
\label{thm:mise0}
Under conditions in Theorem~\ref{thm:deltanx}, and assume that $p_{xxy}(x,y_{\hbox {\tiny $M$}})$ and $p_{yyy}(x,y_{\hbox {\tiny $M$}})$ are square integrable, then,  when $U$ is ordinary smooth,  
$$\textrm{MISE}(\hat y_{\hbox {\tiny $M0$}})= O\{(h_1^2+h_2^2)^2\}+O\left(\frac{1}{nh_1^{1+2b}h_2^3}\right);$$
when $U$ is super smooth, 
$$\textrm{MISE}(\hat y_{\hbox {\tiny $M0$}})= O\{(h_1^2+h_2^2)^2\}+O\left\{\frac{\exp\left(2h_1^{-b}/{d_2}\right)}{nh_1^{1-2b_2}h_2^3}\right\}.$$
\end{them}

Moving to the uniform error, $\Delta_n$, it is helpful to note that, by (\ref{eq:Delta2}), (\ref{eq:keyconnection}), and Lemma~\ref{l:bias0}, one has 
\[\Delta_n(x) = |\{p_{yy}(x, y_{\hbox {\tiny $M$}})\}^{-1}|\left[|\hat p_y(x, y_{\hbox {\tiny $M$}})-E\{\hat p_y(x, y_{\hbox {\tiny $M$}})\}|+O(h_1^2+h_2^2)\right]+ o_{\hbox {\tiny $P$}}(1).\]
Thus
\begin{equation*}
\Delta_n = \sup_{x\in \mathscr{X}} |\{p_{yy}(x, y_{\hbox {\tiny $M$}})\}^{-1}||\hat p_y(x, y_{\hbox {\tiny $M$}})-E\{\hat p_y(x, y_{\hbox {\tiny $M$}})\}|+O(h_1^2+h_2^2)+ o_{\hbox {\tiny $P$}}(1). 
\end{equation*}
Following similar methods in \citet{Gine02} and \citet{Einmahl05}, one can establish the following result regarding $\Delta_n$.
\begin{them}
\label{thm:uniferr0}
Under conditions in Theorem~\ref{thm:deltanx} and (CP2), for ordinary smooth $U$, 
$$\Delta_n=O(h_1^2+h_2^2)+O_{\hbox {\tiny $P$}}\left(\sqrt{\frac{\log n}{nh_1^{1+2b}h_2^3}}\right);$$ 
and, for super smooth $U$, 
$$\Delta_n=O(h_1^2+h_2^2)+O_{\hbox {\tiny $P$}}\left(\sqrt{\frac{\exp(h_1^{-b}/d_2)\log n}{nh_1^{1-2b_2}h_2^3}}\right).$$ 
\end{them}
The uniform error rate associated with the mode estimator with $h_1=h_2=h$ considered in \citet{Chen2016} in the absence of measurement error is $O(h^2)+O_{\hbox {\tiny $P$}}\{\sqrt{\log n/(nh^4)}\}$. Compared with their result, Theorem~\ref{thm:uniferr0} suggests the inevitable compromise in convergence rate due to measurement error. 

\subsection{Convergence rates associated with $\hat y_{\hbox {\tiny $M1$}}(x)$}
\label{s:rate1}
With $g(x,y)=p(y|x)$, we present in Appendix F the bias analysis and in Appendix G the variance analysis of $\hat p_y(y_{\hbox {\tiny $M$}}|x)$. Results from these analyses are summarized in the following two lemmas. 
\begin{lem}
\label{l:bias1} 
When $U$ is ordinary smooth, assume (CK4), (CK5), (CX1), and $(nh_1^{1+2b}h_2^3)^{-1/2}=O(h_1^4h_2^{-1}+h_2^3)$; when $U$ is super smooth, assume (CK4), (CK6), (CX1), (CP1), and $\exp (h_1^{-b}/d_2)(nh_1^{1-2b_2} h_2^3)^{-1/2}=O(h_1^4h_2^{-1}+h_2^3),$ then one has 
\begin{eqnarray}
 \textrm{Bias}\{\hat p_y(y_{\hbox {\tiny $M$}}|x) \}
& = & f_{\hbox {\tiny $X$}}^{-1}(x) \left[ \left\{0.5p_{xxy}(x,y_{\hbox {\tiny $M$}})- f'_{\hbox {\tiny $X$}}(x)p_{xy}(y_{\hbox {\tiny $M$}}|x)\right\}\mu_2^{(1)}h_1^2\right. \nonumber \\
& & \left.+0.5p_{yyy}(x,y_{\hbox {\tiny $M$}})h_2^2\right] +O(h_1^4h_2^{-1}+h_2^3),\label{eq:bias1}
\end{eqnarray}
where $f_{\hbox {\tiny $X$}}(x)$ is the pdf of $X$ and $f'_{\hbox {\tiny $X$}}(x)=(d/dx)f_{\hbox {\tiny $X$}}(x)$. 
\end{lem}
\begin{lem}
\label{l:var1}
Under the conditions in Lemma~\ref{l:bias1}, when $U$ is ordinary smooth, if $nh_1^{1+2b}h_2^3\to \infty$, then 
\begin{equation}
\textrm{Var}\left\{ \hat p_y(y_{\hbox {\tiny $M$}}|x)  \right\}
=\frac{\eta_0 f_{\hbox {\tiny $W,Y$}}(x,y_{\hbox {\tiny $M$}})}{4\sqrt{\pi}c^2 nh_1^{1+2b} h_2^3 f^2_{\hbox {\tiny $X$}}(x)}+o\left(\frac{1}{nh_1^{1+2b}h_2^3}\right); \label{eq:var1ord}
\end{equation}  
and, when $U$ is super smooth, if $nh_1^{1-2b_2}h_2 \exp(-2h_1/d_2)\to \infty$, then 
\begin{equation}
\textrm{Var}\left\{ \hat p_y(y_{\hbox {\tiny $M$}}|x)  \right\}
\le \frac{C\exp\left(2h_1^{-b}/d_2\right)}{nh_1^{1-2b_2}h_2^3f^2_{\hbox {\tiny $X$}}(x)}+O\left\{\frac{\exp\left(2h_1^{-b}/d_2\right)}{nh_1^{1-2b_2}h_2}\right\}, \label{eq:var1sup}
\end{equation}	
where $C$ is some finite positive constant. 
\end{lem}

Although the order of $\textrm{Bias}\{\hat p_y(y_{\hbox {\tiny $M$}}|x)\}$ in (\ref{eq:bias1}) and that of $\textrm{Bias}\{\hat p_y(x, y_{\hbox {\tiny $M$}})\}$ in (\ref{eq:bias0}) are both $O(h_1^2+h_2^2)$, the dependence of the dominating term on $f_{\hbox {\tiny $X$}}^{-1}(x)$ shown in (\ref{eq:bias1}) indicates that estimating $\hat p_y(y_{\hbox {\tiny $M$}}|x)$ at the (diminishing) tail of the $X$-distribution can be challenging. The residual term in (\ref{eq:bias1}) suggests that we need $h_1^4h_2^{-1}\to 0$ in order for $\hat p_y(y_{\hbox {\tiny $M$}}|x) $, and thus for $\hat y_{\hbox {\tiny $M1$}}$, to be consistent. A sufficient condition for $h_1^4h_2^{-1}\to 0$ is to have $h_1\to 0$ faster than $h_2\to 0$ as $n\to \infty$. This may indicate that a sensible bandwidth selection procedure tends to choose $h_1<h_2$, which is indeed observed in our simulation study when we apply the data-driven bandwidth selection method described in Section~\ref{s:implementation}.

Comparing (\ref{eq:var1ord}) with (\ref{eq:var0ord}), one can see that the dominating variance of $\hat p_y(y_{\hbox {\tiny $M$}}|x)$ can be higher than that of $\hat p_y(x, y_{\hbox {\tiny $M$}})$, and the former can be large at the (diminishing) tail of $f_{\hbox {\tiny $X$}}(x)$. This suggests that estimating $p_y(y_{\hbox {\tiny $M$}}|x)$ via $\hat p_y(y_{\hbox {\tiny $M$}}|x)$, and thus estimating $y_{\hbox {\tiny $M$}}(x)$ via $\hat y_{\hbox {\tiny $M1$}}(x)$, will be subject to high uncertainty if data surrounding $x$ are scarce. 

Based on these bias and variance results, we establish the following theorem regarding the pointwise error rate of $\hat y_{\hbox {\tiny $M1$}}(x)$. 
\begin{them}
\label{thm:pwe1}
Under conditions in Lemma~\ref{l:bias1}, Lemma~\ref{l:var1}, (CK2) and (CK9), when $U$ is ordinary smooth, 
\begin{eqnarray}
\Delta_n(x) & = & 
\frac{|\left\{p_{xxy}(x,y_{\hbox {\tiny $M$}})-2f'_{\hbox {\tiny $X$}}(x)p_{xy}(y_{\hbox {\tiny $M$}}|x)   \right\}\mu_2^{(1)}h_1^2+p_{yyy}(x, y_{\hbox {\tiny $M$}})h_2^2|}{-2f_{\hbox {\tiny $X$}}(x)p_{yy}(y_{\hbox {\tiny $M$}}|x)} \nonumber \\
& & +O_{\hbox {\tiny $P$}}\left(\sqrt{\frac{1}{nh_1^{1+2b}h_2^3}}\right)+O(h_1^4h_2^{-1}+h_2^3);
\label{eq:pwe1ord}
\end{eqnarray}
and, when $U$ is super smooth, 
\begin{eqnarray}
\Delta_n(x) & = & 
\frac{|\left\{p_{xxy}(x,y_{\hbox {\tiny $M$}})-2f'_{\hbox {\tiny $X$}}(x)p_{xy}(y_{\hbox {\tiny $M$}}|x)   \right\}\mu_2^{(1)}h_1^2+p_{yyy}(x, y_{\hbox {\tiny $M$}})h_2^2|}{-2f_{\hbox {\tiny $X$}}(x)p_{yy}(y_{\hbox {\tiny $M$}}|x)} \nonumber \\
& & +O_{\hbox {\tiny $P$}}\left\{\frac{\exp(h_1^{-b}/d_2)}{\sqrt{nh_1^{1-2b_2}h_2^3}}\right\} +O(h_1^4h_2^{-1}+h_2^3). \label{eq:pwe1sup}
\end{eqnarray}  
\end{them}

Following the same line of arguments as those leading to $\textrm{MISE}(\hat y_{\hbox  {\tiny $M0$}})$ in Section~\ref{s:rate0}, we show the same convergence rate for $\textrm{MISE}(\hat y_{\hbox  {\tiny $M1$}})$ as those stated in Theorem~\ref{thm:mise0} under slightly different conditions. Now we need to assume all conditions stated in Lemmas~\ref{l:bias1} and \ref{l:var1}, (CP2), and that $p_{xxy}(x, y_{\hbox {\tiny $M$}})$, $p_{yyy}(x, y_{\hbox {\tiny $M$}})$, and $p_{xy}(y_{\hbox {\tiny $M$}}|x)$ are square integrable. 

Finally, for the uniform error $\Delta_n$ associated with $\hat y_{\hbox {\tiny $M1$}}$, using the elaboration of $\hat p_y(y_{\hbox {\tiny $M$}}|x)$ in Appendix F (Section F.1 to be specific), one can see that the dominating term of $\hat p_y(y_{\hbox {\tiny $M$}}|x)$ is simply $\hat p_y(x, y_{\hbox {\tiny $M$}})$ divided by $f_{\hbox {\tiny $X$}}(x)$. Hence, the convergence rate of $\Delta_n$ associated with $\hat y_{\hbox {\tiny $M1$}}$ is the same as that of $\Delta_n$ associated with $\hat y_{\hbox {\tiny $M0$}}$ stated in Theorem~\ref{thm:uniferr0} under the same set of conditions, in addition to the assumption that $f_{\hbox {\tiny $X$}}(x)$ is bounded away from zero over the range of $x$ of interest, $[x_{\hbox {\tiny $L$}}, x_{\hbox {\tiny $U$}}]$. 

\subsection{Asymptotic optimal bandwidths}
\label{s:opth}
With the asymptotic error rates of the proposed mode estimators provided in Sections~\ref{s:rate0} and \ref{s:rate1}, the asymptotic optimal (in some sense) bandwidths are readily available. In particular, taking MISE as the metric to optimize w.r.t. $\bh=(h_1, h_2)^\T$, we show that, for both proposed mode estimators, the optimal rate of MISE($\hat y_{\hbox {\tiny $M$}}$) is of order $O(h_1^4)$ for both ordinary smooth $U$ and super smooth $U$. The orders of the asymptotic optimal $h_1$ and $h_2$ (as $n \to \infty$) are given in a corollary next, where ``$\asymp$" refers to ``tending to zero or infinity at the same rate." Note that explicit expressions of the asymptotic optimal $\bh$ are not available except for $\hat y_{\hbox {\tiny $M0$}}$ when $U$ is ordinary smooth.  
\begin{coro}
\label{cor:optimalh}
Under conditions in Theorem~\ref{thm:mise0}, when $U$ is ordinary smooth of order $b$, the asymptotic optimal bandwidths for $\hat y_{\hbox {\tiny $M0$}}$ satisfy $h_1=r_1h_2$ and $h_2=r_2 n^{-1/(2b+8)}$, where 
\begin{eqnarray*}
r_1 & = & \left\{\frac{(b-1)I_1 + \sqrt{(b-1)^2 I_1^2+3(2b+1)I_2I_3}}{3\mu_2^{(1)} I_2}\right\}^{1/2}, \\
r_2 & = & \left\{\frac{3 \eta_0 I_4}{4\sqrt{\pi}c^2 r_1^{2b+1} (r_1^2 \mu_2^{(1)}I_1+I_3)}\right\}^{1/(2b+8)},
\end{eqnarray*}
in which 
\begin{eqnarray*}
I_1 & = &\int_\mathscr{X} p^{-2}_{yy}(x, y_{\hbox {\tiny $M$}})p_{xxy}(x, y_{\hbox {\tiny $M$}})p_{yyy}(x, y_{\hbox {\tiny $M$}})dx, \\
I_2 & = &\int_\mathscr{X} p^{-2}_{yy}(x, y_{\hbox {\tiny $M$}})p^2_{xxy}(x, y_{\hbox {\tiny $M$}})dx, \\
I_3 & = &\int_\mathscr{X} p^{-2}_{yy}(x, y_{\hbox {\tiny $M$}})p^2_{yyy}(x, y_{\hbox {\tiny $M$}})dx,\\
I_4 & = &\int_\mathscr{X} p^{-2}_{yy}(x, y_{\hbox {\tiny $M$}})f_{\hbox {\tiny $W,Y$}}(x, y_{\hbox {\tiny $M$}})dx.
\end{eqnarray*}
When $U$ is super smooth of order $b$, the asymptotic optimal bandwidths for $\hat y_{\hbox {\tiny $M0$}}$ satisfy $h_1 \asymp h_2^{2/(b+2)}$ and $\exp(2h_1^{-b}/d_2)/h_1^{3b/2-2b_2+6} \asymp n$. The rates of the asymptotic optimal $h_1$ and $h_2$ for $\hat y_{\hbox {\tiny $M1$}}$ are the same as those for $\hat y_{\hbox {\tiny $M0$}}$ under each type of $U$. The corresponding optimal rate of MISE($\hat y_{\hbox {\tiny $M$}}$) is of order $O(h_1^4)$ under each type of $U$ for both $\hat y_{\hbox {\tiny $M0$}}$ and $\hat y_{\hbox {\tiny $M1$}}$.
\end{coro}

These rates of the asymptotic optimal bandwidths for mode estimators are also the rates of the asymptotic optimal bandwidths for the corresponding density derivative estimators. This is not surprising considering the connection between the proposed mode estimators and density derivative estimators pointed out in Section~\ref{s:prelim}. For instance, when $U$ follows a Laplace distribution, which is ordinary smooth of order $b=2$, the asymptotic optimal $h_1$ and $h_2$ are of order $O(n^{-1/12})$ for $\hat p_y(x, y)$, and the corresponding optimal MISE of $\hat p_y(x, y)$ is of order $O(h_1^4)=O(n^{-1/3})$. In the absence of measurement error, \citet[][Theorem 3]{Chacon2011} showed that the asymptotic optimal bandwidths for the kernel-based estimator of $p_y(x, y)$ is of order $O(n^{-1/4})$, and the corresponding optimal MISE of the density derivative estimator is of order $O(h_1^2)=O(n^{-1/2})$. This comparison highlights that measurement error inflate the optimal MISE rate of density derivative estimators, and also lead to much larger optimal bandwidths.    

\section{Bandwidth selection}
\label{s:implementation}
The choice of bandwidths can noticeably affect finite sample performance of almost all kernel-based estimators. The explicit expression of the asymptotic optimal $\bh$ for $\hat y_{\hbox {\tiny $M0$}}$ in Corollary~\ref{cor:optimalh} is not ready to use for choosing bandwidths given a finite sample until reliable estimators of unknown quantities, such as $I_1$, $I_2$, and $I_3$, are available. In this section we present a strategy of choosing $\bh$ that mostly follows the idea of incorporating cross validation (CV) and simulation extrapolation \citep[SIMEX,][]{Cook1994} proposed by \citet{Delaigle08}. This strategy is based on the CV method of choosing bandwidth for estimating conditional density in the absence of measurement error developed by \cite{Fan.Yim2004} and \cite{Hall.etal2004}. These authors constructed a CV criterion based on the weighted integrated squared error (ISE) associated with a kernel-based estimator of $p(y|x)$, $\tilde p(y|x)$, defined by  
\begin{eqnarray*}
\textrm{ISE} & = & \int_{\mathscr{Y}} \int_{\mathscr{X}} \{ \tilde{p}(y|x)-p(y|x) \}^2 f_{\hbox {\tiny $X$}}(x)\omega(x)dxdy \\
& = & \int_{\mathscr{Y}} \int_{\mathscr{X}} \tilde{p}(y|x)^2 f_{\hbox {\tiny $X$}}(x)\omega(x)dxdy - 2 \int_{\mathscr{Y}} \int_{\mathscr{X}} \tilde{p}(y|x) p(y|x) f_{\hbox {\tiny $X$}}(x)\omega(x) dxdy \\
& & + \int_{\mathscr{Y}} \int_{\mathscr{X}}p(y|x)^2 f_{\hbox {\tiny $X$}}(x)\omega(x)dxdy,
\end{eqnarray*}
where $\omega(\cdot)$ is a nonnegative weight function used to avoid estimating $p(y|x)$ at an $x$ around which data are scarce. Given the observed data $\bX$, a reasonable choice of $\omega(\cdot)$ is simply $\omega(x) = I(x\in [x_{\hbox {\tiny $L$}}, x_{\hbox {\tiny $U$}}])$, where $x_{\hbox {\tiny $L$}}$ and $x_{\hbox {\tiny $U$}}$ are the 2.5th and 97.5th percentile of $\bX$, respectively, and $I(\cdot)$ is the indicator function. Noting that the third term in the above elaboration of ISE does not depend on $\bh$ and thus can be ignored when minimizing ISE with respect to $\bh$, they proposed the following estimator of the first two terms as a CV criterion, 
\begin{equation}\label{cv:density}
\textrm{CV}(\tilde{p},  \bh, \bX, \bY, \omega) = \frac{1}{n}\sum_{j=1}^{n}\omega(X_j)\int_{\mathscr{Y}}\tilde{p}_{-j}(y|X_j)^2dy - \frac{2}{n}\sum_{j=1}^{n}\omega(X_j)\tilde{p}_{-j}(Y_j|X_j),
\end{equation} 
where $\tilde{p}_{-j}(y|X_j)$ is the estimate of $p(y|X_j)$ based on data $(\bX, \bY)$ excluding $(X_j, Y_j)$, for $j=1, \ldots, n$. It is worth pointing out that, if one uses the kernel density estimator of $p(y|x)$ with the kernel associated with $y$, i.e., $K_2(\cdot)$, being the standard normal pdf, then the integral in (\ref{cv:density}) can be derived explicitly. 

Clearly, the components in (\ref{cv:density}) that involve $X_j$ cannot be evaluated in the presence of measurement error. To account for measurement error in $\bX$ (but not in $\bY$), we first set $h_2$ at $\hat{h}_2=1.06s_{\hbox {\tiny $Y$}}n^{-1/5}$ according to the normal reference rule \citep{Silverman1986}, where $s_{\hbox {\tiny $Y$}}$ is the sample standard deviation of $\bY$; then we adopt the CV-SIMEX method to find an approximation of 
\begin{equation}
h_1=\argmin_{h_1>0}\textrm{CV}(\hat{p}, \bh, \bX, \bY, \omega),
\label{eq:h1}
\end{equation}
where $\hat p$ denotes the estimate of $p(y|x)$ based on $(\bW, \bY)$. Implementation of the CV-SIMEX method involves the following steps. 
\begin{description}
\item[Step 1:] Generate $B$ sets of further contaminated data according to $\bW_b^* = \bW+\bU_b^*$, where $\bU_b^{*}=\{U_{b,j}^*\}_{j=1}^n$ are i.i.d. from $f_{\hbox {\tiny $U$}}(u)$, for $b=1, \ldots, B$. 
\item[Step 2:] Viewing $\bW$ as the unobserved true covariate values, and $\bW_b^*$ as the error-contaminated surrogate of $\bW$, find 
\begin{equation}
h_1^{*}=\argmin_{h_1>0} \frac{1}{B}\sum_{b=1}^B\textrm{CV}(\hat{p}_b^*, \bh, \bW, \bY, \tilde{\omega}),
\label{eq:h1*}
\end{equation}
where $\hat{p}_b^*$ is the estimate of $p(y|x)$ based on $(\bW_{b}^*, \bY)$, and $\tilde{\omega}(w)=I(w\in [w_{\hbox {\tiny $L$}}, w_{\hbox {\tiny $U$}}])$, with $w_{\hbox {\tiny $L$}}$ and $w_{\hbox {\tiny $U$}}$ being the 2.5th and 97.5th percentile of $\bW$, respectively.
\item[Step 3:] Generate another $B$ sets of further contaminated data, $\bW_{b}^{**} = \bW_b^{*}+\bU_b^{**}$, where $\bU_b^{**}=\{U_{b,j}^{**}\}_{j=1}^n$ are i.i.d. from $f_{\hbox {\tiny $U$}}(u)$, for $b=1, \ldots, B$. 
\item[Step 4:] Viewing $\bW_b^*$ as the unobserved true covariate values, and $\bW_b^{**}$ as the the error-contaminated surrogate of $\bW_b^*$, find 
\begin{equation}
h_1^{**}=\argmin_{h_1>0} \frac{1}{B}\sum_{b=1}^B \textrm{CV}(\hat{p}_b^{**}, \bh, \bW^*_b, \bY, \tilde{\omega}_b),
\label{eq:h1**}
\end{equation}
where $\hat p_b^{**}$ is the estimate of $p(y|x)$ based on $(\bW_{b}^{**}, \bY)$, and $\tilde{\omega}_b(w)=I(w\in [w^*_{\hbox {\tiny $Lb$}}, w^*_{\hbox {\tiny $Ub$}}])$, with $w^*_{\hbox {\tiny $Lb$}}$ and $w^*_{\hbox {\tiny $Ub$}}$ being the 2.5th and 97.5th percentile of $\bW_b^*$, respectively.
\item[Step 5:] Set $\hat h_1=h_1^{*2}/h_1^{**}$ as the final choice of $h_1$ for estimating $p(y|x)$ based on $(\bW, \bY)$. 
\end{description}

The rationale behind the CV-SIMEX method is that the way $h_1^*$ in (\ref{eq:h1*}) relates to $h_1$ in (\ref{eq:h1}) is similar to the way $h_1^{**}$ (\ref{eq:h1**}) relates to $h_1^*$. In particular, \cite{Delaigle08} showed that, as $\textrm{Var}(U)\to 0$, $\log(h_1)-\log(h_1^*)\approx \log(h_1^*)-\log(h_1^{**})$, and thus $h_1 \approx h_1^{*2}/h_1^{**}$, which suggests Step 5. We then set $h_1$ at $\hat h_1$ when estimating local modes using $(\bW, \bY)$. For the local constant mode estimator $\hat y_{\hbox {\tiny $M0$}}(x)$, $\hat p(y|x)$ in (\ref{eq:h1}) is $\hat p(x,y)/\hat f_{\hbox {\tiny $X$}}(x)$, where $\hat p(x,y)$ is given in (\ref{eq:jointpdfest}) and $\hat f_{\hbox {\tiny $X$}}(x)$ is the deconvoluting density estimator of $f_{\hbox {\tiny $X$}}(x)$ as in \citet{Stefanski90}. When considering the local linear mode estimator $\hat y_{\hbox {\tiny $M1$}}(x)$, $\hat p(y|x)$ in (\ref{eq:h1}) is given by (\ref{eq:cpdfme}). 

\section{Empirical evidence}
\label{s:empirical}
In this section we first present simulation studies to demonstrate the performance of the two proposed mode estimators, and compare them with the mode estimator resulting from naively applying the method in \citet{Chen2016} to error-contaminated data. Then we apply these methods to a real data example. In \citet{Chen2016}, it is assumed that $h_1=h_2$. We do not impose this constraint when implementing their method for a fair comparison with our methods. 

\subsection{Simulation design}
\label{s:simulation}
In the simulation experiment, we consider the following two true model configurations: 
\begin{enumerate}
	\item[(C1)] $[Y|X=x] \sim 0.5N\left(m(x)-2\sigma(x), \, 2.5^2\sigma^2(x)\right)+0.5N\left(m(x), \, 0.5^2\sigma^2(x)\right)$, where $m(x)=x+x^2$, $\sigma(x)=0.5+e^{-x^2}$, $X\sim \mathrm{Uniform}(-2,2)$, $U\sim\mathrm{Laplace}(0, \sigma_u/\sqrt{2})$. In this case, $p(y|x)$ is unimodal with $ M(x) \approx \{m(x)\}$, $\forall x\in [-2, \, 2]$.
	\item[(C2)] $[Y|X=x] \sim 0.5N\left(m_1(x), \, 0.5^2\right)+0.5N\left(m_2(x), \, 0.5^2\right)$, where $m_1(x)=x+x^2$, $m_2(x)=m_1(x)-6$, $X\sim \mathrm{Uniform}(-2,2)$, and $U\sim\mathrm{Laplace}(0, \sigma_u/\sqrt{2})$. In this case, $p(y|x)$ is bimodal with $M(x) \approx \{m_1(x), \, m_2(x)\}$, $\forall x\in [-2, \, 2]$.
\end{enumerate}
Under each true model configuration, we vary $\textrm{Var}(U)=\sigma^2_u$ to achieve the reliability ratio $\lambda=\textrm{Var}(X)/\{\textrm{Var}(X)+\sigma_u^2\}$ equal to 0.75, 0.85, and 0.95. Given each of the six simulation settings, we generate $500$ Monte Carlo (MC) replicates, each of sample size $n=500$, from the true model of $(W, Y)$. When implementing our proposed methods, we choose $K_1(\cdot)$ of which the Fourier transform is $\phi_{K_1}(t) = (1-t^2)^3I(t\in [-1, 1])$, and choose $K_2(\cdot)$ to be the standard normal pdf. For the method in \citet{Chen2016}, both $K_1(\cdot)$ and $K_2(\cdot)$ are the standard normal pdfs.

To focus on comparing different mode estimators without being distracted by data-driven bandwidth selection, we first use approximated theoretical optimal bandwidths for each method to mitigate the confounding effect of data-driven bandwidth selection on the estimation quality. Denote by $\hat M(x)$ a mode set estimator generically. Given a candidate $\bh$, we obtain $\hat M(x)$ for a sequence of grid points in $[x_{\hbox {\tiny $L$}}, \, x_{\hbox {\tiny $U$}}]$, $\{x_k=x_{\hbox {\tiny $L$}}+k\Delta\}_{k=1}^\mathcal{M}$, where $\Delta$ is the partition resolution, and $\mathcal{M}$ is the largest integer no larger than $(x_{\hbox {\tiny $U$}}-x_{\hbox {\tiny $L$}})/\Delta$. Then the approximated theoretical optimal $\bh$ associated with $\hat M(x)$ is obtained by minimizing with respect to $\bh$ the empirical ISE, $\ISE=\sum_{k=0}^{\mathcal{M}}\{\textrm{Haus}(\hat M(x_k), \, M(x_k))\}^2\Delta,$
where $\textrm{Haus}(S_1, S_2)$ denotes the Hausdorff distance between sets $S_1$ and $S_2$, which is defined by $\textrm{Haus}(S_1, S_2)=\inf \{ r: \, S_1\subset S_2\oplus r, \, S_2\subset S_1\oplus r \}$, in which $S_\ell\oplus r=\{ x: \, \left(\inf_{y\in S_\ell} |x-y|\right) \leq r \}$, for $\ell=1, 2$. Simply put, $\hat M(x)$ and $M(x)$ are close according to the Hausdorff distance if and only if every point in either set is close to some point in the other set, where the closeness of two points is assessed by the Euclidean distance. 

For any given finite sample, besides the choice of $\bh$, the starting values one uses in the mean-shift algorithm also influence $\textrm{Haus}(\hat M(x), \, M(x))$. A starting mode too far from the majority of the data cloud around $x$ can cause numerical trouble in this iterative algorithm, and thus we suggest exercising great care in choosing starting values. One way that works well in our simulation study to set starting values is as follows. Given $x$ at which $M(x)$ is of interest, define an index set $I_x=\{j:\, |W_j-x|<e\}$, where $e$ is a positive small value chosen so that the number of elements in $I_x$ is relatively large, say, 30. Then the starting values for estimating $M(x)$ via the mean-shift algorithm are set to be the percentiles of $\{Y_j: \, j\in I_x \}$ equally spaced between the 10th and 90th percentiles. For example, if one chooses to start with three initial values for an $x$, then one may set the starting values to be the 10th, 50th, and $90$th percentiles of $\{Y_j:\, j\in I_x \}$. To avoid missing a mode, the number of starting values, denoted by $N$, can be slightly bigger than one's visual impression of the number of clusters of the observed data cloud.   

\subsection{Simulation results}
\label{s:results}
Table~\ref{Sim1LapLap500:table} shows the MC average of ISE of each mode (set) estimate across 500 MC replicates and the associated standard error under each simulation setting, with $[x_{\hbox {\tiny $L$}}, \, x_{\hbox {\tiny $U$}}]=[-2,2]$ and $N=4$. In terms of both ISE and variability, our mode estimates, $\hat M_0(x)$ and $\hat M_1(x)$, outperform the naive mode estimate, denoted by $\hat M_{\hbox {\tiny $N$}}(x)$, in all six settings. And the improvement of our estimates over the naive estimate is more substantial when the error contamination is more severe (i.e., for smaller $\lambda$). In addition, the local linear estimate $\hat M_1(x)$ performs much better than the local constant estimate $\hat M_0(x)$ under (C1), while the advantage of the former is less obvious under (C2). In fact, $\hat M_1(x)$ deteriorates, although still outperforms $\hat M_{\hbox {\tiny $N$}}(x)$, faster than $\hat M_0(x)$ does as $\lambda$ decreases. The boxplots of ISEs and several estimated mode curves from each method are given in Appendix H, which clearly show that there are more outliers for $\hat M_1(x)$ compared to $\hat M_0(x)$. 
This comparison between the two proposed mode estimators indicates that the local linear estimator may be more suitable when $p(y|x)$ is unimodal, and can be subject to more numerical instability when applying to data from a multimodal distribution. This is reminiscent of a remark in \cite{Einbeck2006}, who recommended use the local linear mode estimator (in the absence of measurement error in their study) ``only for the case of functional dependence, i.e. where the mode is unique." 

\begin{table}[h]
	\caption{Averages of ISE across 500 MC replicates using approximated theoretical optimal bandwidths. Numbers in parentheses are ($10\times$ standard errors) associated with the averages\label{Sim1LapLap500:table}}
	
	\begin{adjustbox}{width=0.9\textwidth}
\small
\centering
		\begin{tabular}{cccccccc}
			\hline\noalign{\smallskip}
			& \multicolumn{3}{c}{(C1)} & & \multicolumn{3}{c}{(C2)}\\  
			\cline{2-4}\cline{6-8}
			$\lambda$ & 0.75 & 0.85   & 0.95	&& 0.75 & 0.85 & 0.95\\
			\hline
				 $\hat M_{\hbox {\tiny $N$}}(x)$ & 1.50 (0.21) & 1.05 (0.15)   &  0.64 (0.10) & & 1.52 (0.29) & 0.81 (0.12)   	& 0.34 (0.05)\\
		 $\hat M_0(x)$ & 1.15 (0.17) & 0.88 (0.13)   & 0.62 (0.10) && 0.91 (0.16) & 0.57 (0.09)   & 0.30 (0.04)\\
		 $\hat M_1(x)$ & 0.43 (0.10) & 0.32 (0.07)   & 0.22 (0.05) && 0.93 (0.31) & 0.51 (0.13)   & 0.21 (0.04)\\
			\hline
		\end{tabular}
		\end{adjustbox}
\end{table}

Acknowledging the fact that the approximated theoretical optimal $\bh$ is not available in practice, we carry out a second round of simulation under the same six settings, with $h_2$ fixed at $\hat{h}_2$ for all three methods, $h_1$ used in our estimators chosen via the CV-SIMEX method with $B=15$, and $h_1$ for the naive method chosen via naive CV as if $\bW$ were $\bX$ in (\ref{eq:h1}). Table~\ref{Sim1LapLap500SIMEX-dens:table} shows the MC averages of ISE of the three considered estimates across 500 MC replicates, along with the corresponding standard errors, with $[x_{\hbox {\tiny $L$}}, x_{\hbox {\tiny $U$}}]=[-1.8, 1.8]$. One can see that, under (C1), the proposed estimates still outperform the naive estimate. However, this is not as clear-cut under (C2). It appears that the number of modes in an estimated mode set (for any of these methods) can be sensitive to $h_2$, and a smaller $h_2$ tends to give a bigger estimated mode set, creating more estimated local modes that can be far away from the true modes. As pictorial demonstration of these results, boxplots of these ISEs and several estimated mode curves from each method are provided in Appendix H.

\begin{table}[h]
\caption{Averages of ISE across 500 MC replicates using $\bh$ chosen by the CV-SIMEX method. Numbers in parentheses are ($10 \times$ standard errors) associated with the averages\label{Sim1LapLap500SIMEX-dens:table}}
\begin{adjustbox}{width=0.9\textwidth}
\small
\centering
\begin{tabular}{*{8}{c}}
\hline
& \multicolumn{3}{c}{(C1)} & & \multicolumn{3}{c}{(C2)}\\  
\cline{2-4}\cline{6-8}
$\lambda$ & 0.75 & 0.85 & 0.95 && 0.75 & 0.85 & 0.95\\
\hline
$\hat M_{\hbox {\tiny $N$}}(x)$ & 0.83 (0.19) & 0.51 (0.12)   &  0.25 (0.06) && 1.20 (0.51) & 0.51 (0.11)  & 0.21 (0.03)\\
$\hat M_0(x)$ & 0.61 (0.15) & 0.42 (0.11)   & 0.24 (0.05) && 0.68 (0.37) & 0.44 (0.39)   & 0.33 (0.40)\\
$\hat M_1(x)$ & 0.44 (0.12) & 0.29 (0.08)   & 0.18 (0.07) && 1.07 (0.54) & 0.53 (0.29)   & 0.35 (0.60)\\
\hline
\end{tabular}
\end{adjustbox}
\end{table}

\begin{figure}[h]
	\centering
	\subfigure[]{ \includegraphics[width=2.2in, height=2in]{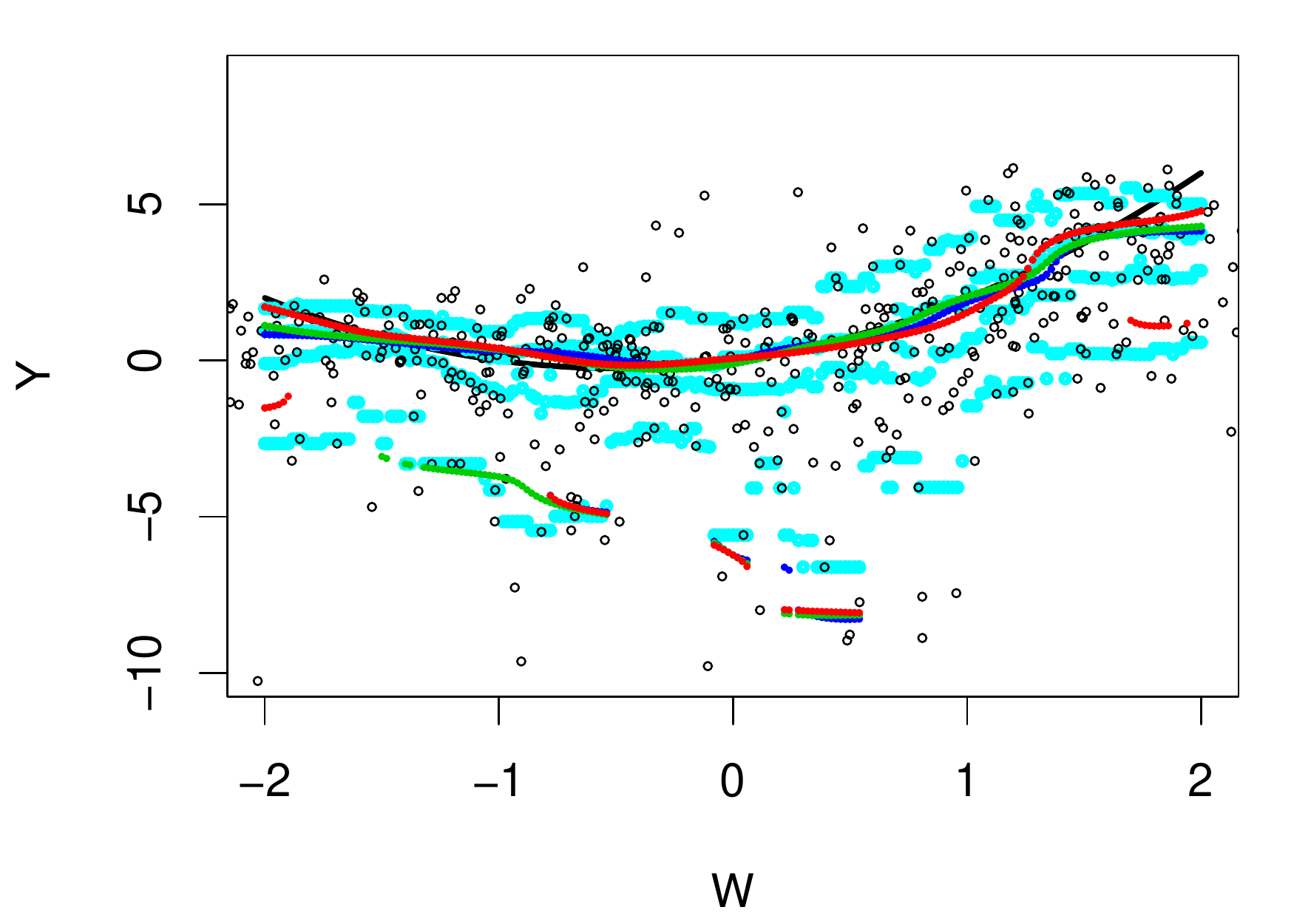} }
	\subfigure[]{ \includegraphics[width=2.2in, height=2in]{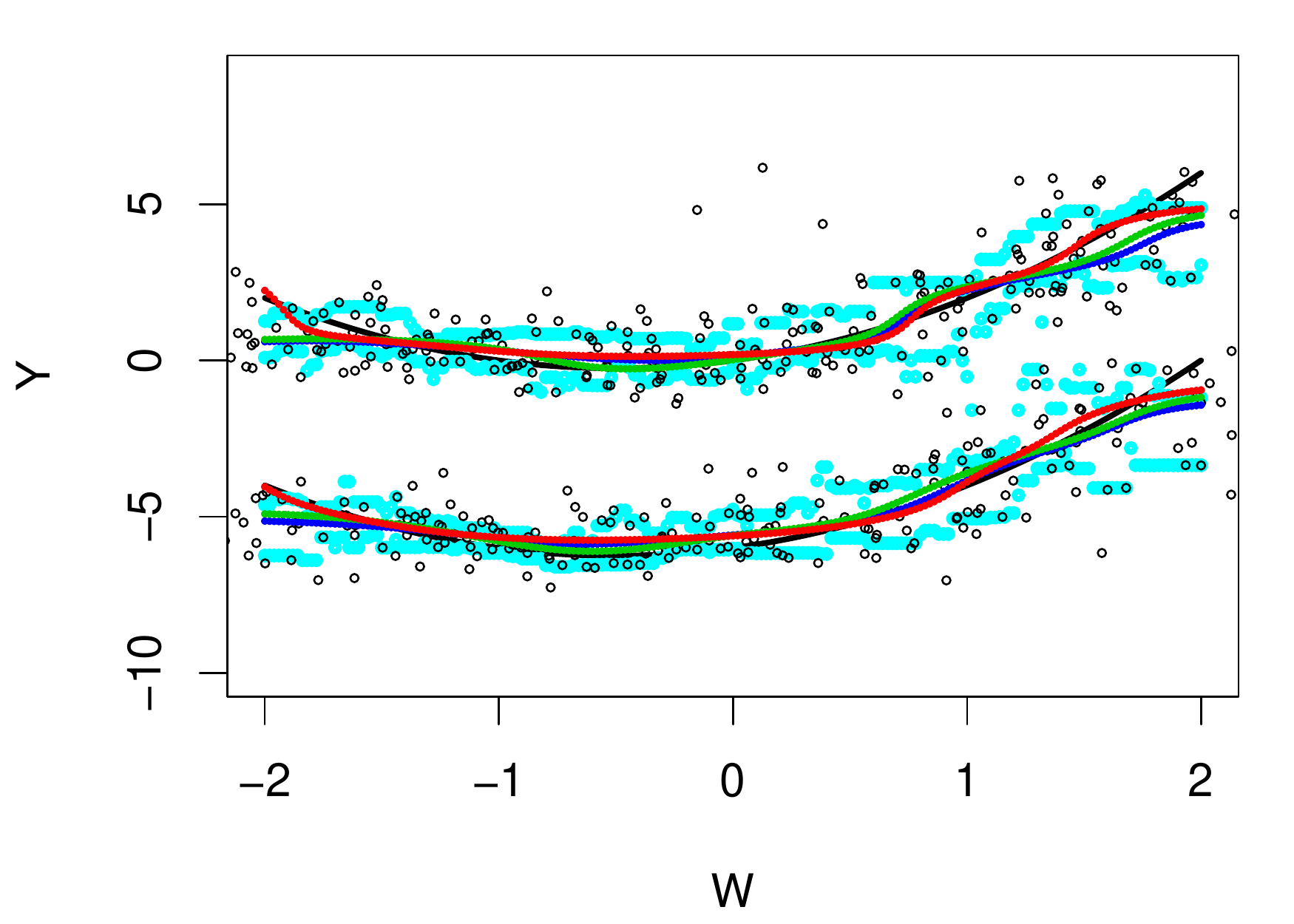} }
	\caption{Estimated mode curves using one data set generated from each of (C1) (panel (a)) and (C2) (panel (b)), incorporating the data-driven bandwidth selection and data-dependent starting values (in turquoise). Each panel contains the truth $M(x)$ (black lines),  $\hat M_{\hbox {\tiny $N$}}(x)$ (blue lines), $\hat M_0(x)$ (green lines), and $\hat M_1(x)$ (red lines).}
	\label{Sim:OneSample}
\end{figure}

Besides the choice of $h_2$, as pointed out in Section~\ref{s:simulation}, the starting values for the mean-shift algorithm also affect finite sample performance of these estimators. We use the data-dependent starting values described in Section~\ref{s:simulation} in the first round of simulations where the approximated theoretical optimal $\bh$ is used. With the data-driven $\bh$ used in the estimates as in the second round of simulations, the quality of an estimate is even more sensitive to the choice of starting values. Figure~\ref{Sim:OneSample} depicts the estimated mode curves from the three methods based on one simulated data set under each of (C1) and (C2), with the data-dependent starting values highlighted in turquoise. From there one can see that, if one starts at a starting mode value far away from the truth, the mean-shift algorithm can fail to converge or/and result in an inferior mode estimate. To avoid the interaction effects between the data-driven bandwidth selection and the data-dependent starting values, allowing us to focus on assessing the performance of the data-driven bandwidth selection, we set the starting values to be $\{m(x)\pm 0.5\}$ under (C1), and $\{m_1(x)\pm 0.5, \, m_2(x)\pm 0.5\}$ under (C2) when estimating $M(x)$ in the second round of simulations that produce Table~\ref{Sim1LapLap500SIMEX-dens:table}.
 
\subsection{Dietary data}
For illustration purposes, we consider estimating local modes of the food frequency questionnaire (FFQ) intake given one's long-term usual intake using dietary data. The data set to be analyzed contains the FFQ intake, measured as percent calories from fat ($Y$), and six 24-hour food recalls from $271$ subjects in the Women's Interview Survey of Health. The covariate of interest, the long-term usual intake ($X$), cannot be observed directly. A common practice in epidemiology studies is to use data from 24-hour food recalls to construct a surrogate ($W$) of the true covariate. For instance, \citet{Liang2005} used the average of two 24-hour food recalls from a subject as $W$ and studied the mean FFQ intake conditioning on $X$ and other error-free covariates; \citet{Wang2012} used the average of six 24-hour food recalls as $W$ and estimated conditional quantiles of the FFQ intake. All intake values are on the log scale in these studies. We followed the construction of $W$ in \citet{Wang2012}, associated with which the estimated reliability ratio is 0.737. 

\begin{figure}[h]
	\centering
	\includegraphics[width=3in, height=2.3in]{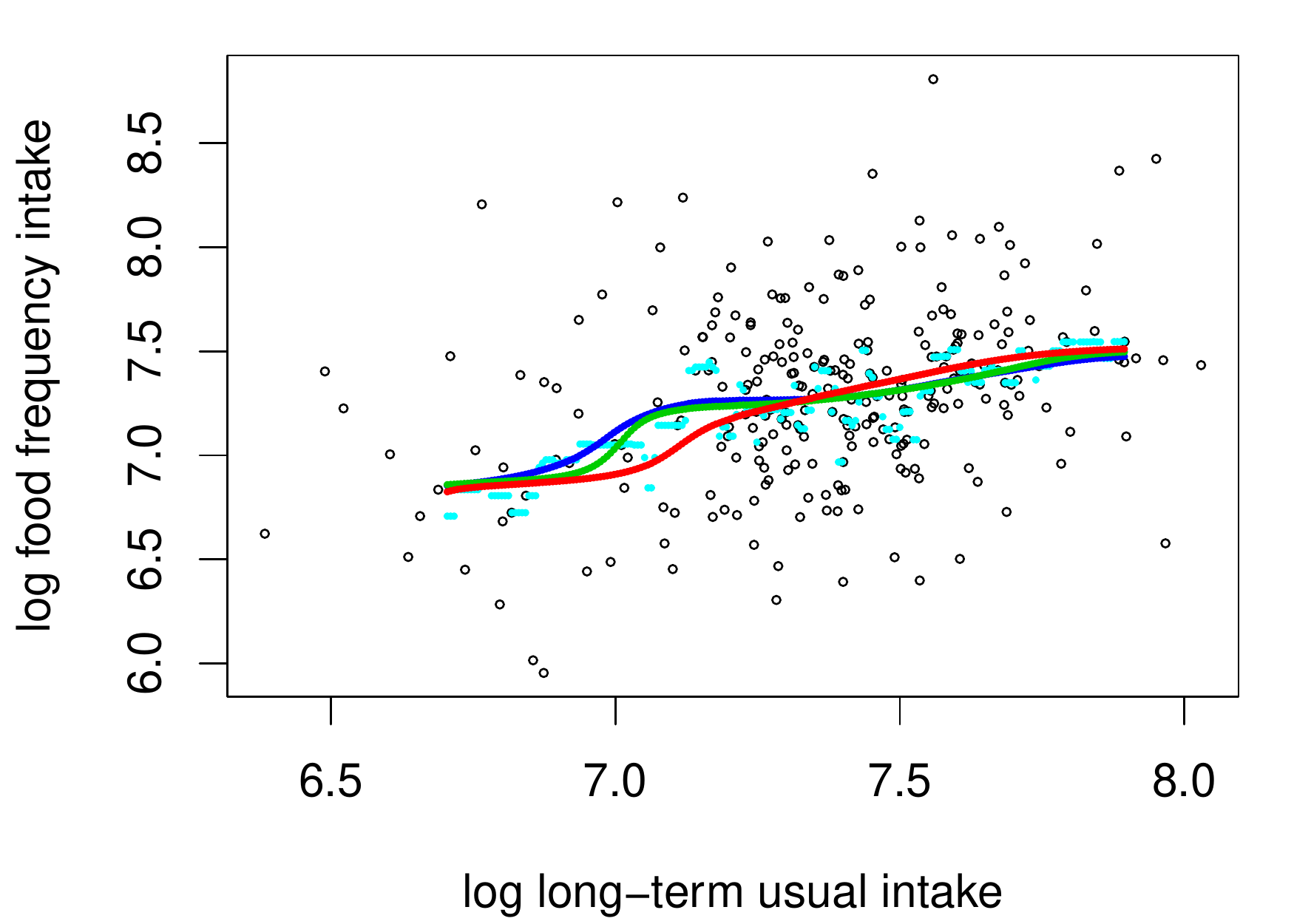}
\caption{The dietary data (black circles) and three estimated mode curves: the naive estimate $\hat M_{\hbox {\tiny $N$}}(x)$ (blue line), and our proposed estimates accounting for measurement error, $\hat  M_0(x)$ (green line) and $\hat M_1(x)$ (red line). Turquoise points are the data-dependent starting values for the mean-shift algorithm.}
	\label{RealData:curve}
\end{figure}

Figure~\ref{RealData:curve} presents the naive estimated mode curve and the two non-naive estimated mode curves from the two proposed methods. Both visual inspection of the scatter plot and the mean-shift algorithm seem to suggest a unimodal $p(y|x)$. Empirical evidence from simulation experiments suggest that this is the scenario where the local linear mode estimator $\hat M_1(x)$ can substantially improve over the naive estimator $\hat M_{\hbox {\tiny $N$}}(x)$, and the local constant mode estimator $\hat M_0(x)$ can also correct $\hat M_{\hbox {\tiny $N$}}(x)$ to some extent. The discrepancy between the three estimated mode curves at the lower segments in Figure~\ref{RealData:curve} can be due to bias correction from the two non-naive estimates, with the correction from $\hat M_1(x)$ more noticeable than that from $\hat M_0(x)$.

\section{Discussion}
\label{s:discussion}
The study presented in this article fills in an important gap in the measurement error literature by providing mode estimation in the presence of measurement error. We rigorously study the asymptotic properties of the proposed mode estimators and develop a data-driven bandwidth selection method. This line of research leads us to more interesting open questions that we have started to investigate upon the completion of this project. 

An immediate extension of this research is to allow the response prone to measurement error, with or without measurement error in covariates. An easy revision of the current estimator of the conditional (or joint) density is to replace $K_2(\cdot)$ with the corresponding deconvoluting kernel. The complication then, at least from the implementation standpoint, is that one will lose the mean-shift algorithm updating formula because the deconvoluting kernel for $Y$ is no longer radially symmetric, the key feature of $K_2(\cdot)$ that leads to the updating formula in all existing mean-shift algorithm applications. A different mode seeking algorithm is needed in this case.

An even more involved problem arises from our development of data-driven bandwidth selection methods. In this study, we employed the CV-SIMEX method to select the bandwidth in the $x$-direction, $h_1$, with the bandwidth in the $y$-direction, $h_2$, fixed at the normal reference that only depends on the response data. We conjecture that a more sensible way to choose these two bandwidths is to choose them jointly according to some CV criterion tailored for mode estimation, as opposed to choosing them in two separate steps using two different CV criteria that are designed for density estimation. It is unclear at this point how to implement such joint selection while bearing in mind that $h_1$ and $h_2$ play very different roles, with one relating to an error-prone predictor and the other corresponding to an error-free response. \citet{Chen2016} assumed $h_1=h_2=h$ and then chose $h$ to minimize the volume of the estimated prediction set, a statistic constructed to strive for a balance between the number of estimated local modes and the distance between the estimated mode and $\bY$. Following their idea, one may incorporate in the CV-SIMEX method the following CV criterion defined by (in the absence of measurement error) 
\begin{equation*}
	\textrm{CV}(\tilde M, \bh,\bX, \bY, \omega) = \frac{1}{n}\sum_{j=1}^{n} d\left(Y_j, \, \tilde M_{-j}(X_j)\right) \tilde{N}_{-j}(X_j) \omega(X_j),
\end{equation*} 
where $d(x, S)=\inf_{y\in S}|x-y|$ for a set $S$, $\tilde M (x)$ represents the estimated mode set at $x$ based on $(\bX,\bY)$, $\tilde{N}(x)$ is the number of distinct elements in $\tilde M (x)$, and $\omega(\cdot)$ is some weight function. However, theoretical justification of this CV criterion has not been established, and we did not see improvement from this bandwidth selection method over our current version of CV-SIMEX method in the simulation study (not shown here). 

\begin{figure}[h]
	\centering
	\includegraphics[height=1.8in, width=3in]{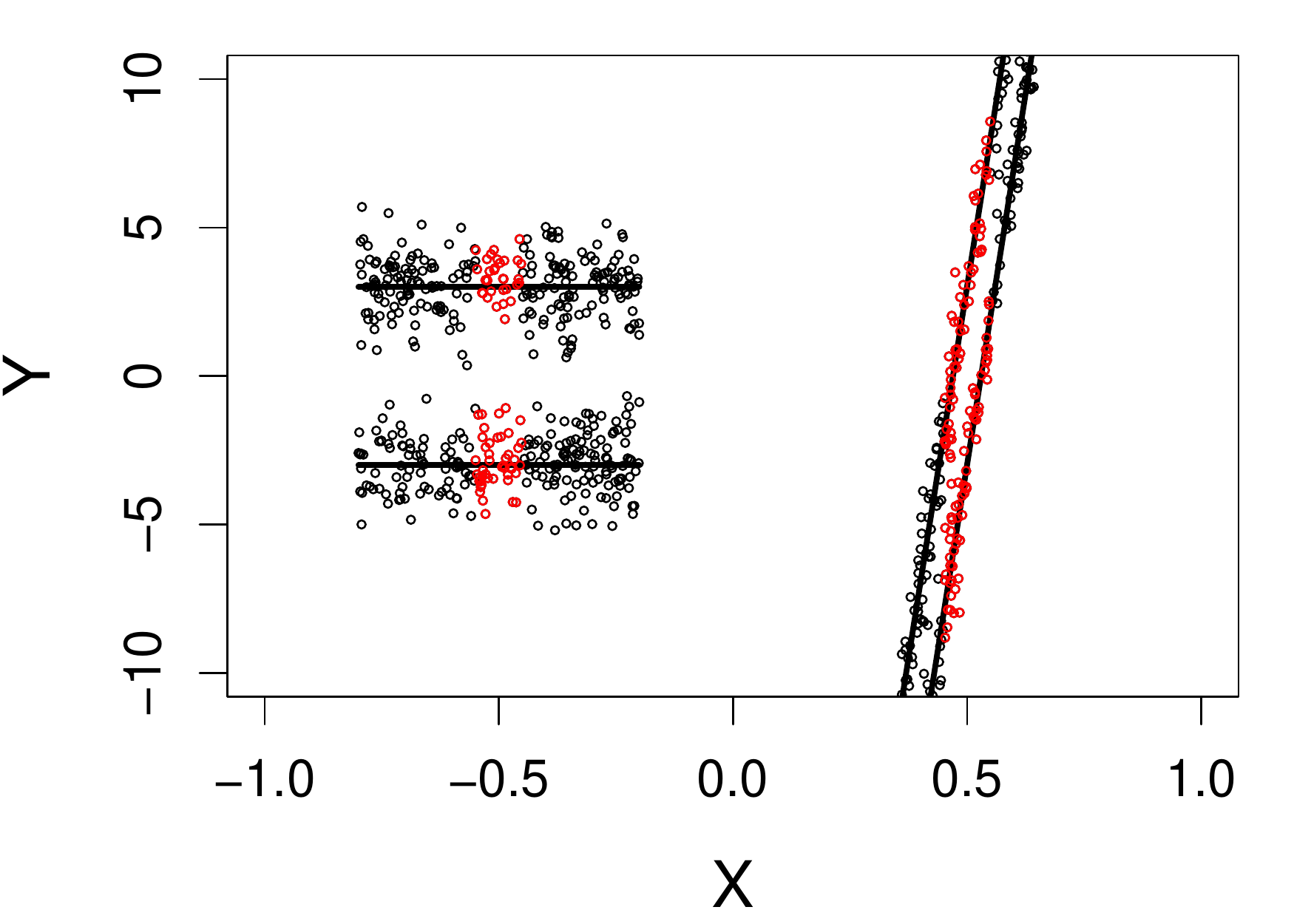} 
	\caption{\label{f:steep} Illustration of two pairs of mode curves, one pair being steeper than the other pair as functions of $x$. The red data points fall within a window of width 0.1 in the $x$-direction centering at $x=-0.5$ for the left pair of curves and $x=0.5$ for the right pair of curves.}
\end{figure}

Besides the need for a new CV criterion for the purpose of mode estimation, we also believe that mode estimation, with or without measurement error, can benefit from using bandwidths that depend on $x$. We gain this intuition from simulation study with multimodal $p(y|x)$, where we encountered more difficulty in estimating modes when multiple true mode curves are steeper and close to each other. This difficulty is expected and can be illustrated by Figure~\ref{f:steep}, where two pairs of curves are shown, with the left pair flat and the right pair steep (as functions of $x$). Fixing at an $x$, the separation (in $y$-direction) between two curves within each pair is the same in this figure, and the variability of $Y$ around each mode curve is also the same. But, within a given window (of a fixed width) in the $x$-direction, the data points (in red in Figure~\ref{f:steep}) surrounding the two steep curves are much harder to be separated into two clusters compared to the (red) data points around the two flat mode curves. And indeed in our simulation experiments (not shown here), the $\bh$ that works well for identifying the flat pair of mode curves does poorly in revealing the steep pair of the mode curves, and vice versa. Hence, if the multiple mode curves associated with $p(y|x)$ show different steepness and different amount of separation along the $y$-direction over different regions along the $x$-direction, it seems more sensible to apply different bandwidths along $\mathscr{X}$. 

In light of the need for variable bandwidths, the asymptotic optimal global bandwidths provided in Corollary~\ref{cor:optimalh} may seem irrelevant for the purpose of bandwidth selection, although they are theoretically valuable because they lead to optimal rates of MISE of the proposed mode estimators. Whether or not these optimal rates reach the minimax rate (given a well-defined class of mode estimators and a class of distributions or mode functions) remains an open problem. We conjecture that the key to solving this problem lies in the  minimax convergence rate of MISE associated with nonparametric density derivative estimators in the presence of measurement error, which itself is an open problem that we will tackle next. 
 
\setcounter{equation}{0}
\setcounter{figure}{0}
\renewcommand{\theequation}{A.\arabic{equation}}
\renewcommand{\thefigure}{A.\arabic{figure}}
\renewcommand{\thesection}{A.\arabic{section}}

\section*{Appendix A: Technical conditions}
Here we provide technical conditions that are needed at different parts of the theoretical development for different estimators considered in the main article. 

\subsection*{(I) Conditions on the joint probability density $p(x, y)$}
\begin{description}
\item[(CP1)] The joint density $p(x,y)$ is four times continuously differentiable with all partial derivatives bounded in absolute value by a finite positive constant $C_p$. 
\item[(CP2)] For $(x,y)\in \mathscr{X}\times \mathscr{Y}$ where $p_y(x,y)=0$, there exists a finite positive constant $\lambda_2$ such that $|p_{yy}(x,y)|>\lambda_2$.  
\end{description}  
Condition (CP1) is an ordinary smoothness condition. Condition (CP2) is a sharpness condition imposed on all critical points and implies no saddle points for $p(x, y)$. 

\subsection*{(II) Conditions on kernels $K_1(t)$ and $K_2(t)$}
\begin{description}
\item[(CK1)] $K_\ell(-t)$ is an even function and $\int K_\ell(t)dt=1$, for $\ell=1, 2$.
\item[(CK2)] $K_2(t)$ is four times continuous differentiable with all derivatives bounded in absolute value, and $\int \{K^{(k)}_2(t)\}^2 dt<\infty$, $\int t^2 K^{(k)}_2(t)dt<\infty$, for $k=0, 1, 2$.
\item[(CK3)] $\sup_t|\phi_{\hbox {\tiny $K_1$}(t)}/\phi_{\hbox {\tiny $U$}}(t/h_1)|<\infty$ and $\int |\phi^{(k)}_{\hbox {\tiny $K_1$}(t)}/\phi_{\hbox {\tiny $U$}}(t/h_1)|dt< \infty$, for $k=0, 1, 2$.
\item[(CK4)] $|\phi_{\hbox {\tiny $K_1$}}(t)|_\infty<\infty$, $|\phi'_{\hbox {\tiny $K_1$}}(t)|_\infty<\infty$.
\item[(CK5)] $\int (|t|^b+|t|^{b-1}) \{|\phi_{\hbox {\tiny $K_1$}}(t)|+|\phi'_{\hbox {\tiny $K_1$}}(t)|\} dt < \infty$, and 
$\int |t|^4|\phi_{\hbox {\tiny $K_1$}}(t)| dt < \infty$.
\item[(CK6)] $|\phi_{\hbox {\tiny $K_1$}}(t)|$ is supported on $[-1, 1]$.
\item[(CK7)] $\phi_{\hbox {\tiny $K_1$}}(t)$ is even and real.
\item[(CK8)] $\phi^{(k)}_{\hbox {\tiny $K_1$}}(t)$ is not identically 0, for $k=0, 1, 2$.
\item[(CK9)] The class of functions defined by $$\mathscr{K}_4= \cup_{k=0}^4\{\bv\mapsto K_{\hbox {\tiny $U,0$}}\{(v_1-x)/h_1\}K^{(k)}_2\{(v_2-y)/h_2\}: \, \bv=(v_1, v_2)^\T \in \mathbb R^2\}$$ is a VC-type class \citep{Wellner96}, that is, there exist $A$, $v>0$, and a constant envelop $\tau$ such that $\sup_{Q} N(\mathscr{K}_4, \mathscr{L}^2(Q), \tau \epsilon) \le (A/\epsilon)^v$, where $N(T, d_{\hbox {\tiny $T$}}, \epsilon)$ is the $\epsilon$-covering number for a semi-metric space $(T, d)$ and $Q$ is any probability measure. 
\end{description}
Define $\mu^{(\ell)}_k=\int t^k K_\ell (t) dt$ for $k=0, 1, \ldots$. Then (CK1) implies that $\mu^{(\ell)}_k=0$ for all odd $k$ and $\mu^{(\ell)}_0=1$, for $\ell=1, 2$. Condition (CK2) and (CK5) are needed to derive the uniform error rates of the proposed mode estimator. With $k=0$, (CK3) gives equation (1.2) in \citet{Stefanski90}, the assumption necessary for a well-behaved deconvoluting kernel $K_{\hbox {\tiny $U,0$}}(t)$. Condition (CK3) with $k=1, 2$ and (CK8) are needed to derive the mean and variance of $\hat p_y(y|x)$. Conditions (CK4) and (the first half of) (CK5) are needed in Lemma B.4 in \citet{Delaigle09}, which we use  to derive $\textrm{Var}\{\hat g_y(x,y)\}$ when $U$ is ordinary smooth. Conditions (CK4) and (CK6) are needed in Lemma B.9 in \citet{Delaigle09}, which we evoke to derive $\textrm{Var}\{\hat g_y(x,y)\}$ when $U$ is super smooth. Condition (CK7), along with (CU1) given later, are imposed so that $K_{\hbox {\tiny $U,0$}}(t)$ is real. Finally, (CK9) is needed to obtain the uniform consistency of the kernel-based estimators involved in the study. 

\subsection*{(III) Conditions on measurement error $U$} 
\begin{description}
\item[(CU1)] $\phi_{\hbox {\tiny $U$}}(t)\ne 0$, $\forall t$, and it is an even function. 
\item[(CU2)] $|\phi'_{\hbox {\tiny $U$}}(t)|_\infty< \infty$.
\end{description}
Condition (CU1) is needed for a well-defined real-valued $K_{\hbox {\tiny $U,\ell$}}(t)$, for $\ell=0, 1, \ldots$, and (CU2) is imposed in Lemma B.4 in \citet{Delaigle09}.

\subsection*{(IV) Conditions on the density of X, $f_{\hbox {\tiny $X$}}(x)$:} 
\begin{description}
\item[(CX1)] $f_{\hbox {\tiny $X$}}(x)>0$, $\forall x\in \mathscr{X}$; $f_{\hbox {\tiny $X$}}(x)$ is twice differentiable and $f^{(k)}_{\hbox {\tiny $X$}}(x)$ is bounded in absolute value by a finite positive constant $C_f$, for $k=0, 1, 2$.
\end{description}
This condition is needed for deriving the mean and variance of $\hat p_y(y|x)$.

\setcounter{equation}{0}
\setcounter{figure}{0}
\renewcommand{\theequation}{B.\arabic{equation}}
\renewcommand{\thefigure}{B.\arabic{figure}}
\renewcommand{\thesection}{B.\arabic{section}}

\section*{Appendix B: Relevant existing lemmas}
In deriving $\textrm{Var}\{\hat g_y(x,y)\}$  we evoke Lemma B.4, Lemma B.6 (for ordinary smooth $U$) and Lemma B.9 (for super smooth $U$) in \citet{Delaigle09}. For completeness, these lemmas are restated next. 
\begin{description}
\item[Lemma B.4:] Assume that, for $\ell=\ell_1, \ell_2$, $\|\phi_{\hbox {\tiny $K$}}^{(\ell)}\|_{\infty}<\infty$, $\|\phi_{\hbox {\tiny $K$}}^{(\ell+1)}\|_{\infty}<\infty$, $\|\phi'_{\hbox {\tiny $U$}}\|_\infty <\infty$, $\int (|t|^b+|t|^{b-1})\{|\phi_{\hbox {\tiny $K$}}^{(\ell)}|+|\phi_{\hbox {\tiny $K$}}^{(\ell+1)}|\}dt<\infty$, and $\int |t|^b |\phi_{\hbox {\tiny $K$}}^{(\ell)}|dt<\infty$, then, for a bounded function $g$,  
\begin{eqnarray*}
 & & \lim_{n\to \infty} h^{2b} \int K_{\hbox {\tiny $U$},\ell_1}(v)K_{\hbox {\tiny $U$},\ell_2}(v)g(x-hv)dv \\
&=& i^{-\ell_1-\ell_2}(-1)^{-\ell_2}\frac{g(x)}{c^2}\frac{1}{2\pi}\int |t|^{2b}\phi_{\hbox {\tiny $K$}}^{(\ell_1)}(t) \phi_{\hbox {\tiny $K$}}^{(\ell_2)}(t)dt.
\end{eqnarray*}
\item[Lemma B.9:] Suppose that $\phi_{\hbox {\tiny $K$}}(t)$ is supported on $[-1, 1]$, and, for $\ell=\ell_1$ and $\ell_2$, $\|\phi^{(\ell)}_{\hbox {\tiny $K$}}(t) \|_{\infty}<\infty$. Then $|\int_{-\infty}^\infty K_{\hbox {\tiny $U$}, \ell_1}(v) K_{\hbox {\tiny $U$}, \ell_2}(v) dv| \le Ch^{2b_2}\exp(2h^{-b}/d_2)$, where $b_2=b_0I(b_0<1/2)$.
\end{description}

\setcounter{equation}{0}
\setcounter{figure}{0}
\renewcommand{\theequation}{C.\arabic{equation}}
\renewcommand{\thefigure}{C.\arabic{figure}}
\renewcommand{\thesection}{C.\arabic{section}}
\section*{Appendix C: Asymptotic bias of $\hat p_{y}(x,y)$}
Assuming (CK3), it is shown that $E[K_{\hbox{\tiny $U,0$}} \{(W-x)/h_1\}|X]=K_1\{(X-x)/h_1\}$ \citep{Carroll88, Stefanski90}. Hence, $E\{\hat p_y(x,y)\}=E\{\tilde p_y(x,y)\}$, thus $\textrm{Bias}\{\hat p_y(x,y)\}=\textrm{Bias}\{\tilde p_y(x,y)\}$. We next focus on deriving $E\{\tilde p_y(x,y)\}$. 

Recall that
$$\tilde p_y(x, y)=\frac{1}{nh_1 h_2^3} \sum_{j=1}^n K_1\left(\frac{X_j-x}{h_1}\right)K_2\left(\frac{Y_j-y}{h_2}\right)(Y_j-y).$$
It follows that
\begin{eqnarray}
E\{\tilde p_y(x,y)\}
& = & \int \int \frac{1}{h_1h_2^2} K_1\left(\frac{u-x}{h_1}\right)K_2\left(\frac{v-y}{h_2}\right)\left(\frac{v-y}{h_2}\right)p(u,v) dudv \nonumber \\
& = & h_2^{-1}\int \int t K_1(s) K_2(t) p(x+h_1s, y+h_2t) ds dt. \label{eq:intK1K2tp}
\end{eqnarray}
Under (CP1), $p(x+h_1s, y+h_2t)$ has the third-order Taylor expansion around $(x, y)$ as follows,
\begin{align}
& p(x,y)+p_x(x,y)h_1s+p_y(x,y)h_2t+\frac{1}{2}\{p_{xx}(x,y)h^2_1s^2+2p_{xy}(x,y)h_1h_2st+\nonumber \\
& p_{yy}(x,y)h_2^2t^2\} +\frac{1}{3!}\{p_{xxx}(x,y)h_1^3s^3 +\frac{3!}{2} p_{xxy}(x,y) h_1^2h_2 s^2t +\frac{3!}{2}p_{xyy}(x,y)h_1h_2^2st^2\nonumber \\
& \left.  + p_{yyy}(x,y)h_2^3t^3\right\}
+r_1 h_1^3s^3+r_2 h_1^2h_2s^2t+r_3 h_1h_2^2st^2+r_4 h_2^3t^3, \label{eq:taylorpxy}
\end{align}
where $r_1$, $r_2$, $r_3$, and $r_4$ approach zero as $h_1, h_2\to 0$, $p_{xx}(x,y)=(\partial^2/\partial x^2)p(x,y)$, $p_{xxy}(x,y)=(\partial^3/\partial x^2 \partial y)p(x,y)$, and other partial derivatives are similarly denoted. Given (CK1), using this third-order Taylor expansion in (\ref{eq:intK1K2tp}) leads to 
\begin{eqnarray*}
E\{\tilde p_y(x,y)\} 
& = & \mu^{(2)}_2 p_y(x,y)+\frac{1}{2}p_{xxy}\mu_2^{(1)}\mu_2^{(2)}h_1^2+\frac{1}{3!}p_{yyy}(x,y) \mu^{(2)}_4 h_2^2\\
& & +r_2\mu_2^{(1)}\mu_2^{(2)}h_1^2+r_4\mu_4^{(2)}h_2^2,
\end{eqnarray*}
which is equal to 
$$p_y(x,y)+0.5\{p_{xxy}(x,y)\mu_2^{(1)}h_1^2+p_{yyy}(x,y) h_2^2\}+r_2\mu_2^{(1)}h_1^2+3 r_4 h_2^2$$
when $K_2(t)$ is the standard normal density, the choice we make in the main article. Hence, 
\begin{equation}
\textrm{Bias}\{\hat p_y(x,y)\}=0.5\{p_{xxy}(x,y)\mu_2^{(1)}h_1^2+p_{yyy}(x,y) h_2^2\}+o(h_1^2+h_2^2). \label{eq:biaspyjoint}
\end{equation}
Setting $y=y_{\hbox {\tiny $M$}}$ gives Lemma 3.1 in the main article. 

\setcounter{equation}{0}
\setcounter{figure}{0}
\renewcommand{\theequation}{D.\arabic{equation}}
\renewcommand{\thefigure}{D.\arabic{figure}}
\renewcommand{\thesection}{D.\arabic{section}}
\section*{Appendix D: Asymptotic variance of $\hat p_{y}(x,y)$}
Recall that 
$$\hat p_y(x, y)=\frac{1}{nh_1 h_2^3} \sum_{j=1}^n K_{\hbox {\tiny $U,0$}}\left(\frac{W_j-x}{h_1}\right)K_2\left(\frac{Y_j-y}{h_2}\right)(Y_j-y).$$ 
It follows that
\begin{align}
\textrm{Var}\{\hat p_y(x,y)\} & = \frac{1}{n h_1^2 h_2^6} \textrm{Var}\left\{K_{\hbox {\tiny $U,0$}}\left(\frac{W_j-x}{h_1}\right)K_2\left(\frac{Y_j-y}{h_2}\right)(Y_j-y)\right\} \nonumber\\
& = \frac{1}{n h_1^2 h_2^6}E\left\{K^2_{\hbox {\tiny $U,0$}}\left(\frac{W_j-x}{h_1}\right)K^2_2\left(\frac{Y_j-y}{h_2}\right)(Y_j-y)^2\right\}-\nonumber \\ 
& \frac{1}{n h_1^2 h_2^6} \left[E\left\{K_{\hbox {\tiny $U,0$}}\left(\frac{W_j-x}{h_1}\right)K_2\left(\frac{Y_j-y}{h_2}\right)(Y_j-y)\right\}\right]^2.  \label{eq:varsummand}
\end{align}
From the bias analysis in Appendix C, we have 
\begin{eqnarray*}
& & E\left\{K_{\hbox {\tiny $U,0$}}\left(\frac{W_j-x}{h_1}\right)K_2\left(\frac{Y_j-y}{h_2}\right)(Y_j-y)\right\}\\
& = &h_1h_2^3[ p_y(x,y)+0.5\{p_{xxy}(x,y)\mu_2^{(1)}h_1^2+p_{yyy}(x,y) h_2^2\}+o(h_1^2+h_2^2)],
\end{eqnarray*}
thus
\begin{align}
&\left[E\left\{K_{\hbox {\tiny $U,0$}}\left(\frac{W_j-x}{h_1}\right)K_2\left(\frac{Y_j-y}{h_2}\right)(Y_j-y)\right\}\right]^2\nonumber\\
= &h_1^2 h_2^6\left[p_y^2(x,y)+p_y(x,y)\left\{p_{xxy}(x,y) \mu_2^{(1)}h_1^2+p_{yyy}(x,y) h_2^2\right\}+o(h_1^2+h_2^2)\right]. \label{eq:meansq}
\end{align}
This suggests that the second term in (\ref{eq:varsummand}) is of order $O(n^{-1})$ if $p_y(x,y)\ne 0$, and it is of order $o\{n^{-1}(h_1^2+h_2^2)\}$ if $p_y(x,y)=0$. We next look into the first term in (\ref{eq:varsummand}). 

With nondifferential measurement error, the joint pdf of $(W, Y)$ is 
\begin{equation}
f_{\hbox {\tiny $W,Y$}}(w, y)=\int p(x,y) f_{\hbox {\tiny $U$}}(w-x)dx. \label{eq:fwy}
\end{equation}
It follows that
\begin{align}
& E \left\{ K^2_{\hbox {\tiny $U,0$}}\left(\frac{W_j-x}{h_1}\right)K^2_2\left(\frac{Y_j-y}{h_2}\right)(Y_j-y)^2 \right\} \nonumber\\
= & h_2^2\int \int K^2_{\hbox {\tiny $U,0$}}\left(\frac{u-x}{h_1}\right)K^2_2\left(\frac{v-y}{h_2}\right)\left(\frac{v-y}{h_2}\right)^2 f_{\hbox {\tiny $W,Y$}}(u, v)dudv \nonumber\\
= & h_1h_2^3 \int \int  K^2_{\hbox {\tiny $U,0$}}(s) K_2^2(t)t^2 f_{\hbox {\tiny $W,Y$}}(x+h_1s, y+h_2t)dsdt \nonumber\\ 
= & h_1h_2^3 \int \int  K^2_{\hbox {\tiny $U,0$}}(s) K_2^2(t)t^2 \int p(r,y+h_2t) f_{\hbox {\tiny $U$}}(x+h_1s-r)drdsdt \nonumber\\
= & h_1h_2^3 \int K^2_{\hbox {\tiny $U,0$}}(s) \int f_{\hbox {\tiny $U$}}(x+h_1s-r) \int K_2^2(t)t^2 p(r,y+h_2t) dtdrds. \label{eq:threelayer}
\end{align}
Using the first-order Taylor expansion of $p(r,y+h_2t)$ around $(r, y)$ in the innermost integral in (\ref{eq:threelayer}) gives
\begin{eqnarray*}
\int K_2^2(t)t^2 p(r,y+h_2t) dt 
& = & \int K_2^2(t)t^2 \{p(r, y)+p_y(r, y) h_2t+O(h_2^2)\} dt \\
& = & p(r, y)\nu_2^{(2)}+p_y(r, y) \nu_3^{(2)}h_2+O(h_2^2),
\end{eqnarray*}
where $\nu_k^{(2)}=\int t^k K^2_2 (t) dt$, for $k=2, 3$.
Putting this elaboration of the innermost integral in (\ref{eq:threelayer}) and using the first-order Taylor expansion of $f_{\hbox {\tiny $U$}}(x+h_1s-r)$ around $x-r$ in (\ref{eq:threelayer}) gives
\begin{eqnarray*}
& & E \left\{ K^2_{\hbox {\tiny $U,0$}}\left(\frac{W_j-x}{h_1}\right)K^2_2\left(\frac{Y_j-y}{h_2}\right)(Y_j-y)^2 \right\} \\
& = & h_1h_2^3 \int  K^2_{\hbox {\tiny $U,0$}}(s) \int  \{ f_{\hbox {\tiny $U$}}(x-r) +f'_{\hbox {\tiny $U$}}(x-r)h_1s+O(h_1^2)\}\times \\
& & \{p(r, y)\nu_2^{(2)}+p_y(r, y) \nu_3^{(2)}h_2+O(h_2^2)\} drds.
\end{eqnarray*}
This reveals the dominating term of the expectation above as
\begin{equation}
h_1h_2^3 \nu_2^{(2)}\int K^2_{\hbox {\tiny $U,0$}}(s) \int f_{\hbox {\tiny $U$}}(x-r) p(r,y) drds
= h_1h_2^3 \nu_2^{(2)}f_{\hbox {\tiny $W,Y$}}(x, y)\int K^2_{\hbox {\tiny $U,0$}}(s) ds. \label{eq:Ku02}
\end{equation}

When $U$ is ordinary smooth of order $b$, by Lemma B.4 in \citet{Delaigle09}, which is repeated under Appendix B above,  (\ref{eq:Ku02}) indicates that 
\begin{equation}
\resizebox{.98 \textwidth}{!}
{$
E \left\{ K^2_{\hbox {\tiny $U,0$}}\left(\frac{W_j-x}{h_1}\right)K^2_2\left(\frac{Y_j-y}{h_2}\right)(Y_j-y)^2 \right\} 
= h_1^{1-2b}h_2^3 c^{-2}\eta_0 \nu_2^{(2)} f_{\hbox {\tiny $W,Y$}}(x,y)+o(h_1^{1-2b}h_2^3),
$} \label{eq:osmooth}
\end{equation}
where $\eta_0=(2\pi)^{-1}\int |t|^{2b} |\phi_{\hbox {\tiny $K_1$}}(t)|^2 dt$. Because (\ref{eq:osmooth}) tends to zero slower than $O(h_1^2h_2^6)$, (\ref{eq:varsummand}) is dominated by the first term there. Hence, assuming $nh_1^{1+2b}h_2^3\to \infty$, we have
$$\textrm{Var}\{\hat p_y(x,y)\}=\frac{\eta_0\nu_2^{(2)} f_{\hbox {\tiny $W,Y$}}(x,y)}{nh_1^{1+2b}h_2^3 c^2}+o\left(\frac{1}{nh_1^{1+2b}h_2^3}\right).$$ Setting $y=y_{\hbox {\tiny $M$}}$ gives the first half of Lemma 3.2 in the main article, where $\nu_2^{(2)}$ is replaced by $1/(4\sqrt{\pi})$ because $K_2(\cdot)$ is the standard normal pdf in the main article. 

When $U$ is super smooth of order $b$, by Lemma B.9 in \citet{Delaigle09}, repeated in Appendix B above, (\ref{eq:Ku02}) suggests that 
\begin{equation}
\resizebox{.98 \textwidth}{!}
{$
E \left\{ K^2_{\hbox {\tiny $U,0$}}\left(\frac{W_j-x}{h_1}\right)K^2_2\left(\frac{Y_j-y}{h_2}\right)(Y_j-y)^2 \right\} 
\le  h_1^{1+2b_2}\exp(2h_1^{-b}/d_2)h_2^3 C\nu_2^{(2)} f_{\hbox {\tiny $W,Y$}}(x,y),
$}
\label{eq:ssmooth}
\end{equation}
where $C$ is a finite positive constant. 
Hence, if $nh_1^{1-2b_2}h_2^3 \exp(-2h_1^{-b}/{d_2})\to \infty$, (\ref{eq:varsummand}) suggests 
\begin{eqnarray*}
\textrm{Var}\{\hat p_y(x,y)\} & \le & \frac{1}{n h_1^2 h_2^6} E\left\{K^2_{\hbox {\tiny $U,0$}}\left(\frac{W_j-x}{h_1}\right)K^2_2\left(\frac{Y_j-y}{h_2}\right)(Y_j-y)^2\right\}\\
& \le & \frac{\exp\left(2h_1^{-b}/{d_2}\right)}{nh_1^{1-2b_2}h_2^3}C\nu_2^{(2)} f_{\hbox {\tiny $W,Y$}}(x,y)+o\left\{\frac{\exp\left(2h_1^{-b}/{d_2}\right)}{nh_1^{1-2b_2}h_2^3}\right\}.
\end{eqnarray*}
Setting $y=y_{\hbox {\tiny $M$}}$ and absorbing $\nu_2^{(2)}=1/(4\sqrt{\pi})$ in $C$ gives the second half of Lemma 3.2 in the main article.

\setcounter{equation}{0}
\setcounter{figure}{0}
\setcounter{section}{0}
\renewcommand{\theequation}{E.\arabic{equation}}
\renewcommand{\thefigure}{E.\arabic{figure}}
\renewcommand{\thesection}{E.\arabic{section}}
\section*{Appendix E: Convergence rate of $\textrm{MISE}(\hat y_{\hbox {\tiny M0}})$}
Assuming interchangeability of integrations, the mean integrated squared error (MISE) of $\hat y_{\hbox {\tiny $M0$}}(x)$ can be decomposed into the sum of two parts, 
$$\textrm{MISE}(\hat y_{\hbox {\tiny $M$}})  
= \int_{\mathscr{X}} \textrm{Bias}^2(\hat y_{\hbox {\tiny $M0$}}) dx + \int_{\mathscr{X}}\textrm{Var}(\hat y_{\hbox {\tiny $M$}})dx.$$
By equation (17) in the main article, one has 
\begin{align}
\textrm{Bias}(\hat y_{\hbox {\tiny $M0$}}) & = -\{p_{yy}(x, y_{\hbox {\tiny $M$}})\}^{-1}E\left\{\hat p_y(x, y_{\hbox {\tiny $M$}})\right\}+E\left\{O\left(\|\hat p_{yy} -p_{yy}  \|_\infty\right)\hat p_y(x, y_{\hbox {\tiny $M$}})\right\}, \label{eq:biasym0} \\
\textrm{Var}(\hat y_{\hbox {\tiny $M0$}}) & = \{p_{yy}(x, y_{\hbox {\tiny $M$}})\}^2\textrm{Var}\left\{\hat p_y(x, y_{\hbox {\tiny $M$}}) \right\}+o(1). \label{eq:varym0}
\end{align} 

To derive the integrated squared bias, using (\ref{eq:biaspyjoint}) in (\ref{eq:biasym0}), one has that, provided that $\|\hat p_{yy} -p_{yy}  \|_\infty$ tends to zero, 
\begin{eqnarray*}
\textrm{Bias}^2(\hat y_{\hbox {\tiny $M$}}) & = & 0.25p_{yy}^{-2}(x, y_{\hbox {\tiny $M$}})\Big\{\left(\mu_2^{(1)}\right)^2 h_1^4p^2_{xxy}(x, y_{\hbox {\tiny $M$}})+h_2^4 p^2_{yyy}(x, y_{\hbox {\tiny $M$}})\\
& & +2 \mu_2^{(1)} h_1^2 h_2^2 p_{xxy}(x, y_{\hbox {\tiny $M$}})p_{yyy}(x, y_{\hbox {\tiny $M$}})\Big\}+o\left\{\left(h_1^2+h_2^2\right)^2\right\}.
\end{eqnarray*}
Under (CP2), and assuming $p_{xxy}(x,y_{\hbox {\tiny $M$}})$ and $p_{yyy}(x,y_{\hbox {\tiny $M$}})$ square integrable, one has 
$$\int_{\mathscr{X}} \textrm{Bias}^2(\hat y_{\hbox {\tiny $M0$}})dx=O\left\{\left(h_1^2+h_2^2\right)^2\right\}.$$ 

To derive the integrated variance, using the variance analysis Appendix D in (\ref{eq:varym0}), one has that, when $U$ is ordinary smooth, 
$$\int_{x\in \mathscr{X}}\textrm{Var}(\hat y_{\hbox {\tiny $M0$}})dx=O\left(\frac{1}{nh_1^{1+2b}h_2^3}\right);$$ 
and, when $U$ is super smooth, 
$$\int_{x\in \mathscr{X}}\textrm{Var}(\hat y_{\hbox {\tiny $M$}})dx=O\left\{\frac{\exp\left(2h_1^{-b}/{d_2}\right)}{nh_1^{1-2b_2}h_2^3}\right\}.$$

Putting the above integrated squared bias of $\hat y_{\hbox {\tiny $M0$}}(x)$ and integrated variance of $\hat y_{\hbox {\tiny $M0$}}(x)$ together gives Theorem 3.2 in the main article.

\section*{Appendix F: Asymptotic bias of $\hat p_{y}(y_{\hbox {\tiny $M$}}|x)$}
\setcounter{figure}{0}
\setcounter{section}{0}
\renewcommand{\theequation}{F.\arabic{equation}}
\renewcommand{\thefigure}{F.\arabic{figure}}
\renewcommand{\thesection}{F.\arabic{section}}
\section{Outline of deriving  $E\{\hat p_y(y_{\hbox {\tiny $M$}}|x)\}$}
\label{s:outline}
Recall that $\hat p_y(y|x)= \be_1^\T \hat \bS^{-1}_n(x) \hat\bT'_n(x,y)$.   
Denote by $[\hat S^{0,0}_n(x), \, \hat S^{0,1}_n(x)]$ the first row of $\hat \bS^{-1}_n(x)$, then one has 
\begin{equation}
\hat p_y(y|x)=\sum_{\ell=0}^1 \hat S^{0,\ell}_n(x)\hat T'_{n, \ell}(x,y).  \label{eq:pyhat2} 
\end{equation} 
The derivations of the dominating bias of $\hat p_y(y|x)$ involve two tasks. First, revealing the dominating terms in $\hat S^{0,\ell}_n(x)$. Under (CX1), (CU1), and (CK1), (CK3)--(CK5), (CK8), \citet{Delaigle09} showed that  
$$ \hat \bS^{-1}=\bS^{-1}f_{\hbox {\tiny $X$}}^{-1}(x) -h_1\bS^{-1}\widetilde{\bS}\bS^{-1}f'_{\hbox {\tiny $X$}}(x)f^{-2}_{\hbox {\tiny $X$}}(x)+O_{\hbox {\tiny $P$}}(h_1^2),$$
where 
\begin{equation*}
\bS = 
\begin{bmatrix}
\mu_0^{(1)} & \mu_1^{(1)} \\
\mu_1^{(1)} & \mu_2^{(1)} 
\end{bmatrix} =
\begin{bmatrix}
1 & 0 \\
0 & \mu_2^{(1)} 
\end{bmatrix},  \,\,
\widetilde \bS =  
\begin{bmatrix}
\mu_1^{(1)} & \mu_2^{(1)} \\
\mu_2^{(1)} & \mu_3^{(1)} 
\end{bmatrix} = 
\begin{bmatrix}
0 & \mu_2^{(1)} \\
\mu_2^{(1)} &  0
\end{bmatrix}. 
\end{equation*}
Elaborating the above result yields 
\begin{equation}
\hat S^{0,0}_n (x) = f^{-1}_{\hbox {\tiny $X$}}(x)+O_{\hbox {\tiny $P$}}(h_1^2), \, \,
\hat S^{0,1}_n (x) = -h_1 f'_{\hbox {\tiny $X$}}(x) f^{-2}_{\hbox {\tiny $X$}}(x) +O_{\hbox {\tiny $P$}}(h_1^2). \label{eq:twoS}
\end{equation}
This completes the first task. 

The second task is to reveal the dominating terms in $\hat T'_{n, \ell}(x,y)$ according to the following decomposition, 
\begin{equation}
\hat T'_{n, \ell}(x,y)=E\left\{\hat T'_{n, \ell}(x,y)\right\}+O_{\hbox {\tiny $P$}}\left[\sqrt{\textrm{Var}\left\{\hat T'_{n, \ell}(x,y)\right\}}\right]. \label{eq:T'meanvar}
\end{equation}
Next, we first look into $E\{\hat T'_{n, \ell}(x,y)\}$, then we study $\textrm{Var}\{\hat T'_{n, \ell}(x,y)\}$, and show that the former dominates that latter under certain conditions. 

\section{Deriving $E\{\hat T'_{n, \ell}(x,y)\}$}
\label{s:meanT'}
\setcounter{equation}{3}
Because 
\begin{equation}
\hat T'_{n,\ell}(x,y) =  \frac{1}{n h_1 h_2^2} \sum_{j=1}^n K_{\hbox {\tiny $U$}, \ell}\left(\frac{W_j-x}{h_1}\right)K_2\left(\frac{Y_j-y}{h_2}\right)\left(\frac{Y_j-y}{h_2}\right), \label{eq:T'}
\end{equation}
and 
$$E\left\{\left.K_{\hbox {\tiny $U$},  \ell}\left(\frac{W_j-x}{h_1}\right)\right\vert X_j \right\} = \left(\frac{X_j-x}{h_1}\right)^\ell K_1\left(\frac{X_j-x}{h_1}\right),$$
we have 
\begin{eqnarray}
& & E\{\hat T'_{n, \ell}(x,y)\}  \nonumber \\
& = & \frac{1}{h_1 h_2^2} E\left\{ K_{\hbox {\tiny $U,\ell$}}\left(\frac{W_j-x}{h_1}\right)K_2\left(\frac{Y_j-y}{h_2}\right)\left(\frac{Y_j-y}{h_2}\right) \right\} \label{eq:mean01}\\
& = & \frac{1}{h_1 h_2^2} E\left\{ \left(\frac{X_j-x}{h_1}\right)^\ell K_1\left(\frac{X_j-x}{h_1}\right)K_2\left(\frac{Y_j-y}{h_2}\right)\left(\frac{Y_j-y}{h_2}\right) \right\} \nonumber \\
& = & \frac{1}{h_1 h_2^2} \int \int \left(\frac{u-x}{h_1}\right)^\ell K_1\left(\frac{u-x}{h_1}\right)K_2\left(\frac{v-y}{h_2}\right)\left(\frac{v-y}{h_2}\right) p(u, v) du dv \nonumber \\
& = & h_2^{-1}\int \int s^\ell t K_1(s)  K_2(t)p(x+h_1s, \, y+h_2t) dsdt. \nonumber 
\end{eqnarray}
Inserting (\ref{eq:taylorpxy}) in the above integrand leads to 
\begin{align}
& E\{\hat T'_{n, \ell}(x,y)\} \nonumber \\
= & h_2^{-1}\int \int s^\ell t K_1(s)  K_2(t) \Big[p(x,y)+p_x(x,y)h_1s+p_y(x,y)h_2t \nonumber\\
+ &\frac{1}{2}\left\{p_{xx}(x,y)h^2_1s^2+2p_{xy}(x,y)h_1h_2st+p_{yy}(x,y)h_2^2t^2\right\}+\frac{1}{3!}\Big\{p_{xxx}(x,y)h_1^3s^3 \nonumber\\
+ &\frac{3!}{2} p_{xxy}(x,y) h_1^2h_2 s^2t +\frac{3!}{2}p_{xyy}(x,y)h_1h_2^2 st^2 + p_{yyy}(x,y)h_2^3 t^3\Big\}+O\left\{r_4(\bh)\right\}\Big] dsdt, \nonumber \\
\label{eq:taylor} 
\end{align}
where $r_4(\bh)=\sum_{r_1, r_2=0, \ldots, 4}^{r_1+r_2=4} h_1^{r_1}h_2^{r_2}$, in which $\bh=(h_1, \, h_2)^\T$.

Focusing on the case with $y=y_{\hbox {\tiny $M$}}$, noting that $p_y(x,y_{\hbox {\tiny $M$}})=0$ and $\mu^{(\ell)}_k=0$ when $k$ is odd, for $\ell=1, 2$, (\ref{eq:taylor}) reduces to 
\begin{equation}
\left\{
\begin{array}{l}
E\{\hat T'_{n, 0}(x,y_{\hbox {\tiny $M$}})\} = 0.5\left\{p_{xxy}(x,y_{\hbox {\tiny $M$}})\mu_2^{(1)}h_1^2+p_{yyy}(x,y_{\hbox {\tiny $M$}})h_2^2\right\}+O\left\{h_2^{-1}r_4(\bh)\right\}, \\
E\{\hat T'_{n, 1}(x,y_{\hbox {\tiny $M$}})\} = p_{xy}(x,y_{\hbox {\tiny $M$}})\mu_2^{(1)}h_1+O\left\{h_2^{-1}r_4(\bh)\right\}.
\end{array}
\right. \label{eq:twoET'}
\end{equation}

\section{Deriving $\textrm{Var}\{\hat T'_{n, \ell}(x,y)\}$}
\label{s:varT'}
\setcounter{equation}{7}
By (\ref{eq:T'}), we have 
\begin{align}
& \textrm{Var}\{\hat T'_{n, \ell}(x,y)\}  \nonumber\\
= & \frac{1}{n h_1^2 h_2^4}\textrm{Var}\left\{K_{\hbox {\tiny $U$}, \ell}\left(\frac{W_j-x}{h_1}\right)K_2\left(\frac{Y_j-y}{h_2}\right)\left(\frac{Y_j-y}{h_2}\right)\right\} \nonumber\\
\le & \frac{1}{n h_1^2 h_2^4}E\left\{K^2_{\hbox {\tiny $U$}, \ell}\left(\frac{W_j-x}{h_1}\right)K^2_2\left(\frac{Y_j-y}{h_2}\right)\left(\frac{Y_j-y}{h_2}\right)^2\right\} \label{eq:var01}\\
= & \frac{1}{n h_1^2 h_2^4}\int \int K^2_{\hbox {\tiny $U$}, \ell}\left(\frac{w-x}{h_1}\right)K^2_2\left(\frac{v-y}{h_2}\right)\left(\frac{v-y}{h_2}\right)^2 f_{\hbox {\tiny $W, Y$}}(w, v)dwdv \nonumber\\
= &  \frac{1}{n h_1^2 h_2^4}\int \int \int K^2_{\hbox {\tiny $U$}, \ell}\left(\frac{w-x}{h_1}\right)K^2_2\left(\frac{v-y}{h_2}\right)\left(\frac{v-y}{h_2}\right)^2 f_{\hbox {\tiny $U$}}(w-u) p(u, v) dudwdv \nonumber\\
= &  \frac{1}{n h_1 h_2^3}\int \int \int K^2_{\hbox {\tiny $U$}, \ell}(s)K^2_2(t)t^2 f_{\hbox {\tiny $U$}}(x+h_1s-u) p(u, y+h_2t) dudsdt \nonumber\\
= & \frac{1}{n h_1 h_2^3}\int \left\{\int K^2_{\hbox {\tiny $U$}, \ell}(s)f_{\hbox {\tiny $U$}}(x+h_1s-u) ds\right\} \left\{\int t^2 K^2_2(t) p(u, y+h_2 t) dt\right\} du.\nonumber \\
 \label{eq:twoinner}
\end{align}
Next, we elaborate the two inner integrals, one w.r.t. $s$ and the other w.r.t. $t$, in (\ref{eq:twoinner}). 

For the inner integral w.r.t. $t$, using the second-order Taylor expansion of $p(u, y+h_2t)$ around $(u, y)$, one has
\begin{eqnarray*}
&  & \int K^2_2(t)t^2 p(u, y+h_2t) dt \\
& = & \int K^2_2(t)t^2\left\{p(u, y)+p_y(u, y)h_2 t+0.5p_{yy}(u, y)h_2^2 t^2+O(h_2^3)\right\} dt.
\end{eqnarray*}
Setting $y=y_{\hbox {\tiny $M$}}$, the above gives
\begin{align}
 & \int K^2_2(t)t^2 p(u, y_{\hbox {\tiny $M$}}+h_2 t) dt \\
= & \int K^2_2(t)t^2\left\{p(u, y_{\hbox {\tiny $M$}})+0.5p_{yy}(u, y_{\hbox {\tiny $M$}})h_2^2 t^2+O(h_2^3)\right\} dt \nonumber\\
= & p(u, y_{\hbox {\tiny $M$}}) \nu^{(2)}_2+0.5 p_{yy}(u, y_{\hbox {\tiny $M$}})\nu_4^{(2)}h_2^2+O(h_2^3). \label{eq:tinner}
\end{align}

When it comes to the inner integral w.r.t $s$ in (\ref{eq:twoinner}), one shall distinguish between ordinary smooth $U$ and super smooth $U$. If $U$ is ordinary smooth of order $b$, under conditions (CK4) and (CK5), Lemma B.4 in \citet{Delaigle09} implies that,
\begin{equation} 
\int K^2_{\hbox {\tiny $U$}, \ell}(s)f_{\hbox {\tiny $U$}}(x+h_1s-u) ds 
 =  h_1^{-2b}c^{-2}\eta_\ell f_{\hbox {\tiny $U$}}(x-u) +o(h_1^{-2b}), \label{eq:sinner1}
\end{equation}
where $\eta_\ell=(2\pi)^{-1}\int |t|^{2b}\{ \phi^{(\ell)}_{\hbox {\tiny $K_1$}} (t)\}^2dt$, for $\ell=0, 1$. 
By (\ref{eq:tinner}) and (\ref{eq:sinner1}), (\ref{eq:twoinner}) is equal to 
\begin{eqnarray}
& & \frac{\eta_\ell}{nh_1h_2^3 c^2 } \int \left\{h_1^{-2b}f_{\hbox {\tiny $U$}}(x-u)+o(h_1^{-2b})\right\}\times \nonumber \\
& & \{p(u, y_{\hbox {\tiny $M$}}) \nu^{(2)}_2+0.5 p_{yy}(u, y_{\hbox {\tiny $M$}})\nu_4^{(2)}h_2^2+O(h^3)\}du \nonumber\\
& = & \frac{\eta_\ell}{nh_1h_2^3 c^2} \left[\left\{p(\cdot, y_{\hbox {\tiny $M$}})*f_{\hbox {\tiny $U$}}\right\}(x) \nu_2^{(2)} h_1^{-2b}
+0.5\left\{p_{yy}(\cdot, y_{\hbox {\tiny $M$}})*f_{\hbox {\tiny $U$}}\right\}(x) \nu_4^{(2)} h_1^{-2b}h_2^2 
\right.\nonumber \\
& & \left.+o(h_1^{-2b})\right],\label{eq:ordtwoinner}
\end{eqnarray}
where ``$*$" is the convolution operator, that is, $\{g(\cdot, y)*f_{\hbox {\tiny $U$}}\}(x)=\int f_{\hbox {\tiny $U$}}(x-u)g(u, y)du$.
Because the dominating term within the square brackets in (\ref{eq:ordtwoinner}) is $$\left\{p(\cdot, y_{\hbox {\tiny $M$}})*f_{\hbox {\tiny $U$}}\right\}(x) \nu_2^{(2)} h_1^{-2b},$$ which is equal to $f_{\hbox {\tiny $W,Y$}}(x, y_{\hbox {\tiny $M$}})\nu_2^{(2)} h_1^{-2b}$  by (\ref{eq:fwy}), (\ref{eq:twoinner}) indicates that, for $\ell=0, 1$,
\begin{equation}
\textrm{Var}\{\hat T'_{n, \ell}(x,y_{\hbox {\tiny $M$}})\} 
 \le  \frac{f_{\hbox {\tiny $W,Y$}}(x, y_{\hbox {\tiny $M$}}) \nu_2^{(2)} \eta_\ell}{nh_1^{1+2b}h_2^3 c^2} +o\left(\frac{1}{nh_1^{1+2b}h_2^3}\right). \label{eq:ordvarT'}
\end{equation}
It follows that, when $U$ is ordinary smooth, the first term in (\ref{eq:T'meanvar}) dominates the second term there if $(nh_1^{1+2b}h_2^3)^{-1/2}=O\{h_2^{-1}r_4(\bh)\}$. 

If $U$ is super smooth of order $b$, by Lemma B.9 in \citet{Delaigle09}, under conditions (CK4) and (CK6), one has $$\int K^2_{\hbox {\tiny $U$}, \ell}(t) dt \le C h_1^{2b_2}\exp (2h_1^{-b}/d_2),$$ where $b_2=b_0I(b_0<0.5)$, and $C$ is some positive finite constant. Hence,
\begin{eqnarray}
& &\textrm{Var}\{\hat T'_{n, \ell}(x,y_{\hbox {\tiny $M$}})\} \nonumber \\
& \le & \frac{1}{nh_1 h_2^3} \int K^2_{\hbox {\tiny $U$}, \ell}(s) \int f_{\hbox {\tiny $U$}}(x+sh_1-u) \{p(u, y_{\hbox {\tiny $M$}}) \nu^{(2)}_2 +0.5 p_{yy}(u, y_{\hbox {\tiny $M$}})\nu_4^{(2)}h_2^2\nonumber \\
& & +O(h_2^3) \}du ds \nonumber \\
& \le & \frac{1}{nh_1 h_2^3} \int K^2_{\hbox {\tiny $U$}, \ell}(s) \int f_{\hbox {\tiny $U$}}(x+sh_1-u) \{C_p \nu^{(2)}_2+O(h_2^2)\} du ds, \textrm{ by (CP1),}\nonumber \\
& \le & \frac{\exp (2h_1^{-b}/d_2)}{nh_1^{1-2b_2} h_2^3}  \{CC_p \nu^{(2)}_2+O(h_2^2)\} \nonumber \\
& =&  \frac{ \exp (2h_1^{-b}/d_2) CC_p \nu_2^{(2)}}{nh_1^{1-2b_2} h_2^3}+O\left\{\frac{\exp (2h_1^{-b}/d_2)}{nh_1^{1-2b_2} h_2}\right\}. 
\label{eq:supvarT'}
\end{eqnarray}
Hence, when $U$ is super smooth, the first term in (\ref{eq:T'meanvar}) dominates the second term there if $\{\exp (2h_1^{-b}/d_2)/(nh_1^{1-2b_2} h_2^3)\}^{1/2}=O\{h_2^{-1}r_4(\bh)\}$.

\section{Concluding Lemma 3.3}
\setcounter{equation}{15}
Based on the mean and variance analysis of $\hat T'_{n, \ell}(x,y_{\hbox {\tiny $M$}})$ in Sections~\ref{s:meanT'} and \ref{s:varT'}, we now reach the conclusion that, if $(nh_1^{1+2b}h_2^3)^{-1/2}=O\{h_2^{-1}r_4(\bh)\}$ when $U$ ordinary smooth, or if $\exp (h_1^{-b}/d_2)/\sqrt{nh_1^{1-2b_2} h_2^3}=O\{h_2^{-1}r_4(\bh)\}$ when $U$ is super smooth, then 
\begin{equation}
\left\{
\begin{array}{l}
\hat T'_{n, 0}(x,y_{\hbox {\tiny $M$}}) = 0.5\left\{p_{xxy}(x,y_{\hbox {\tiny $M$}})\mu_2^{(1)}h_1^2+p_{yyy}(x,y_{\hbox {\tiny $M$}})h_2^2\right\}+O_{\hbox {\tiny $P$}}\left\{h_2^{-1}r_4(\bh)\right\}, \\
\hat T'_{n, 1}(x,y_{\hbox {\tiny $M$}}) = p_{xy}(x,y_{\hbox {\tiny $M$}})\mu_2^{(1)}h_1+O_{\hbox {\tiny $P$}}\left\{h_2^{-1}r_4(\bh)\right\}.
\end{array}
\right. \label{eq:twoT'}
\end{equation}
This completes the second task stated in Section~\ref{s:outline} in order to derive $E\{\hat p_y(y_{\hbox {\tiny $M$}}|x)\}$. 

Using (\ref{eq:twoS}) and (\ref{eq:twoT'}) in (\ref{eq:pyhat2}), we have 
\begin{eqnarray}
& & \hat p_y(y_{\hbox {\tiny $M$}}|x) \nonumber \\
& = & \hat S^{0,0}_n (x) \hat T'_{n, 0}(x,y_{\hbox {\tiny $M$}}) +\hat S^{0,1}_n (x) \hat T'_{n, 1}(x,y_{\hbox {\tiny $M$}}) \nonumber\\
& = &  f_{\hbox {\tiny $X$}}^{-1}(x) \left[ 0.5\left\{p_{xxy}(x,y_{\hbox {\tiny $M$}})\mu_2^{(1)}h_1^2+p_{yyy}(x,y_{\hbox {\tiny $M$}})h_2^2\right\} \right.\nonumber \\
& & \left.- f_{\hbox {\tiny $X$}}^{-1}(x) f'_{\hbox {\tiny $X$}}(x)p_{xy}(x,y_{\hbox {\tiny $M$}})\mu_2^{(1)}h^2_1\right] +O_{\hbox {\tiny $P$}}\left\{h_2^{-1}r_4(\bh)\right\}. \label{eq:dominating}
\end{eqnarray}
Because the dominating term in (\ref{eq:dominating}) is a non-random quantity, this dominating term is also the dominating bias of $\hat p_y(y_{\hbox {\tiny $M$}}|x)$. This proves Lemma 3.3 in the main article.

\setcounter{equation}{0}
\setcounter{figure}{0}
\setcounter{section}{0}
\renewcommand{\thefigure}{G.\arabic{figure}}
\renewcommand{\thesection}{G.\arabic{section}}
\renewcommand{\theequation}{G.\arabic{equation}}

\section*{Appendix G: Asymptotic variance of $\hat p_{y}(y_{\hbox {\tiny $M$}}|x)$}
By (\ref{eq:twoS}), 
\begin{align*}
& \hat p_y(y_{\hbox {\tiny $M$}}|x) \\ 
= & \sum_{\ell=0}^1 \hat S^{0,\ell}_n(x)\hat T'_{n, \ell}(x,y_{\hbox {\tiny $M$}}) \\
= & \left\{f^{-1}_{\hbox {\tiny $X$}}(x)+O_{\hbox {\tiny $P$}}(h_1^2)\right\}\hat T'_{n, 0}(x,y_{\hbox {\tiny $M$}}) +\left\{-h_1 f'_{\hbox {\tiny $X$}}(x) f^{-2}_{\hbox {\tiny $X$}}(x) +O_{\hbox {\tiny $P$}}(h_1^2)\right\}\hat T'_{n, 1}(x,y_{\hbox {\tiny $M$}}) \\
= & \frac{1}{n h_1 h_2^2} \sum_{j=1}^n \Big[ \left\{f^{-1}_{\hbox {\tiny $X$}}(x)+O_{\hbox {\tiny $P$}}(h_1^2)\right\}K_{\hbox {\tiny $U$}, 0}\left(\frac{W_j-x}{h_1}\right)K_2\left(\frac{Y_j-y_{\hbox {\tiny $M$}}}{h_2}\right)\left(\frac{Y_j-y_{\hbox {\tiny $M$}}}{h_2}\right)\\
& +\left\{-h_1 f'_{\hbox {\tiny $X$}}(x) f^{-2}_{\hbox {\tiny $X$}}(x) +O_{\hbox {\tiny $P$}}(h_1^2)\right\} K_{\hbox {\tiny $U$}, 1}\left(\frac{W_j-x}{h_1}\right)K_2\left(\frac{Y_j-y_{\hbox {\tiny $M$}}}{h_2}\right)\left(\frac{Y_j-y_{\hbox {\tiny $M$}}}{h_2}\right)\Big]. 
\end{align*}
Extracting the dominating terms in the above expression reveals that, to find the dominating variance of $\hat p_y(y_{\hbox {\tiny $M$}}|x)$, it suffices to look into the variance of  
\begin{eqnarray}
& & \frac{1}{n h_1 h_2^2} \sum_{j=1}^n \left\{ f^{-1}_{\hbox {\tiny $X$}}(x)K_{\hbox {\tiny $U$}, 0}\left(\frac{W_j-x}{h_1}\right)K_2\left(\frac{Y_j-y_{\hbox {\tiny $M$}}}{h_2}\right)\left(\frac{Y_j-y_{\hbox {\tiny $M$}}}{h_2}\right)\right. \nonumber\\
& & \left.-h_1 f'_{\hbox {\tiny $X$}}(x) f^{-2}_{\hbox {\tiny $X$}}(x) K_{\hbox {\tiny $U$}, 1}\left(\frac{W_j-x}{h_1}\right)K_2\left(\frac{Y_j-y_{\hbox {\tiny $M$}}}{h_2}\right)\left(\frac{Y_j-y_{\hbox {\tiny $M$}}}{h_2}\right)\right\}. \label{eq:keyvar}
\end{eqnarray}
This leads us to study the following variance and covariance, 
\begin{equation}
\textrm{Var} \left\{K_{\hbox {\tiny $U$}, \ell}\left(\frac{W_j-x}{h_1}\right)K_2\left(\frac{Y_j-y_{\hbox {\tiny $M$}}}{h_2}\right)\left(\frac{Y_j-y_{\hbox {\tiny $M$}}}{h_2}\right)\right\}, \textrm{ for $\ell=0,1$,}\label{eq:term12}
\end{equation}
\begin{equation}
\resizebox{.98 \textwidth}{!}
{$
\textrm{Cov} \left\{ K_{\hbox {\tiny $U$}, 0}\left(\frac{W_j-x}{h_1}\right)K_2\left(\frac{Y_j-y_{\hbox {\tiny $M$}}}{h_2}\right)\left(\frac{Y_j-y_{\hbox {\tiny $M$}}}{h_2}\right), 
 K_{\hbox {\tiny $U$}, 1}\left(\frac{W_j-x}{h_1}\right)K_2\left(\frac{Y_j-y_{\hbox {\tiny $M$}}}{h_2}\right)\left(\frac{Y_j-y_{\hbox {\tiny $M$}}}{h_2}\right)\right\}.
$} \label{eq:term3} 
\end{equation}

\section{Deriving the variance in (\ref{eq:term12})}
\setcounter{equation}{3}
For $\ell=0, 1$, 
\begin{eqnarray}
& & \textrm{Var}\left\{K_{\hbox {\tiny $U$}, \ell}\left(\frac{W_j-x}{h_1}\right)K_2\left(\frac{Y_j-y_{\hbox {\tiny $M$}}}{h_2}\right)\left(\frac{Y_j-y_{\hbox {\tiny $M$}}}{h_2}\right)\right\} \nonumber \\
& = & E\left\{K_{\hbox {\tiny $U$}, \ell}^2\left(\frac{W_j-x}{h_1}\right)K_2^2\left(\frac{Y_j-y_{\hbox {\tiny $M$}}}{h_2}\right)\left(\frac{Y_j-y_{\hbox {\tiny $M$}}}{h_2}\right)^2\right\} \label{eq:meansq}\\
& & -\left[E\left\{K_{\hbox {\tiny $U$}, \ell}\left(\frac{W_j-x}{h_1}\right)K_2\left(\frac{Y_j-y_{\hbox {\tiny $M$}}}{h_2}\right)\left(\frac{Y_j-y_{\hbox {\tiny $M$}}}{h_2}\right) \right\}\right]^2. \label{eq:sqmean}
\end{eqnarray}
The expectation in (\ref{eq:meansq}) is considered in Section~\ref{s:varT'} (see (\ref{eq:var01})). From there, we have shown that, if $U$ is ordinary smooth of order $b$, then (relating to (\ref{eq:ordvarT'}))
\begin{eqnarray}
& & E\left\{K_{\hbox {\tiny $U$}, \ell}^2\left(\frac{W_j-x}{h_1}\right)K_2^2\left(\frac{Y_j-y_{\hbox {\tiny $M$}}}{h_2}\right)\left(\frac{Y_j-y_{\hbox {\tiny $M$}}}{h_2}\right)^2\right\}\nonumber \\
& = & h_1^{1-2b} h_2 c^{-2} \nu_2^{(2)} \eta_\ell f_{\hbox {\tiny $W,Y$}}(x,y_{\hbox {\tiny $M$}})+o\left(h_1^{1-2b}h_2\right); \label{eq:meansqord}
\end{eqnarray}
and, if $U$ is super smooth of order $b$, then (relating to (\ref{eq:supvarT'}))
\begin{eqnarray}
& & E\left\{K_{\hbox {\tiny $U$}, \ell}^2\left(\frac{W_j-x}{h_1}\right)K_2^2\left(\frac{Y_j-y_{\hbox {\tiny $M$}}}{h_2}\right)\left(\frac{Y_j-y_{\hbox {\tiny $M$}}}{h_2}\right)^2\right\} \nonumber \\
& \le & h_1^{1+2b_2}h_2\exp\left(2h_1^{-b}/d_2\right)C C_p \nu_2^{(2)}+O\left\{h_1^{1-2b_2}h_2^3\exp\left(2h_1^{-b}/d_2\right)\right\}. \label{eq:meansqsup}
\end{eqnarray}
The expectation in (\ref{eq:sqmean}) is considered in Section~\ref{s:meanT'} (see (\ref{eq:mean01})). By (\ref{eq:twoET'}), this expectation is of order $O\{h_1h_2^2(h_1^2+h_2^2)\}$ when $\ell =0$, and it is of order $O(h_1^2h_2^2)$ when $\ell=1$. Hence, for both $\ell=0, 1$, (\ref{eq:sqmean}) tends to zero faster than (\ref{eq:meansqord}) and (\ref{eq:meansqsup}). 

It follows that that, for ordinary smooth $U$, 
\begin{eqnarray}
& & \textrm{Var}\left\{K_{\hbox {\tiny $U$}, \ell}\left(\frac{W_j-x}{h_1}\right)K_2\left(\frac{Y_j-y_{\hbox {\tiny $M$}}}{h_2}\right)\left(\frac{Y_j-y_{\hbox {\tiny $M$}}}{h_2}\right)\right\}  \nonumber \\
& = & h_1^{1-2b} h_2  \nu_2^{(2)} c^{-2}\eta_\ell f_{\hbox {\tiny $W,Y$}}(x,y_{\hbox {\tiny $M$}})+o\left(h_1^{1-2b}h_2\right); \label{eq:term1ord}
\end{eqnarray}
and, for super smooth $U$, 
\begin{align}
& \textrm{Var}\left\{K_{\hbox {\tiny $U$}, \ell}\left(\frac{W_j-x}{h_1}\right)K_2\left(\frac{Y_j-y_{\hbox {\tiny $M$}}}{h_2}\right)\left(\frac{Y_j-y_{\hbox {\tiny $M$}}}{h_2}\right)\right\} \nonumber  \\
\le & h_1^{1+2b_2}h_2\exp\left(2h_1^{-b}/d_2\right)C C_p \nu_2^{(2)}+O\left\{h_1^{1-2b_2}h_2^3\exp\left(2h_1^{-b}/d_2\right)\right\}. \label{eq:term1sup} 
\end{align}

\section{Deriving the covariance in (\ref{eq:term3})}
\setcounter{equation}{9}
The covariance in (\ref{eq:term3}) is equal to  
\begin{align}
& E\left\{ K_{\hbox {\tiny $U$}, 0}\left(\frac{W_j-x}{h_1}\right)K_{\hbox {\tiny $U$}, 1}\left(\frac{W_j-x}{h_1}\right)K_2^2\left(\frac{Y_j-y_{\hbox {\tiny $M$}}}{h_2}\right)\left(\frac{Y_j-y_{\hbox {\tiny $M$}}}{h_2}\right)^2  \right\}-  \label{eq:meanprod}\\
& E\left\{K_{\hbox {\tiny $U$}, 0}\left(\frac{W_j-x}{h_1}\right) K_2\left(\frac{Y_j-y_{\hbox {\tiny $M$}}}{h_2}\right)\left(\frac{Y_j-y_{\hbox {\tiny $M$}}}{h_2}\right)\right\} \times\nonumber \\
& E\left\{K_{\hbox {\tiny $U$}, 1}\left(\frac{W_j-x}{h_1}\right) K_2\left(\frac{Y_j-y_{\hbox {\tiny $M$}}}{h_2}\right)\left(\frac{Y_j-y_{\hbox {\tiny $M$}}}{h_2}\right)\right\}. \label{eq:prodmean}
\end{align}
In Section~\ref{s:meanT'}, we have shown that the product in (\ref{eq:prodmean}) is of order $O(h_1^3h_2^4)$, which tends to zero faster than (\ref{eq:meanprod}), as to be revealed next. 

When $U$ is ordinary smooth, by Lemma B.4 in \citet{Delaigle09},  
\begin{eqnarray*}
& & E\left\{ K_{\hbox {\tiny $U$}, 0}\left(\frac{W_j-x}{h_1}\right)K_{\hbox {\tiny $U$}, 1}\left(\frac{W_j-x}{h_1}\right)K_2^2\left(\frac{Y_j-y_{\hbox {\tiny $M$}}}{h_2}\right)\left(\frac{Y_j-y_{\hbox {\tiny $M$}}}{h_2}\right)^2  \right\} \\
& = & h_1^{1-2b} h_2 \nu_2^{(2)}c^{-2} \eta_{01} f_{\hbox {\tiny $W,Y$}}(x,y_{\hbox {\tiny $M$}})+o\left(h_1^{1-2b}h_2\right),
\end{eqnarray*}
where $\eta_{01}=\int |t|^{2b}\phi_{\hbox {\tiny $K_1$}}(t)\phi'_{\hbox {\tiny $K_1$}}(t) dt$.
When $U$ is super smooth, by Lemma B.9 in \citet{Delaigle09}, 
\begin{eqnarray*}
& & E\left\{ K_{\hbox {\tiny $U$}, 0}\left(\frac{W_j-x}{h_1}\right)K_{\hbox {\tiny $U$}, 1}\left(\frac{W_j-x}{h_1}\right)K_2^2\left(\frac{Y_j-y_{\hbox {\tiny $M$}}}{h_2}\right)\left(\frac{Y_j-y_{\hbox {\tiny $M$}}}{h_2}\right)^2  \right\} \\
& \le & h_1^{1+2b_2}h_2\exp\left(2h_1^{-b}/d_2\right)C C_p \nu_2^{(2)}+O\left\{h_1^{1+2b_2}h_2^3\exp\left(2h_1^{-b}/d_2\right)\right\}. 
\end{eqnarray*}
Hence, the covariance in (\ref{eq:term3}) and variances in (\ref{eq:term12}) are of the same order. 

\section{Concluding Lemma 3.4}
Since (\ref{eq:term12}) and (\ref{eq:term3}) are of the same order, the variance of (\ref{eq:keyvar}) with $y=y_{\hbox {\tiny $M$}}$ is dominated by 
$$\frac{1}{nh_1^2h_2^4}\textrm{Var}\left\{K_{\hbox {\tiny $U$}, 0}\left(\frac{W_j-x}{h_1}\right)K_2\left(\frac{Y_j-y_{\hbox {\tiny $M$}}}{h_2}\right)\left(\frac{Y_j-y_{\hbox {\tiny $M$}}}{h_2}\right)\right\}.$$ 
Hence, for ordinary smooth $U$, by (\ref{eq:term1ord}) with $\ell=0$, 
\begin{equation*}
\textrm{Var}\left\{ \hat p_y(y_{\hbox {\tiny $M$}}|x)  \right\}
=\frac{\eta_0\nu_2^{(2)} f_{\hbox {\tiny $W,Y$}}(x,y_{\hbox {\tiny $M$}})}{nh_1^{1+2b} h_2^3 c^2 f^2_{\hbox {\tiny $X$}}(x)}+o\left(\frac{1}{nh_1^{1+2b}h_2^3}\right);
\end{equation*}  
and, for super smooth $U$, by (\ref{eq:term1sup}), 
\begin{equation*}
\textrm{Var}\left\{ \hat p_y(y_{\hbox {\tiny $M$}}|x)  \right\}
\le \frac{\exp\left(2h_1^{-b}/d_2\right)C C_p \nu_2^{(2)}}{nh_1^{1-2b_2}h_2^3f^2_{\hbox {\tiny $X$}}(x)}+O\left\{\frac{\exp\left(2h_1^{-b}/d_2\right)}{nh_1^{1-2b_2}h_2}\right\}. 
\end{equation*}	
This proves Lemma 3.4 in the main article. 

\setcounter{equation}{0}
\setcounter{figure}{0}
\setcounter{section}{0}
\renewcommand{\theequation}{H.\arabic{equation}}
\renewcommand{\thefigure}{H.\arabic{figure}}
\renewcommand{\thesection}{H.\arabic{section}}
\section*{Appendix H: Pictorial demonstrations of simulation results}
\begin{figure}[h]
	\centering
	\setlength{\linewidth}{5cm}
	\subfigure[]{ \includegraphics[width=\linewidth]{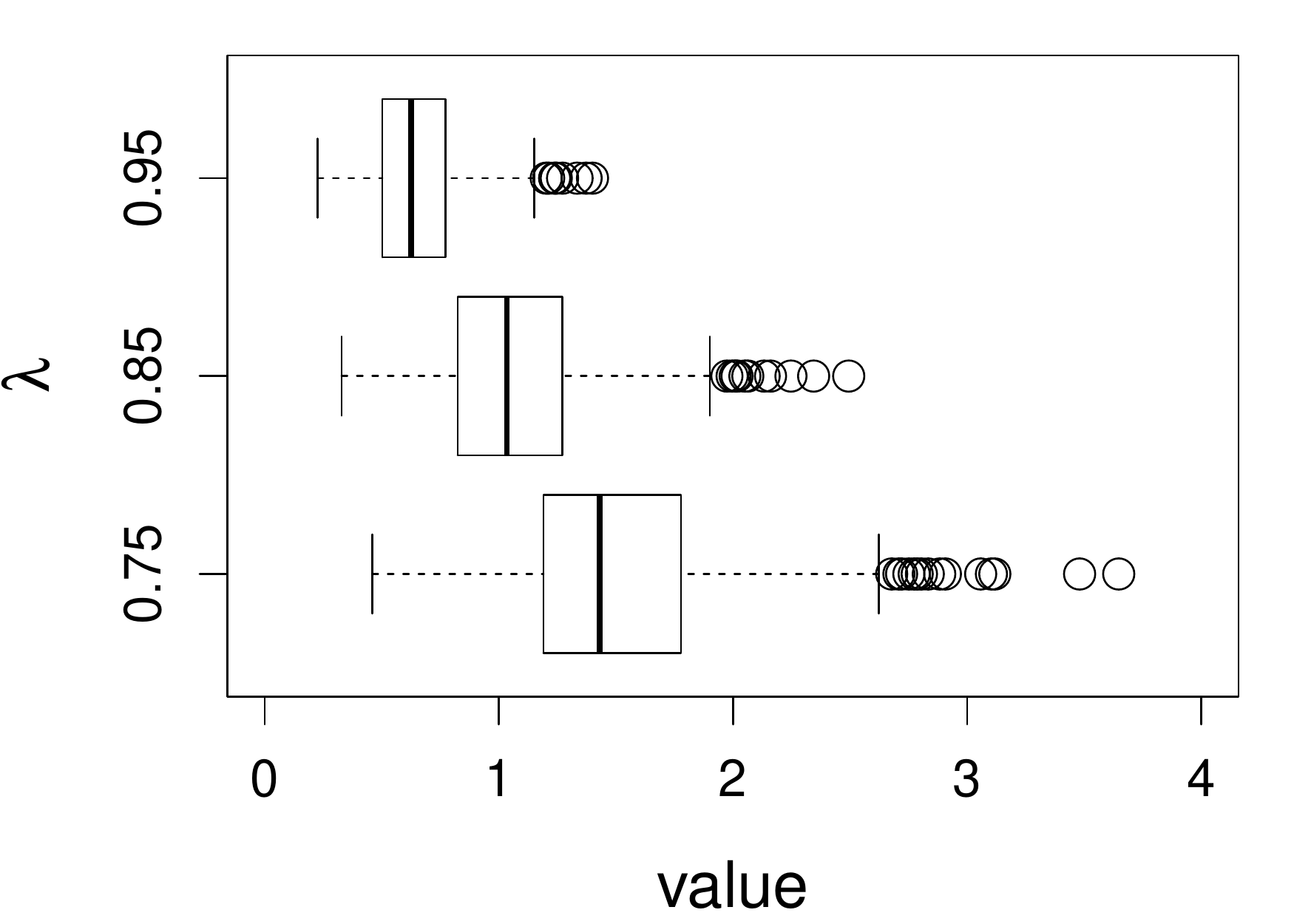} }
	\subfigure[]{ \includegraphics[width=\linewidth]{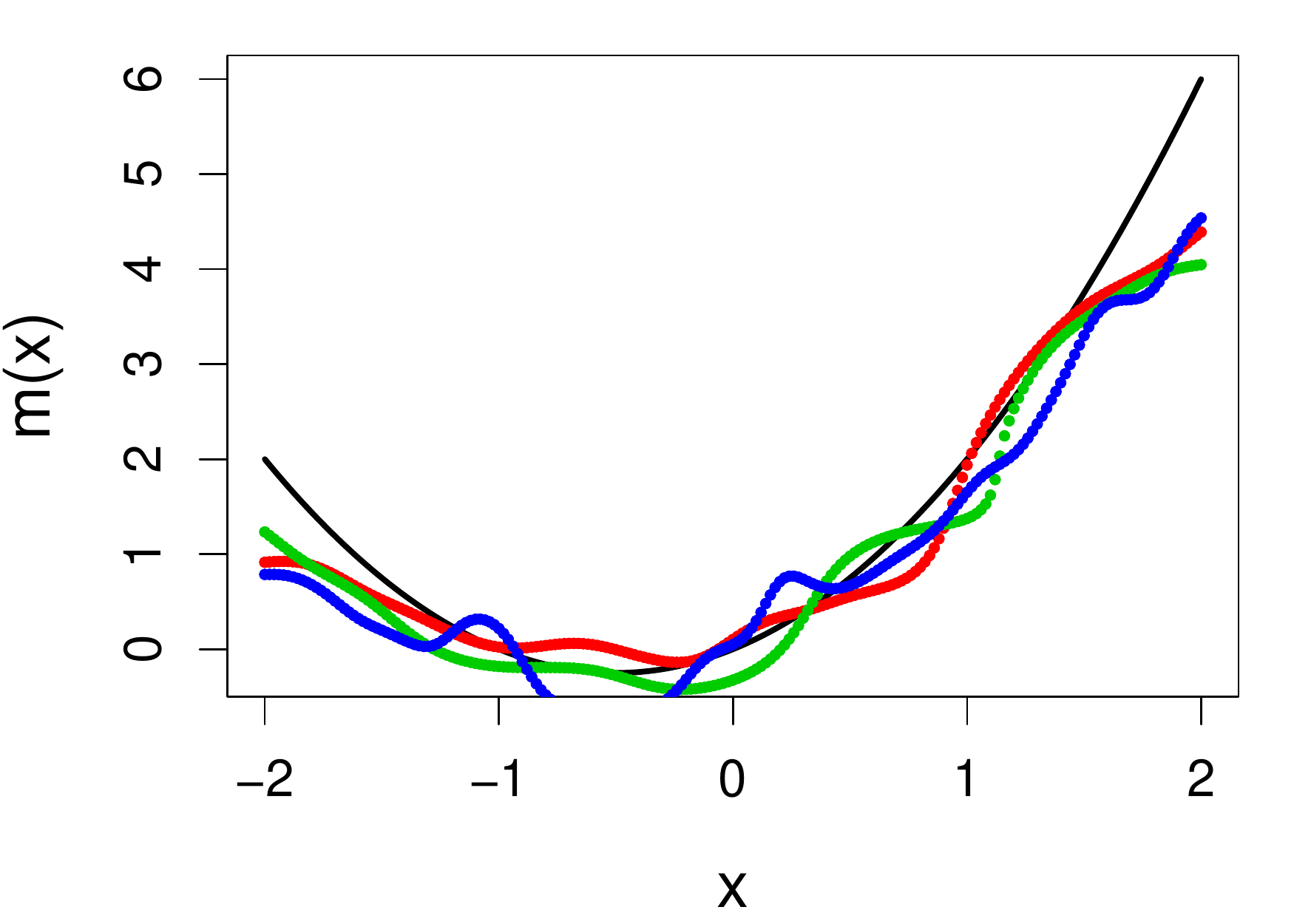} }\\
	\subfigure[]{ \includegraphics[width=\linewidth]{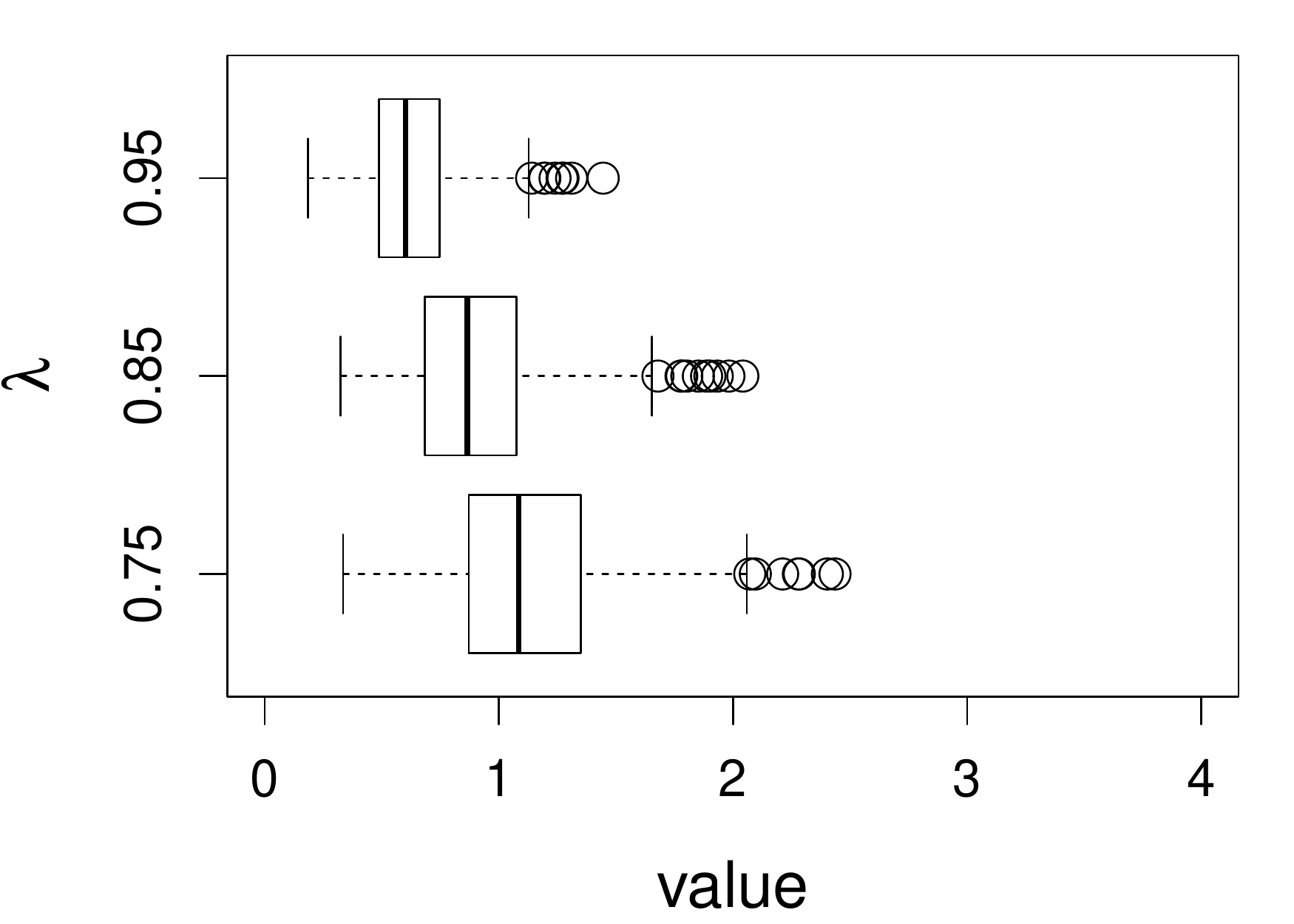} }
	\subfigure[]{ \includegraphics[width=\linewidth]{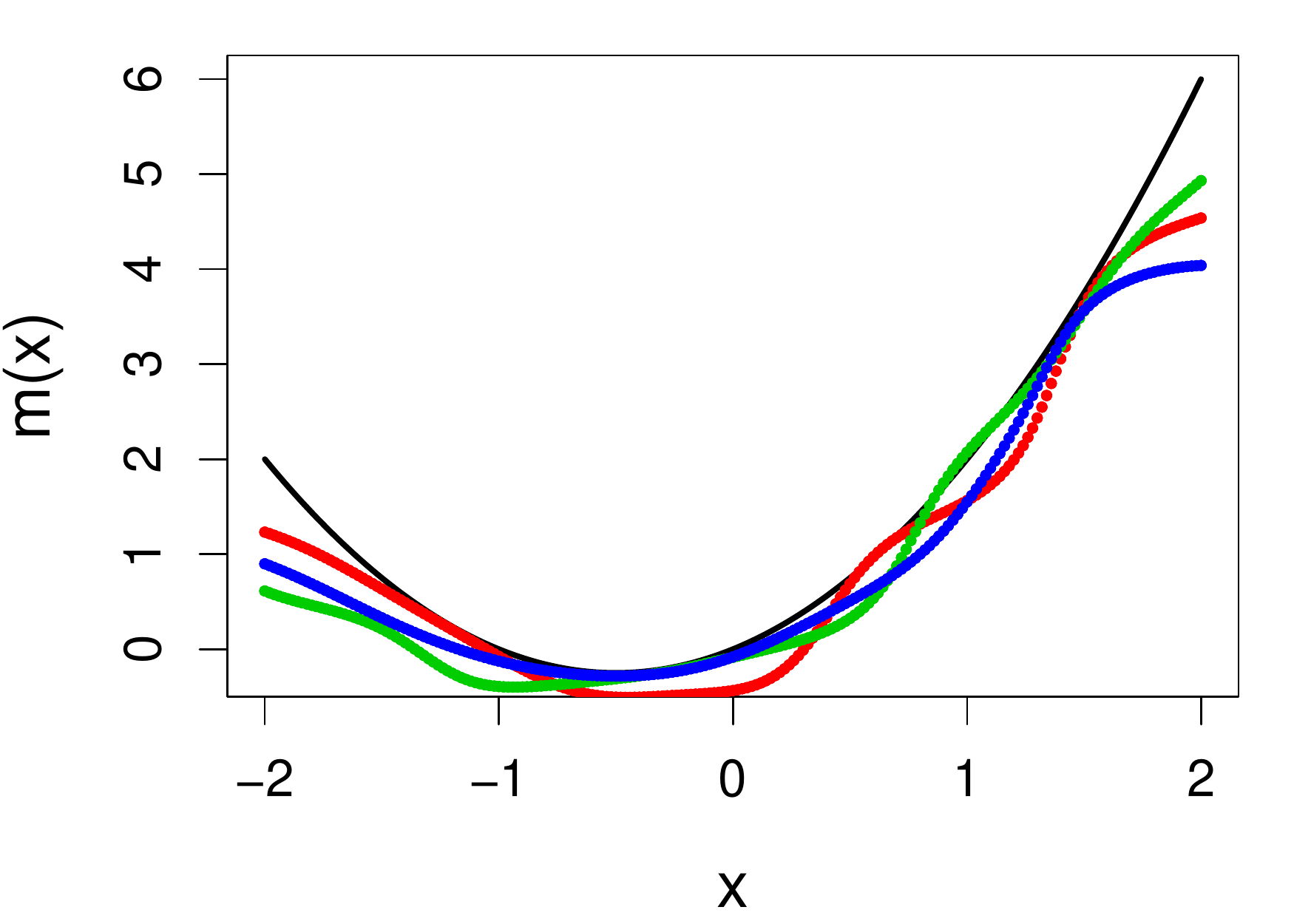} }\\
	\subfigure[]{ \includegraphics[width=\linewidth]{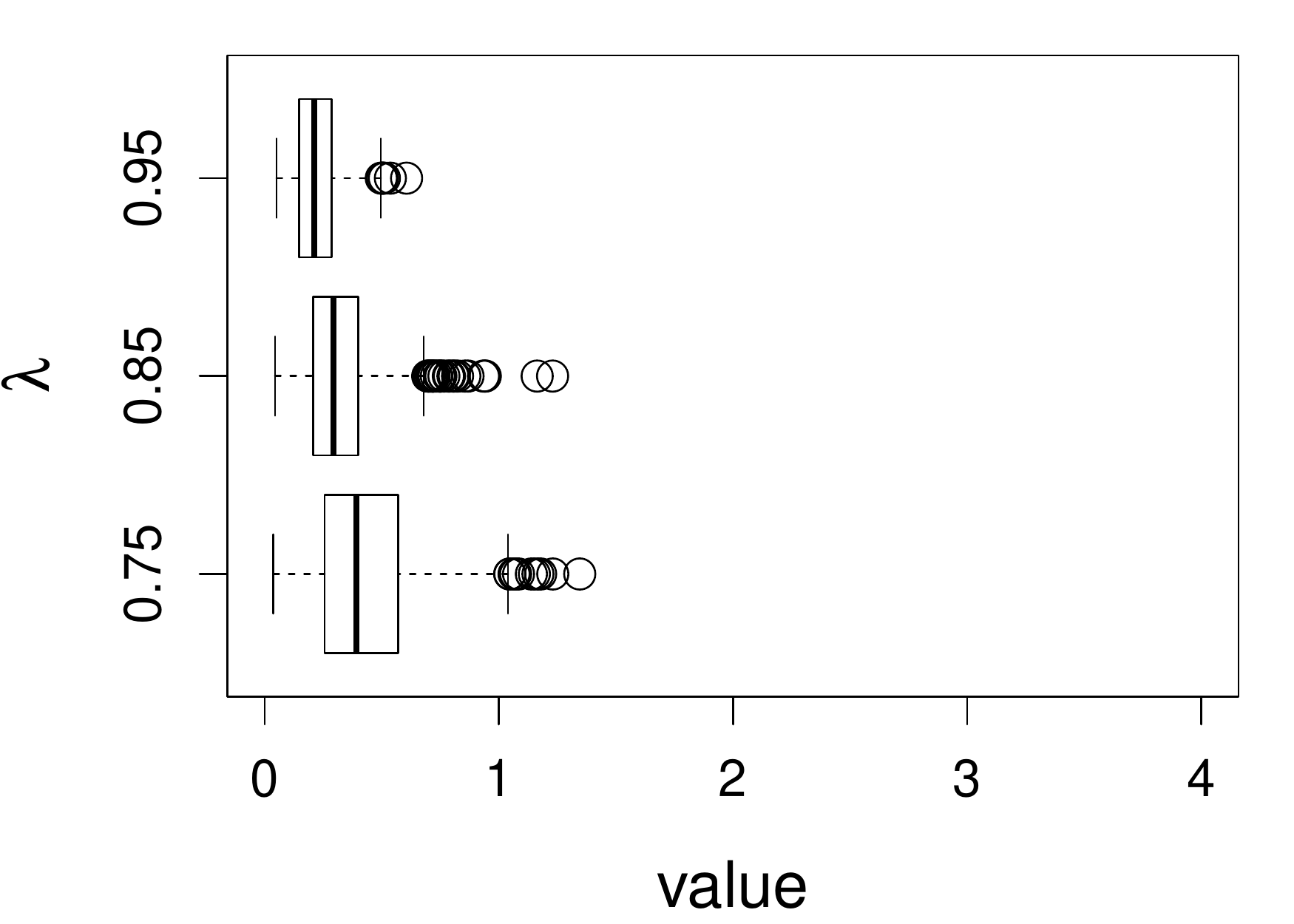} }
	\subfigure[]{ \includegraphics[width=\linewidth]{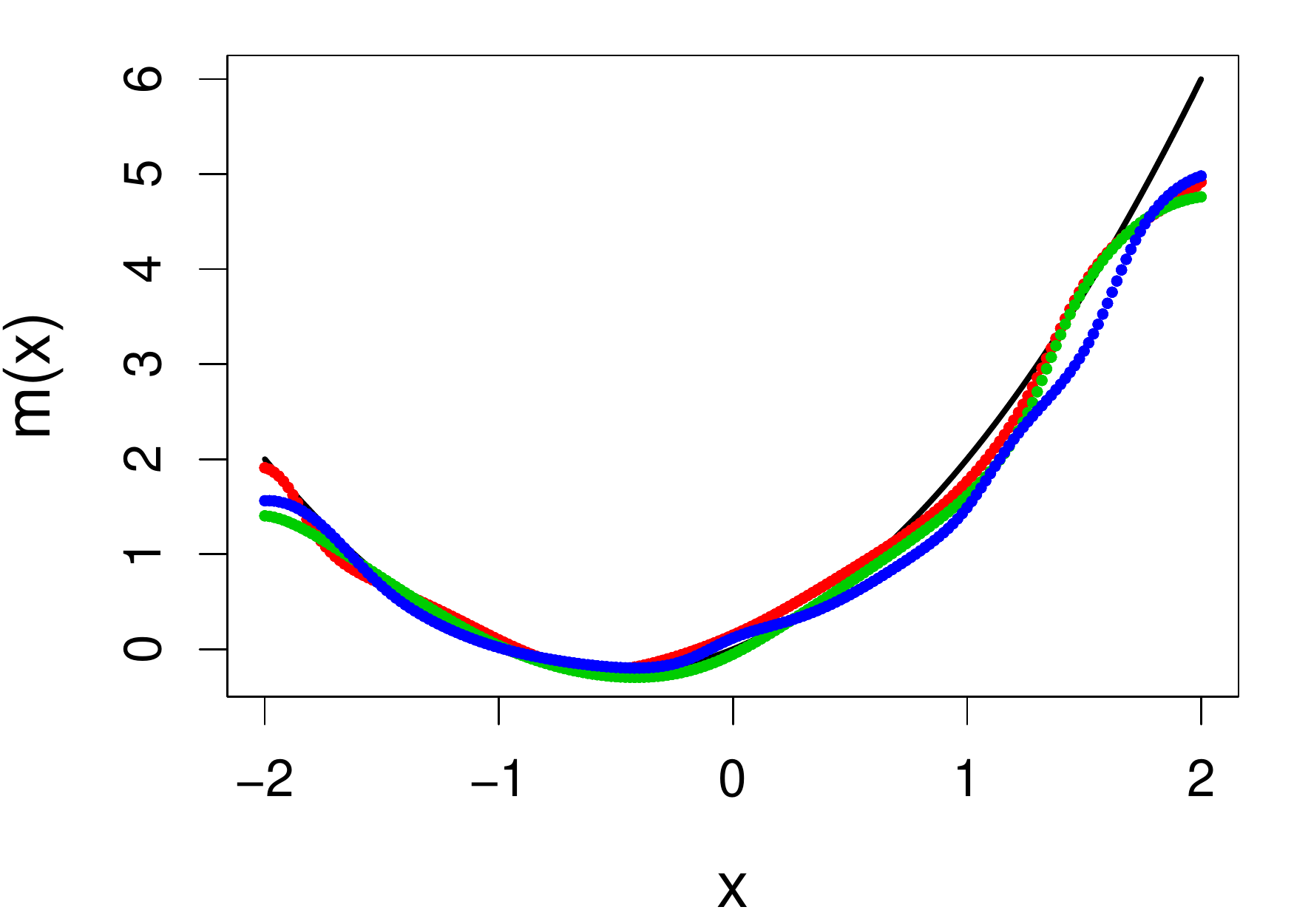} }
	\caption{Results under (C1) using approximated theoretical optimal bandwidths. Panels (a), (c), and (e): boxplots of ISEs versus $\lambda$ for $\hat M_{\hbox {\tiny $N$}}(x)$, $\hat M_0(x)$, and $\hat M_1(x)$, respectively. Panels (b), (d), and (f): estimated mode curves, $\hat M_{\hbox {\tiny $N$}}(x)$, $\hat M_0(x)$, and $\hat M_1(x)$, respectively, when $\lambda=0.85$. In each panel with estimated mode curves associated with an estimator, the black line depicts the true mode curve, the red, green, and blue lines are three estimated mode curves from the same method that yield ISE being the first, second, and third quantiles among the 500 ISEs for that method from the simulation, respectively.}
	\label{Sim1LapLap500:box}
\end{figure}

\begin{figure}
	\centering
	\setlength{\linewidth}{5cm}
	\subfigure[]{ \includegraphics[width=\linewidth]{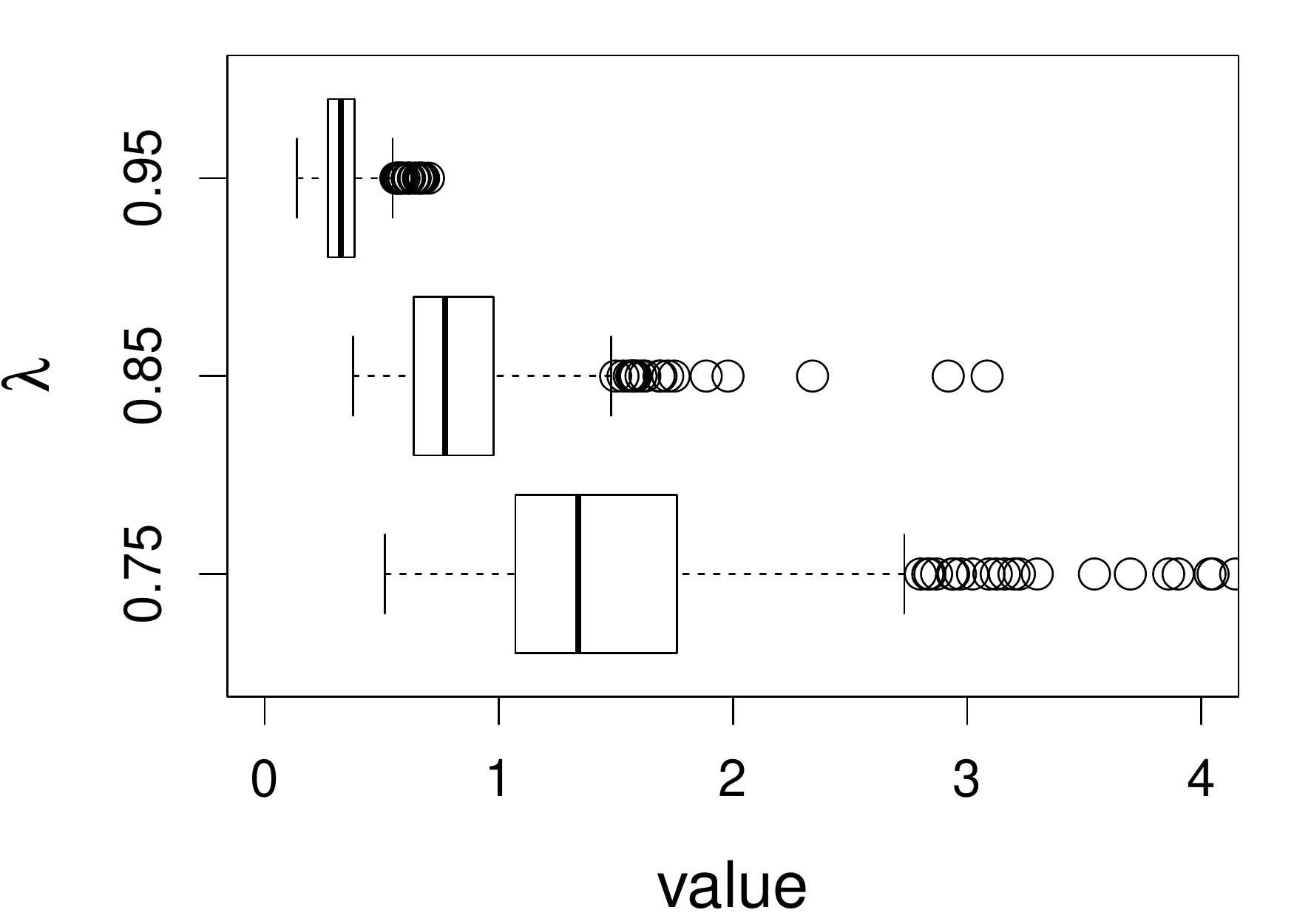} }
	\subfigure[]{ \includegraphics[width=\linewidth]{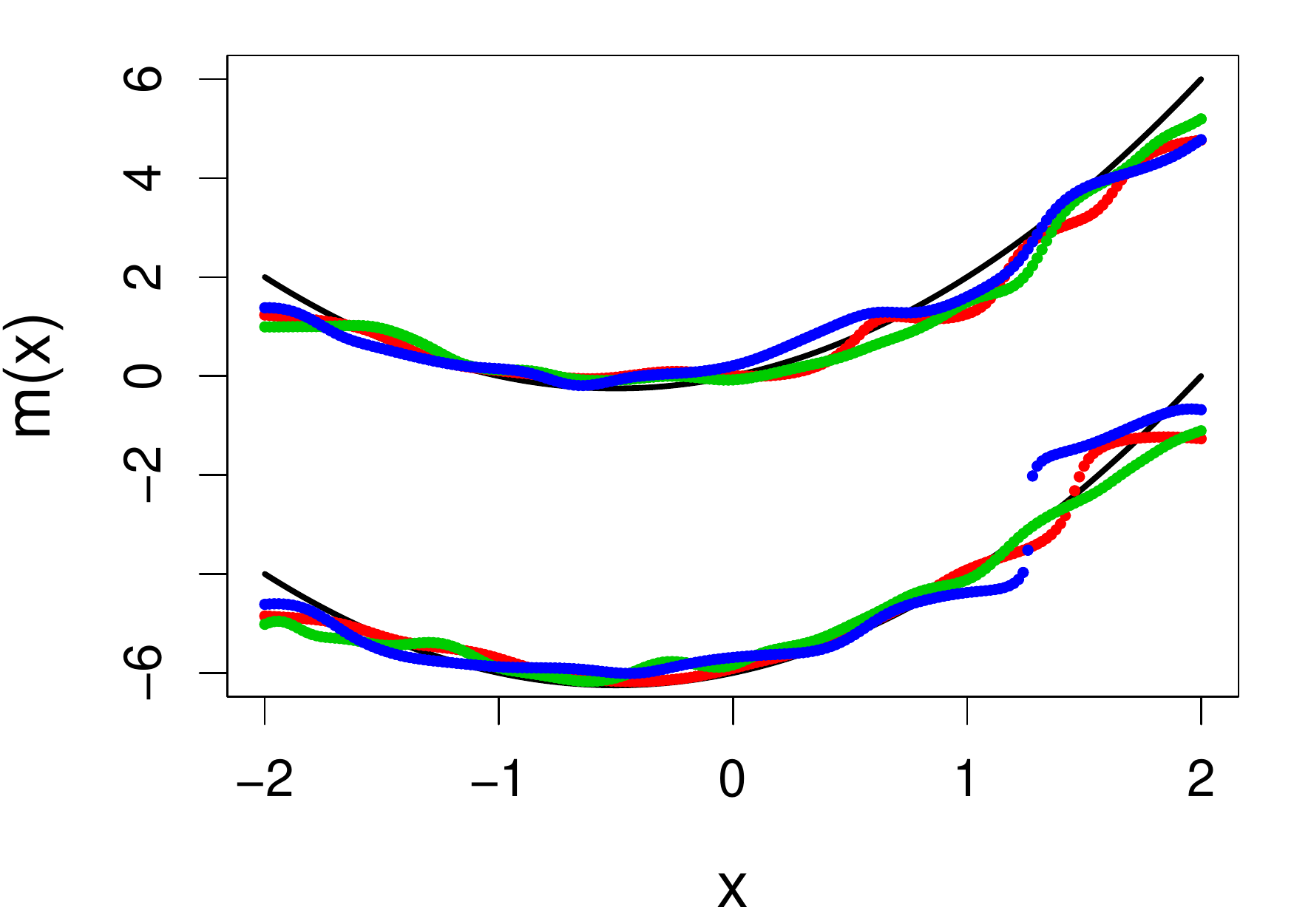} }\\
	\subfigure[]{ \includegraphics[width=\linewidth]{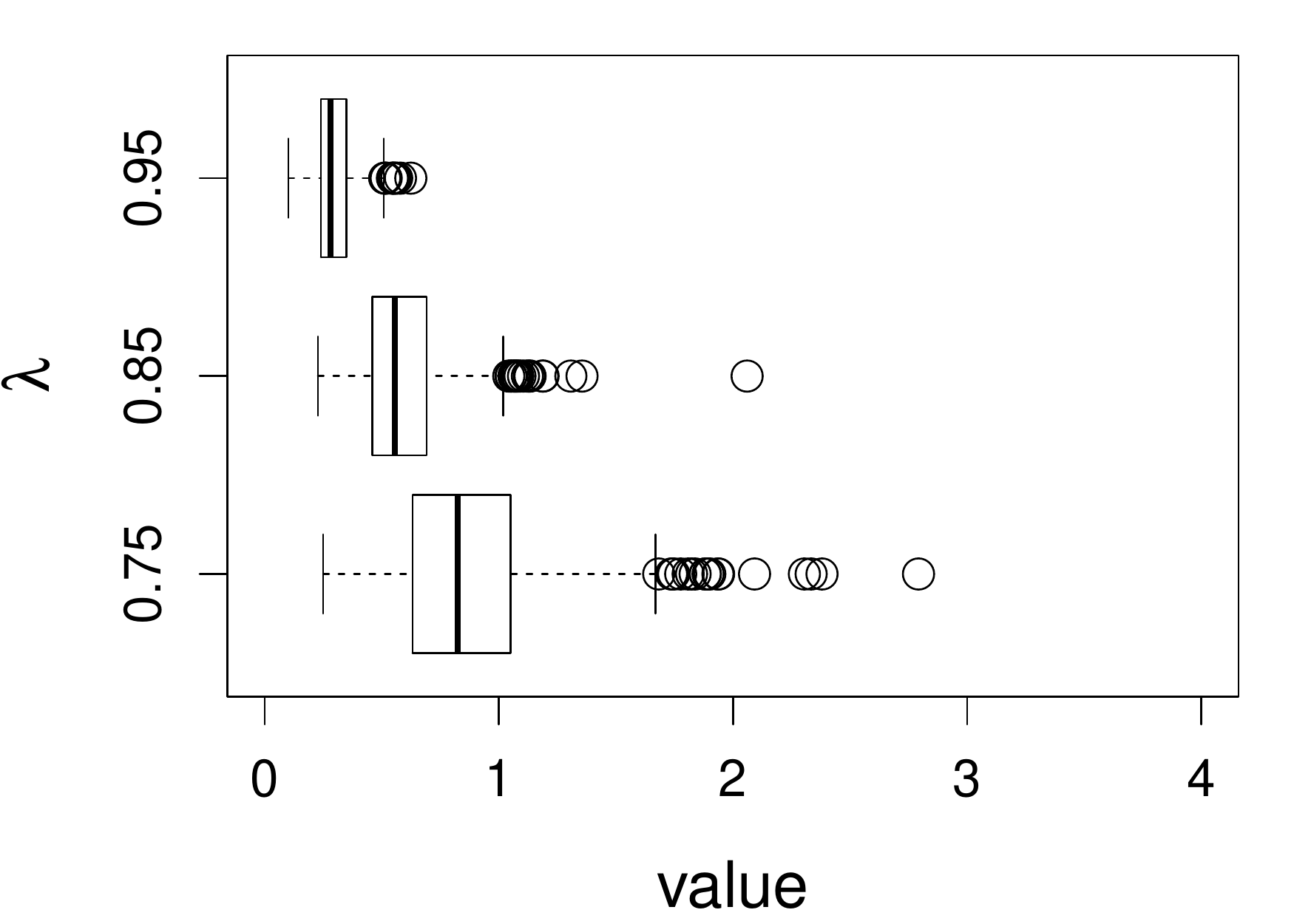} }
	\subfigure[]{ \includegraphics[width=\linewidth]{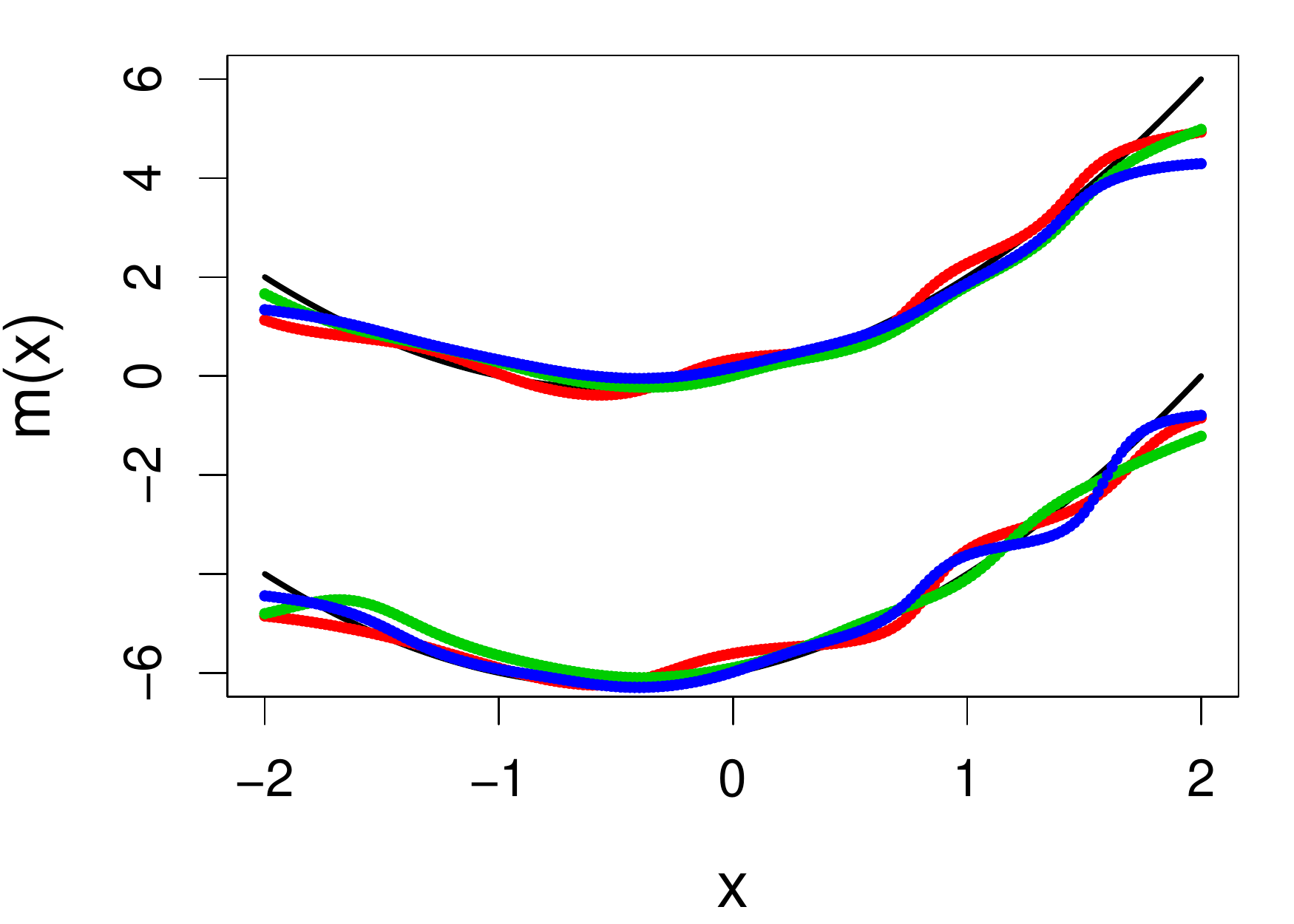} }\\
	\subfigure[]{ \includegraphics[width=\linewidth]{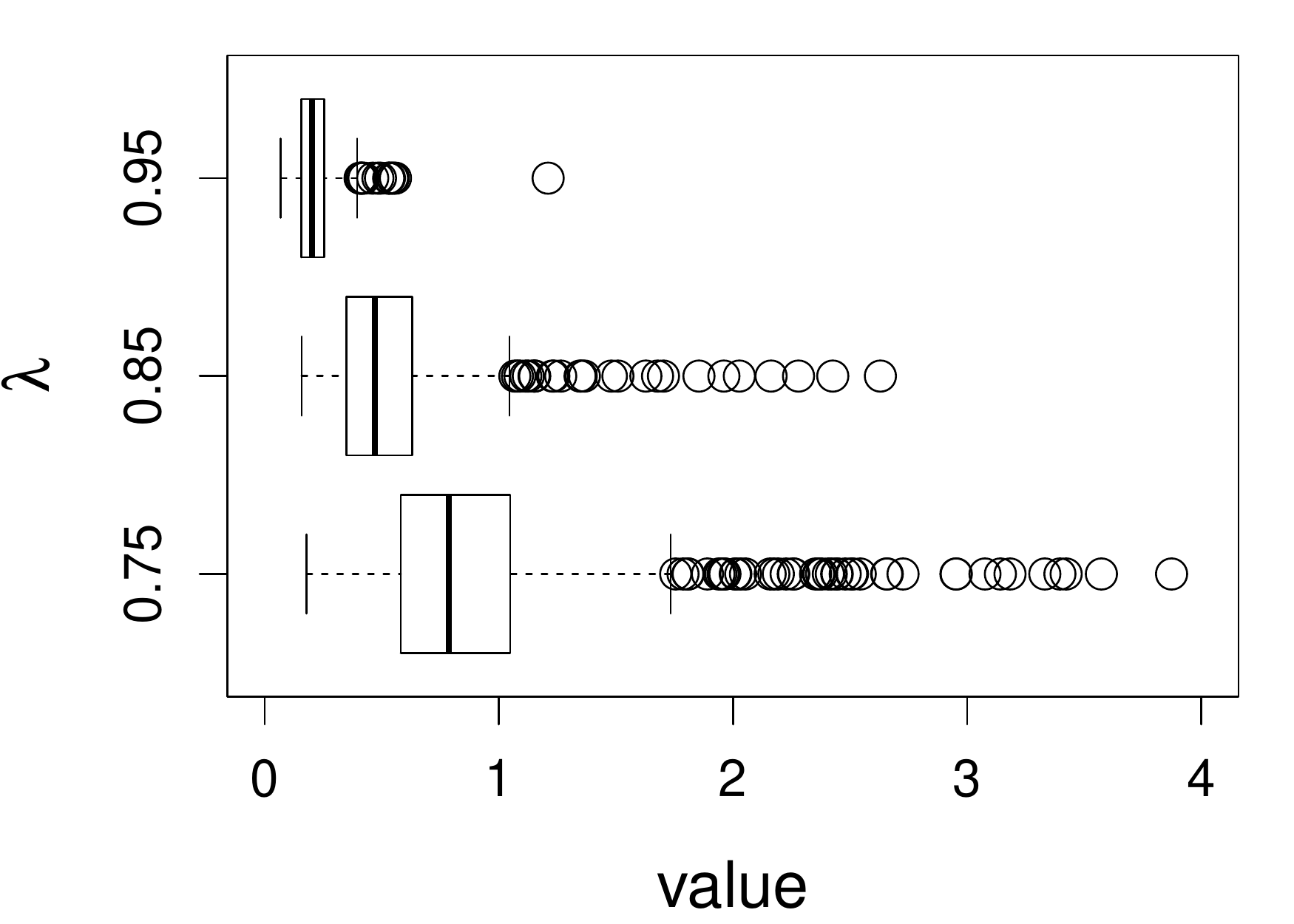} }
	\subfigure[]{ \includegraphics[width=\linewidth]{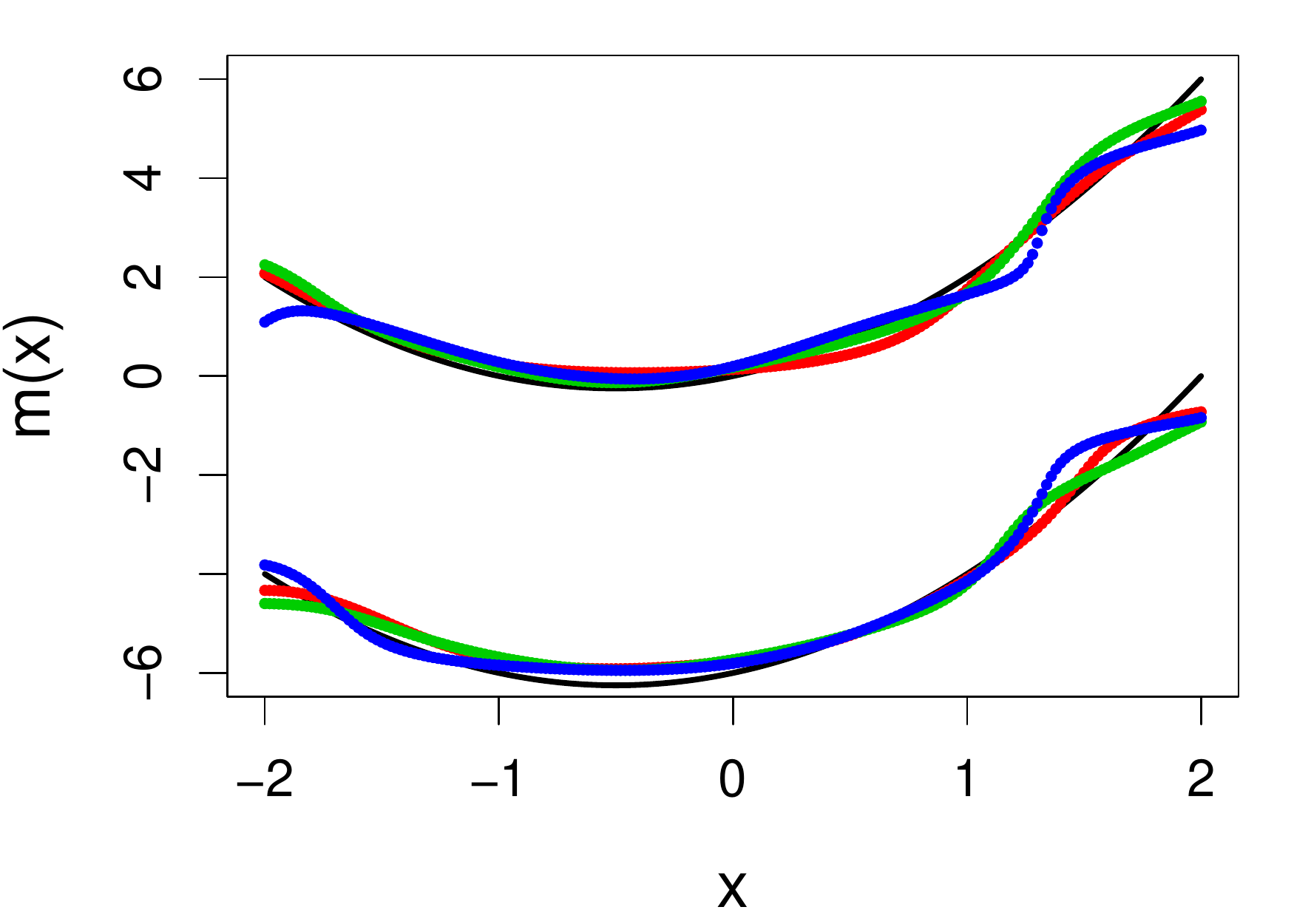} }
	\caption{Results under (C2) using approximated theoretical optimal bandwidths. Panels (a), (c), and (e): boxplots of ISEs versus $\lambda$ for $\hat M_{\hbox {\tiny $N$}}(x)$, $\hat M_0(x)$, and $\hat M_1(x)$, respectively. Panels (b), (d), and (f): estimated mode curves, $\hat M_{\hbox {\tiny $N$}}(x)$, $\hat M_0(x)$, and $\hat M_1(x)$, respectively, when $\lambda=0.85$. In each panel with estimated mode curves associated with an estimator, the black lines depict the true mode curves, the red, green, and blue lines are three estimated mode curves from the same method that yield ISE being the first, second, and third quantiles among the 500 ISEs for that method from the simulation, respectively.}
	\label{Sim2LapLap500:box}
\end{figure}

\begin{figure}
	\centering
	\setlength{\linewidth}{5cm}
	\subfigure[]{ \includegraphics[width=\linewidth]{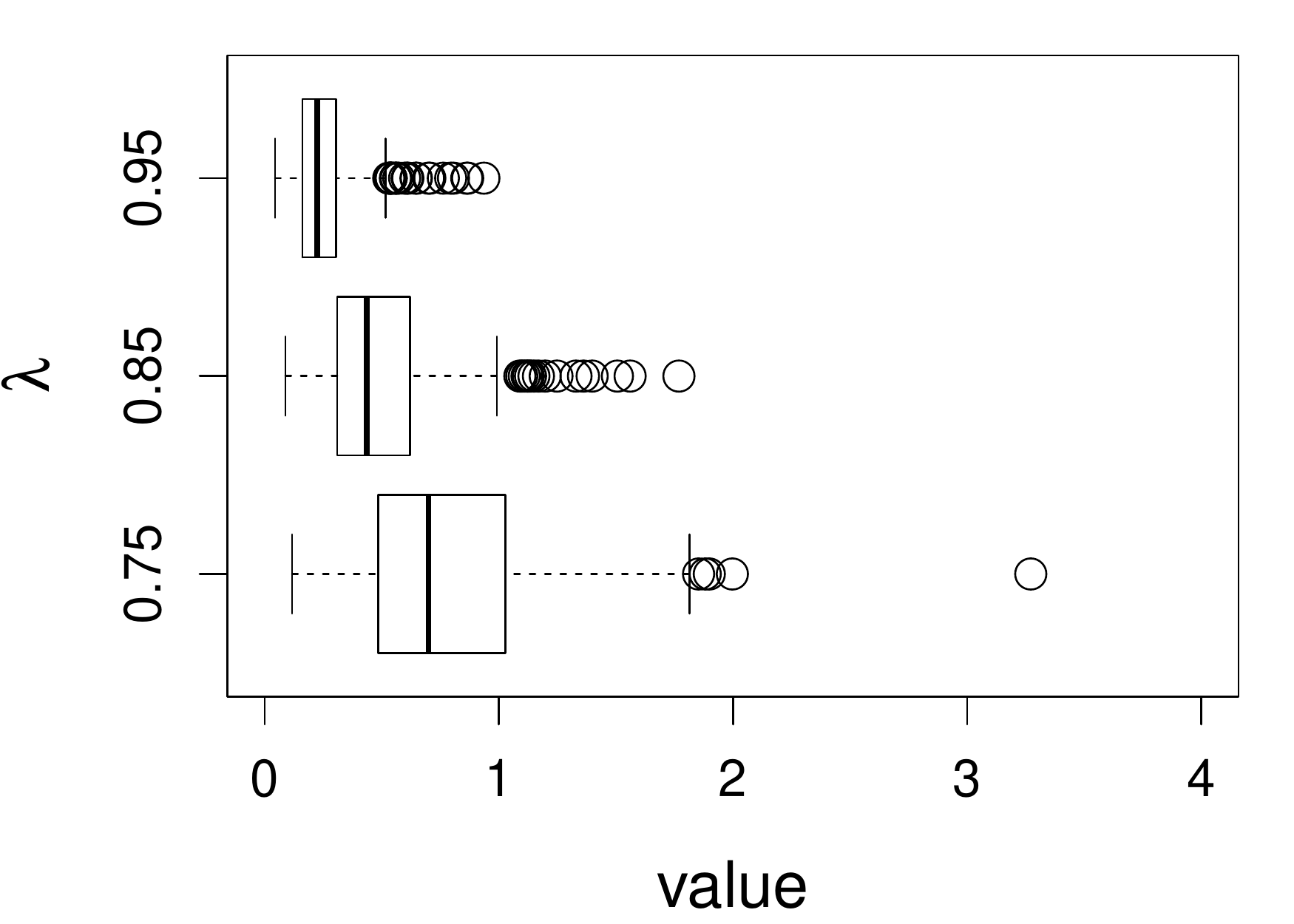} }
	\subfigure[]{ \includegraphics[width=\linewidth]{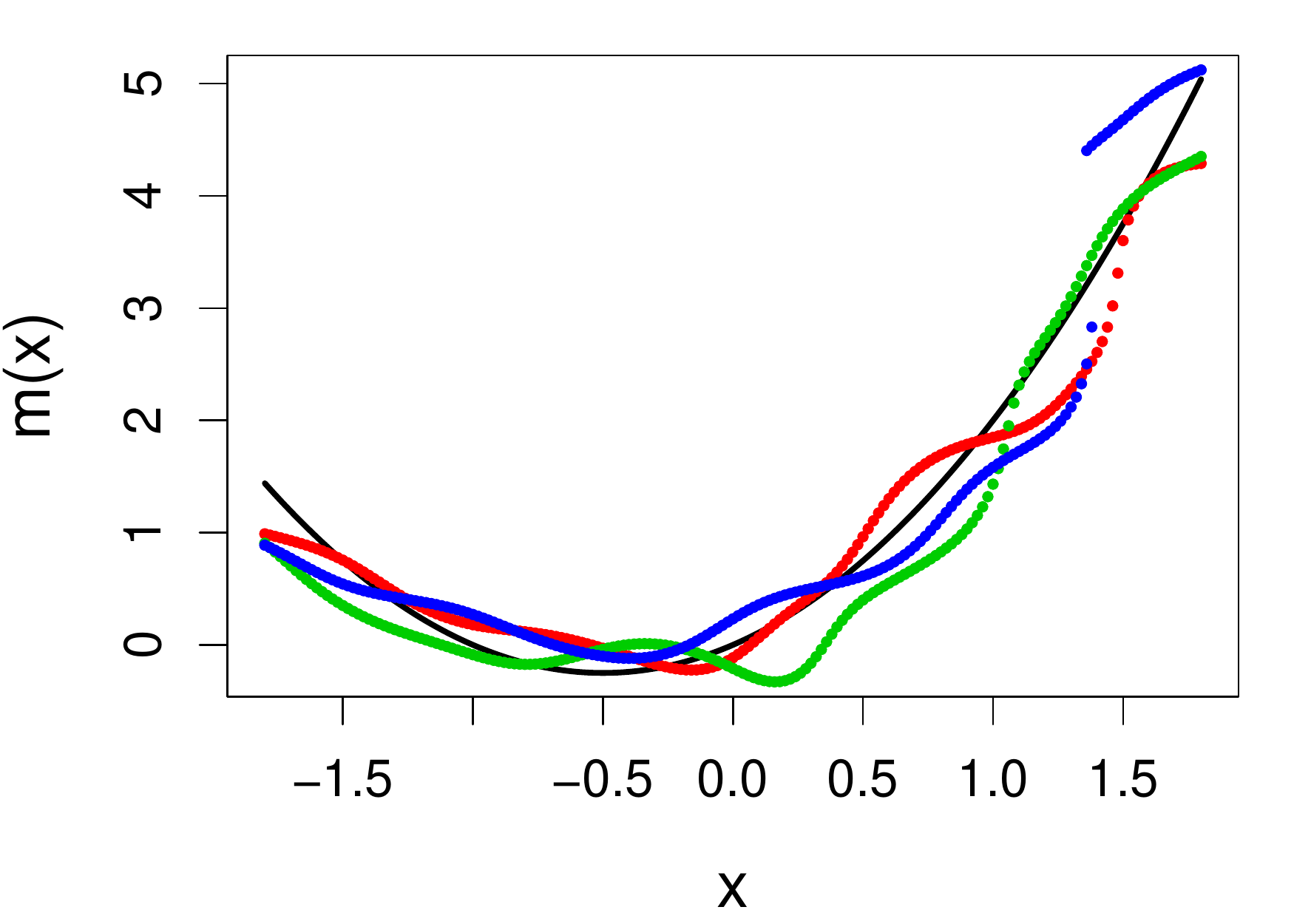} }\\
	\subfigure[]{ \includegraphics[width=\linewidth]{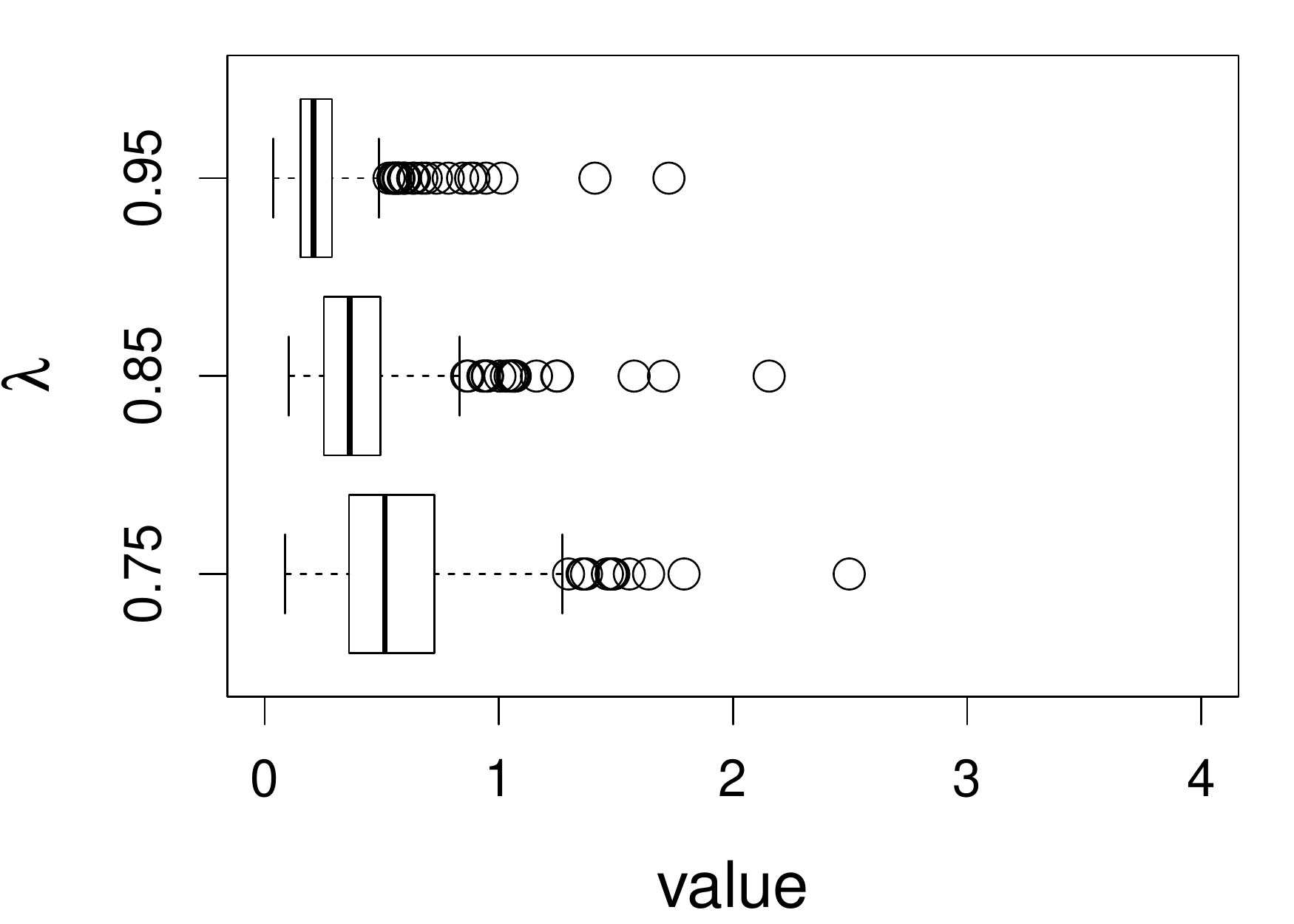} }
	\subfigure[]{ \includegraphics[width=\linewidth]{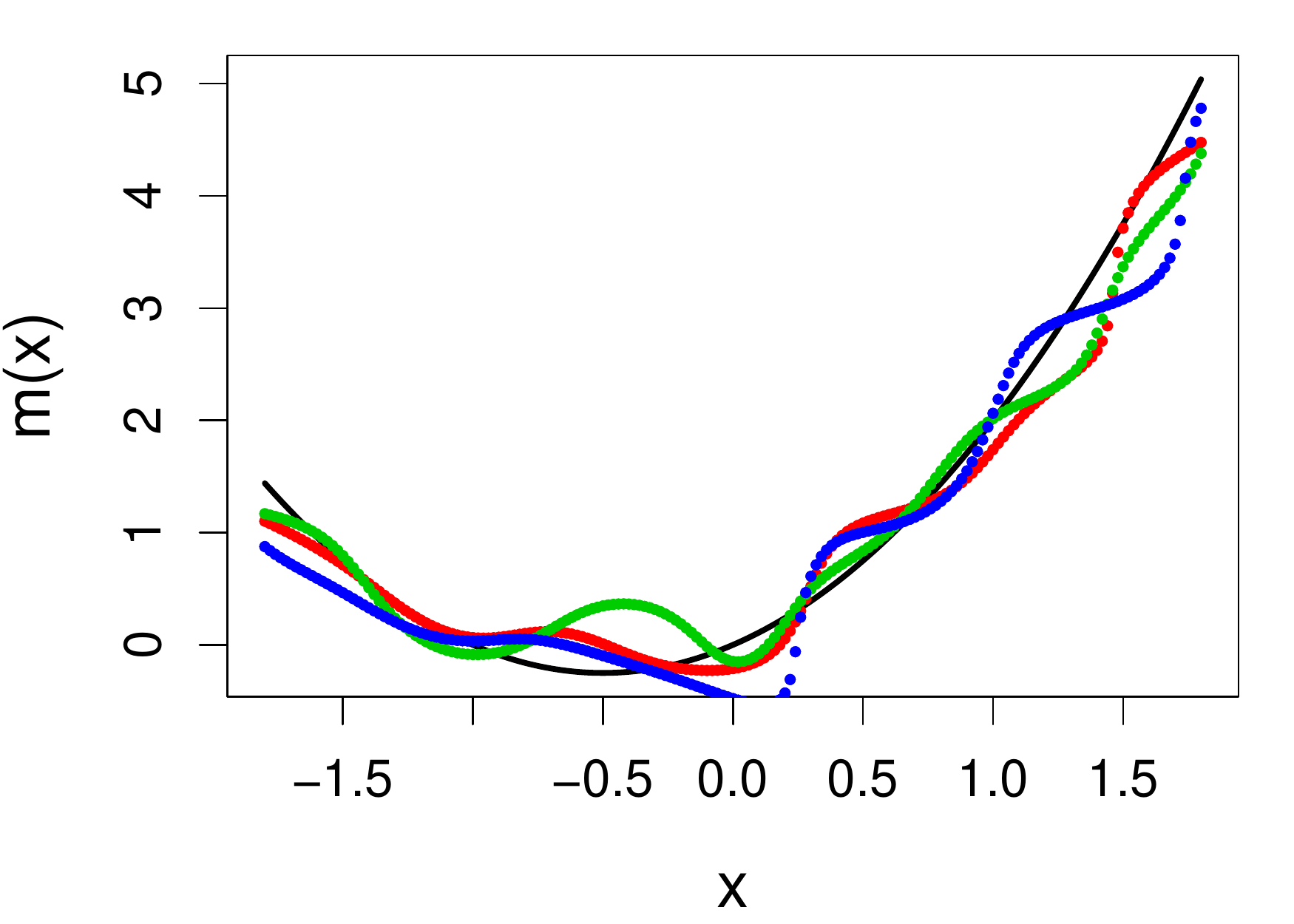} }\\
	\subfigure[]{ \includegraphics[width=\linewidth]{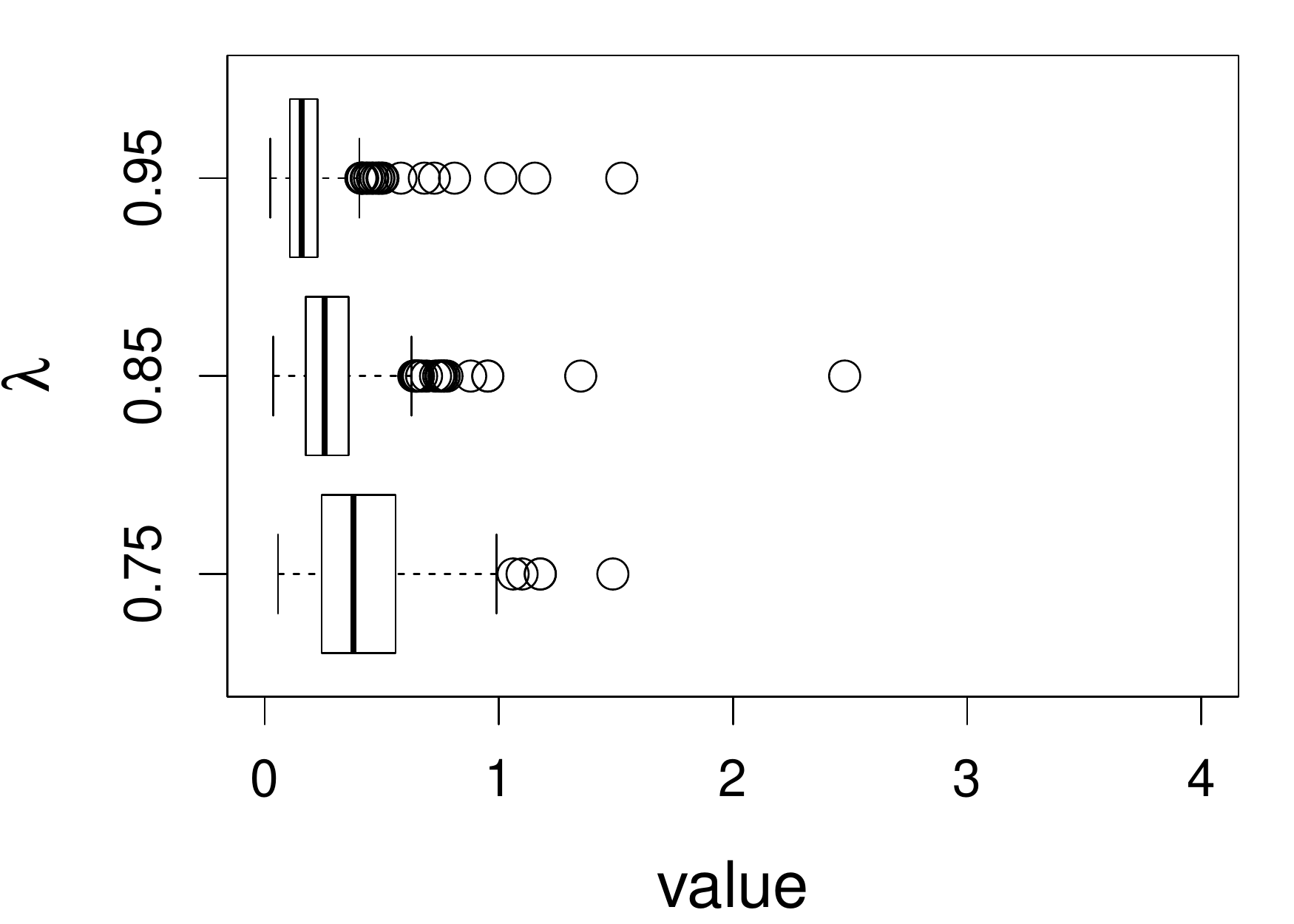} }
	\subfigure[]{ \includegraphics[width=\linewidth]{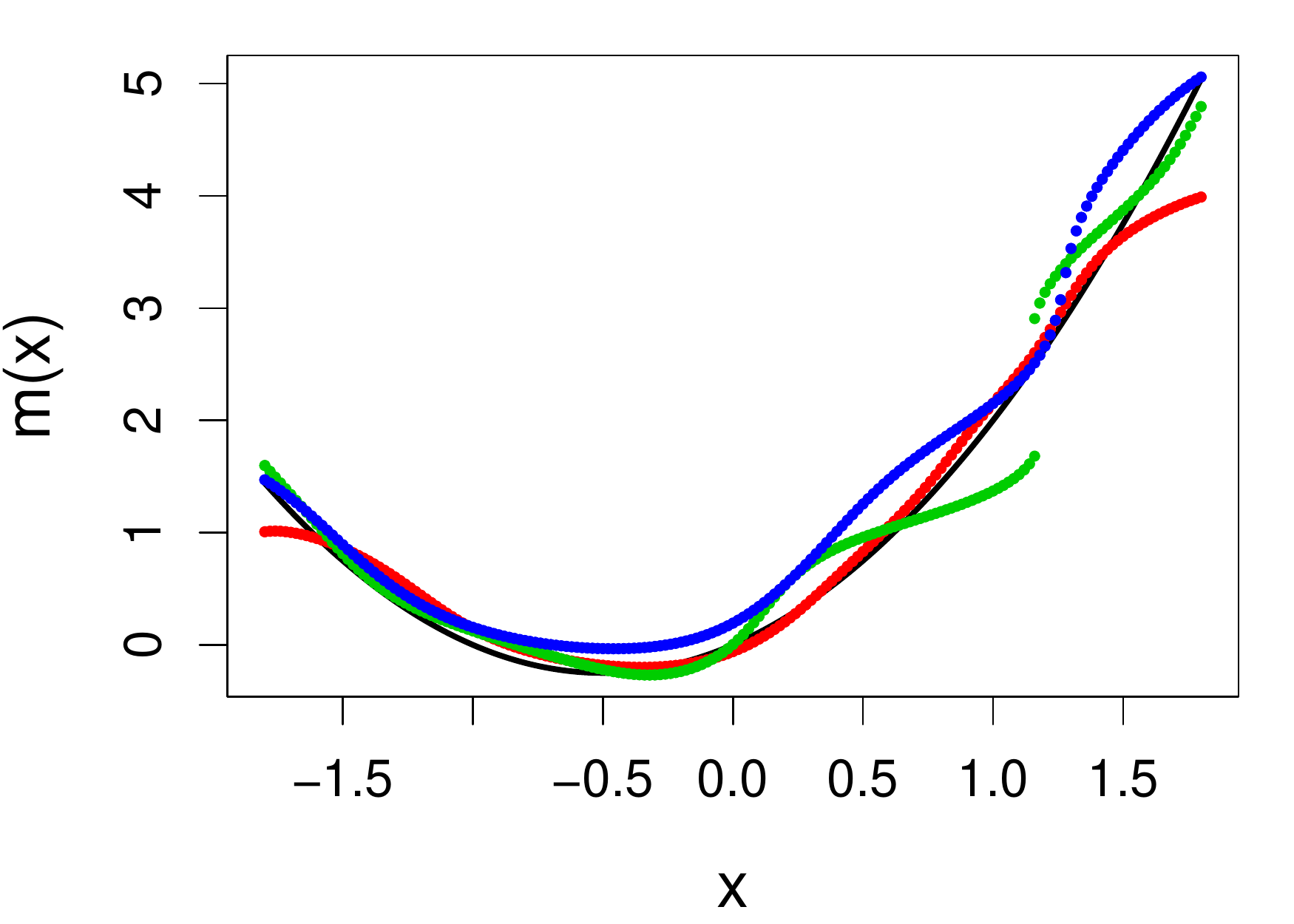} }
	\caption{Results under (C1) using the CV-SIMEX bandwidth selection. Panels (a), (c), and (e): boxplots of ISEs versus $\lambda$ for $\hat M_{\hbox {\tiny $N$}}(x)$, $\hat M_0(x)$, and $\hat M_1(x)$, respectively. Panels (b), (d), and (f): estimated mode curves, $\hat M_{\hbox {\tiny $N$}}(x)$, $\hat M_0(x)$, and $\hat M_1(x)$, respectively, when $\lambda=0.85$. In each panel with estimated mode curves associated with an estimator, the black line depicts the true mode curve, the red, green, and blue lines are three estimated mode curves from the same method that yield ISE being the first, second, and third quantiles among the 500 ISEs for that method from the simulation, respectively.}
	\label{Sim1LapLap500SIMEX-dens:box}
\end{figure}

\begin{figure}
	\centering
	\setlength{\linewidth}{5cm}
	\subfigure[]{ \includegraphics[width=\linewidth]{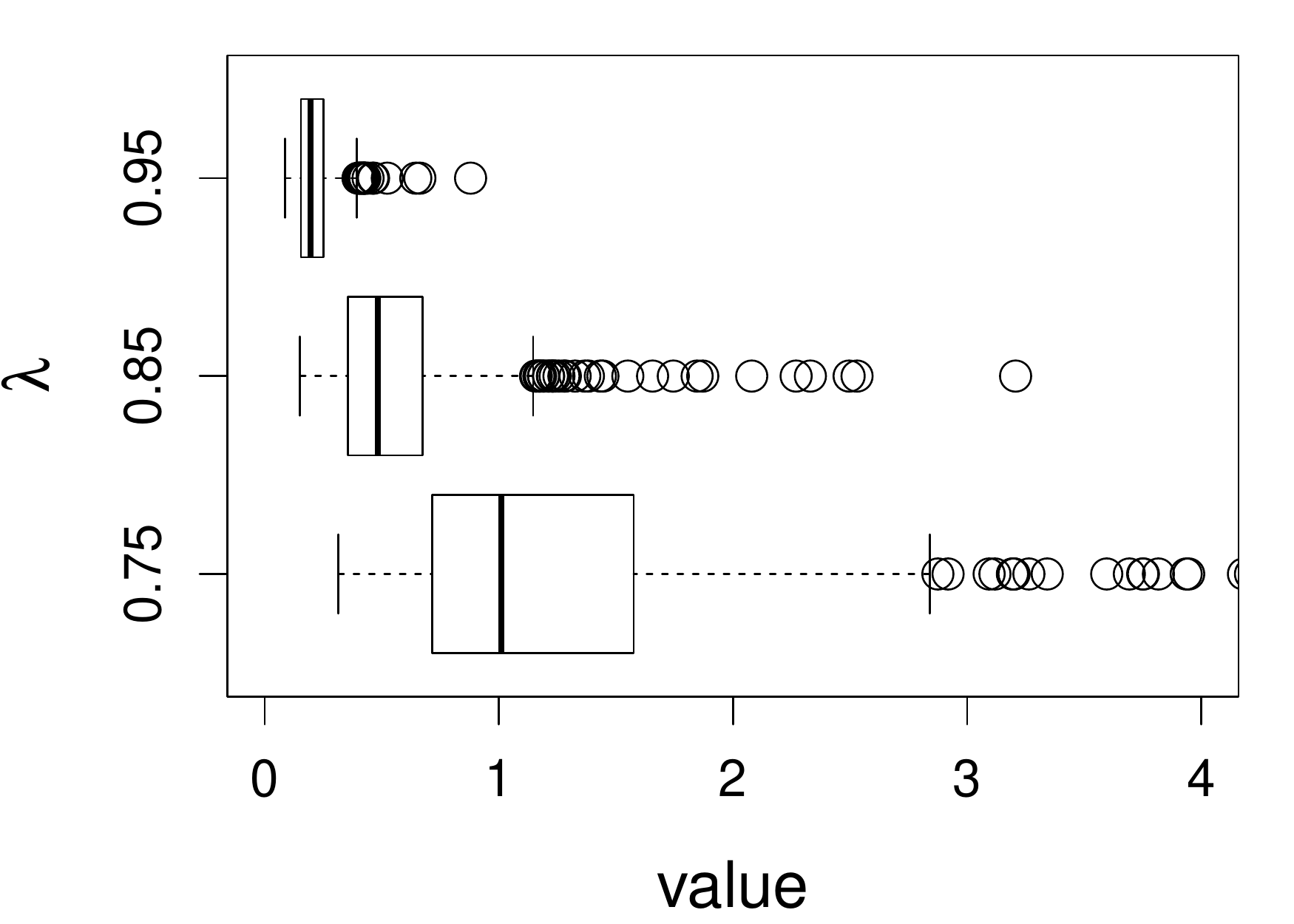} }
	\subfigure[]{ \includegraphics[width=\linewidth]{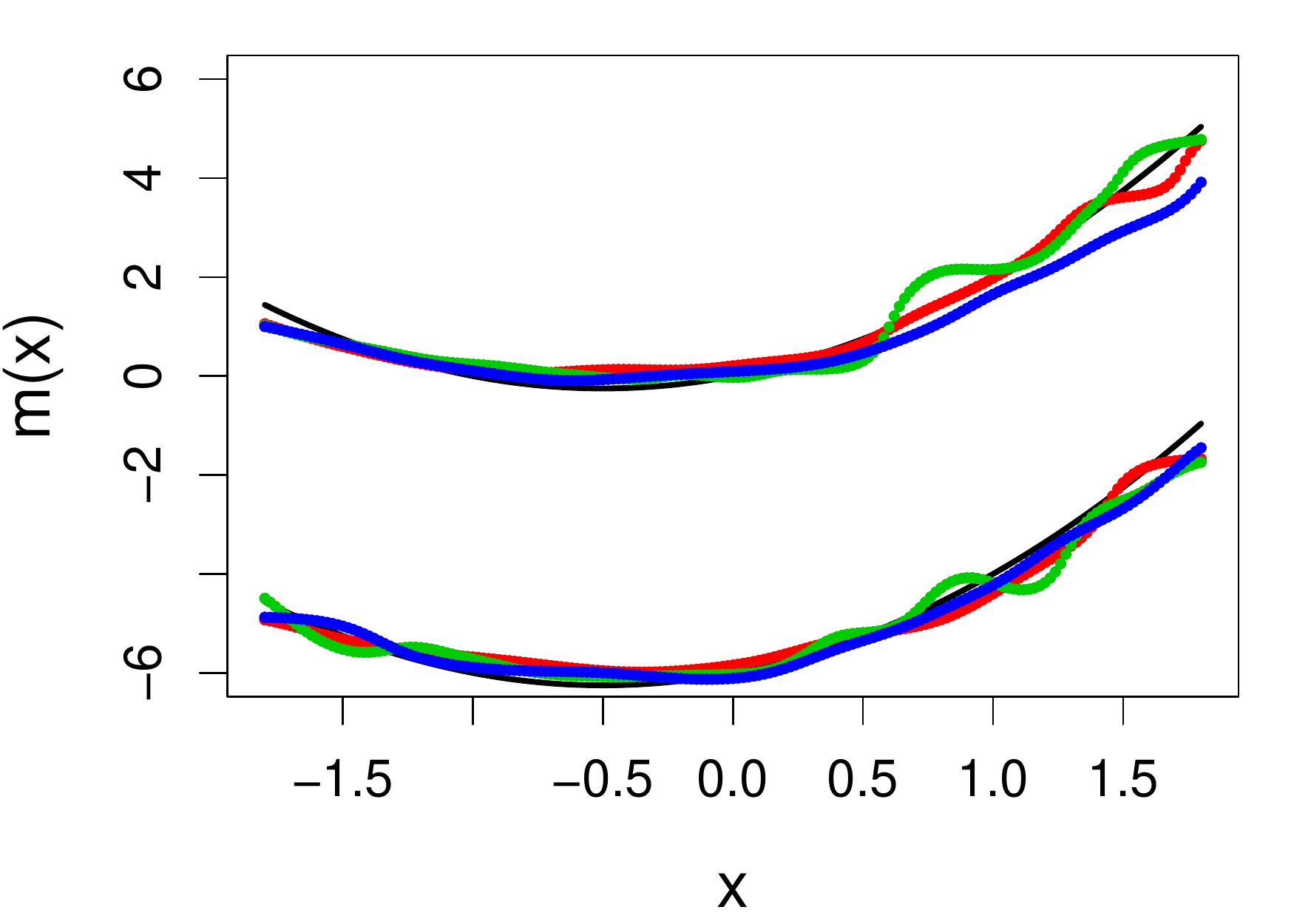} }\\
	\subfigure[]{ \includegraphics[width=\linewidth]{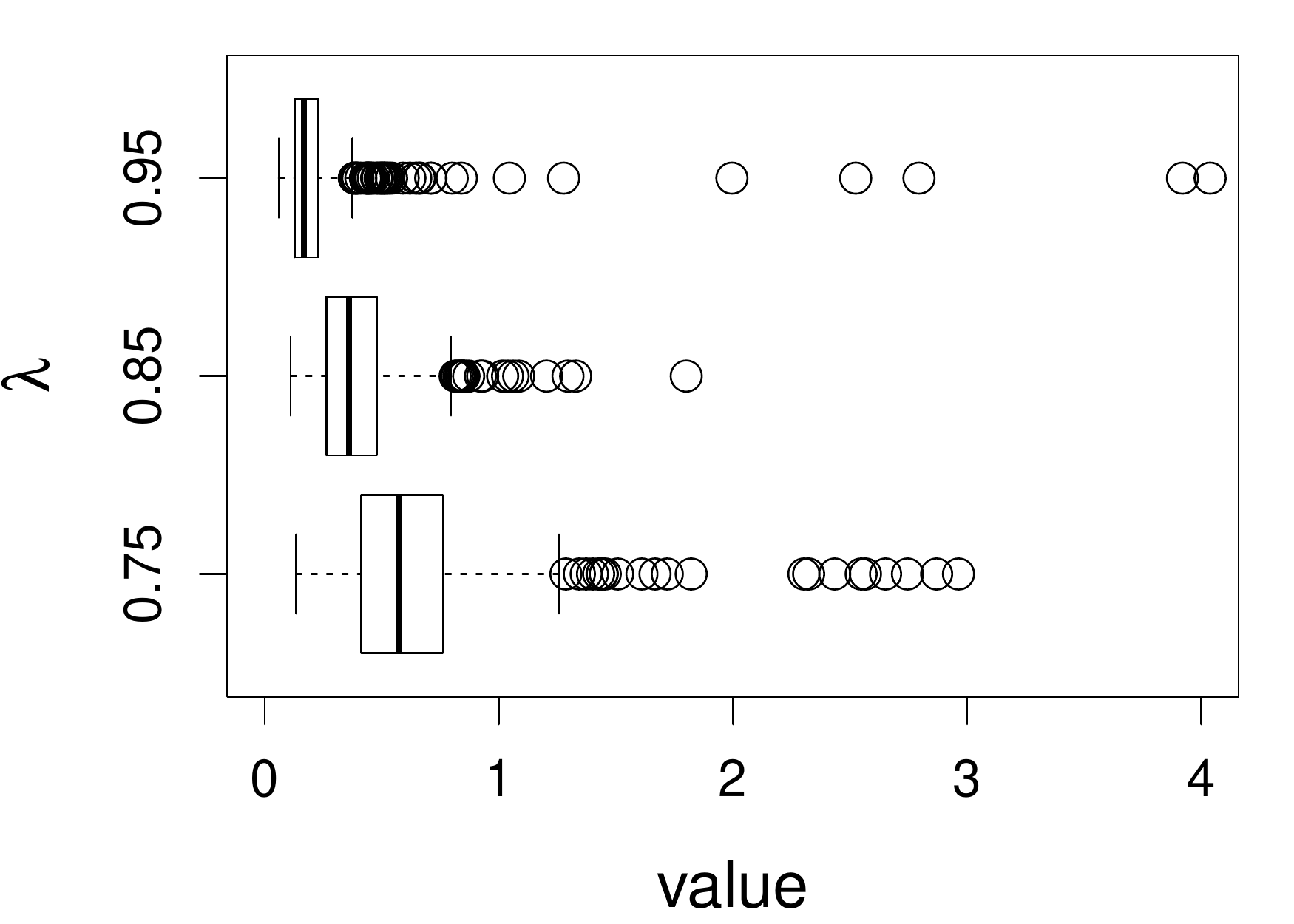} }
	\subfigure[]{ \includegraphics[width=\linewidth]{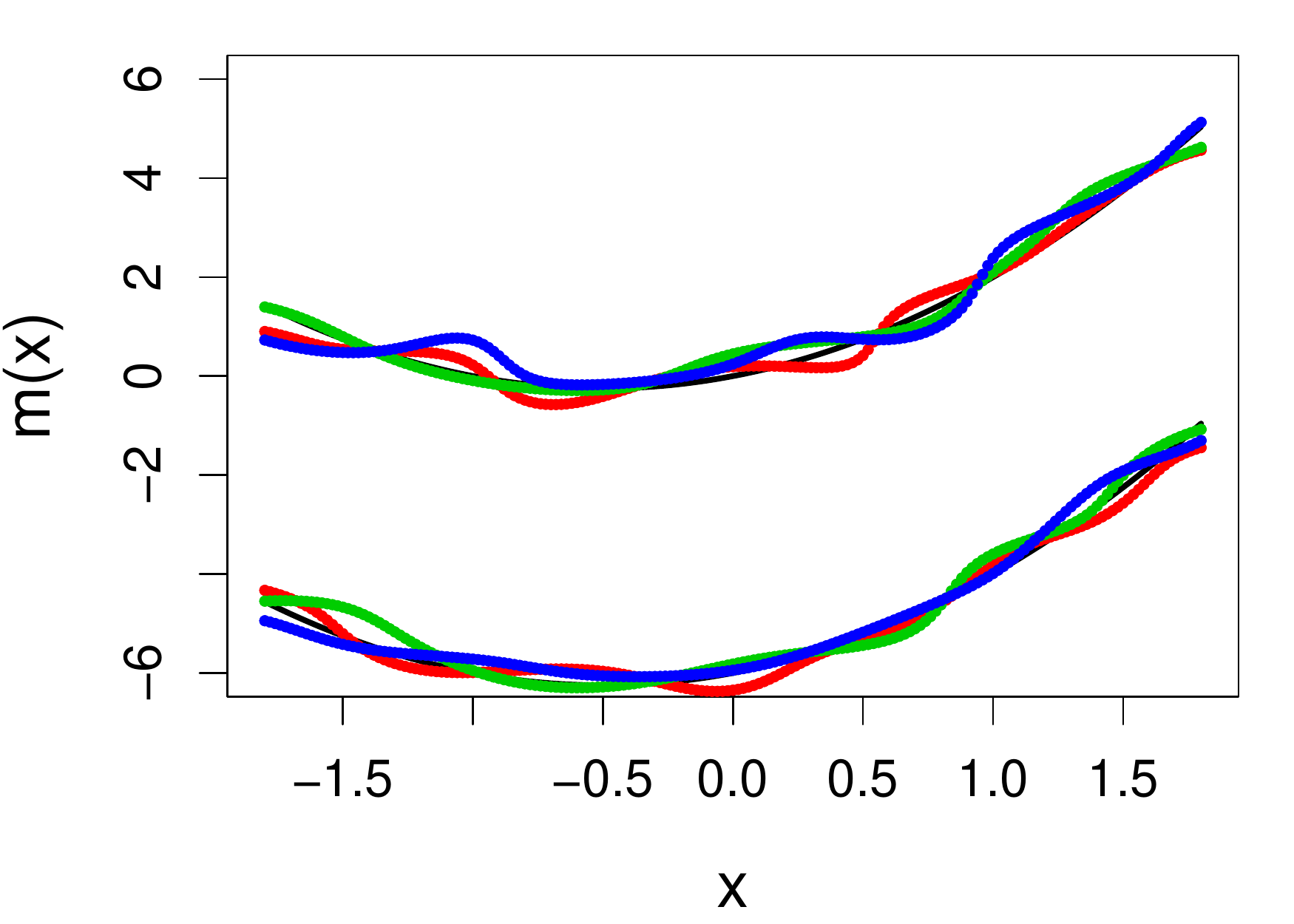} }\\
	\subfigure[]{ \includegraphics[width=\linewidth]{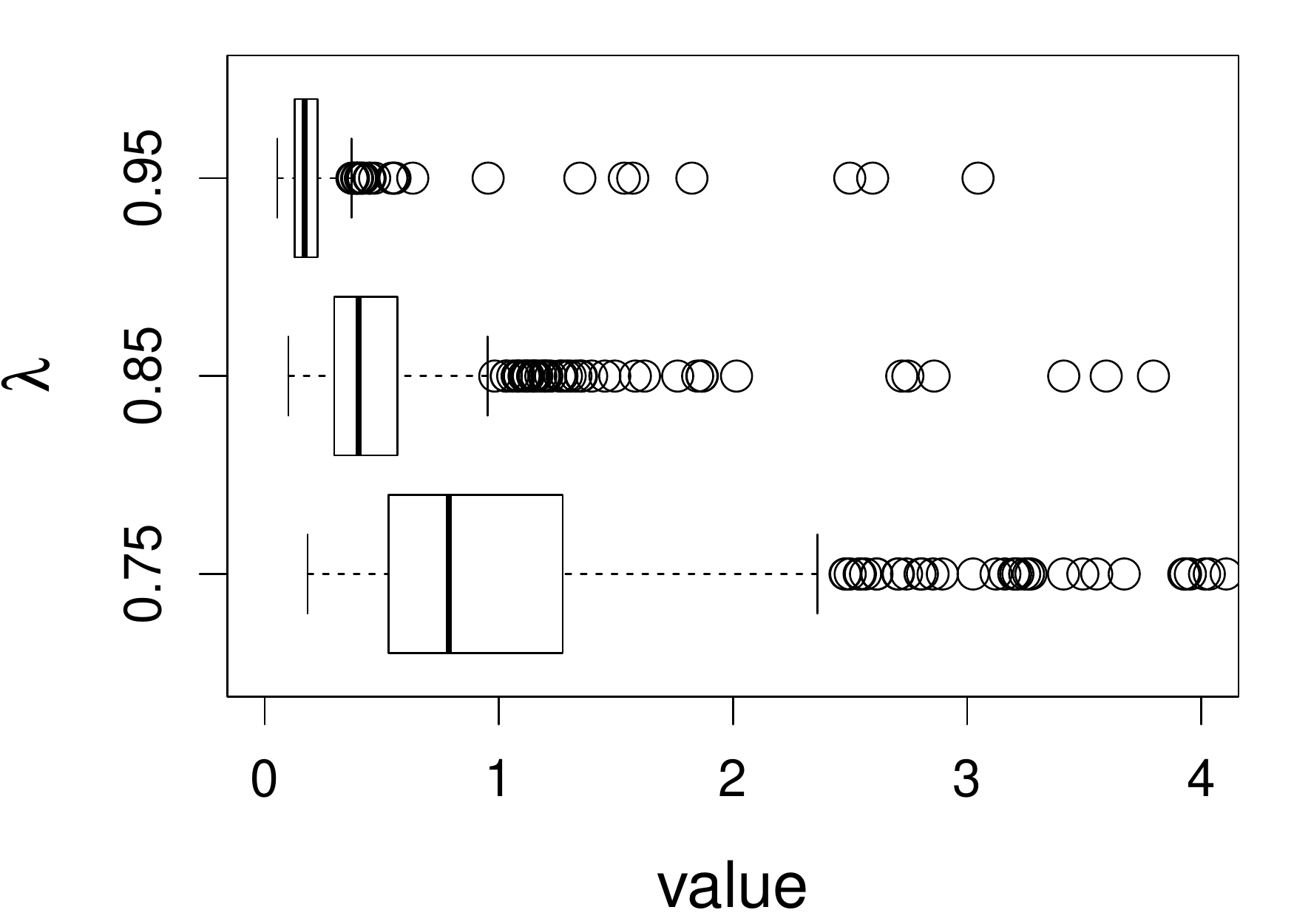} }
	\subfigure[]{ \includegraphics[width=\linewidth]{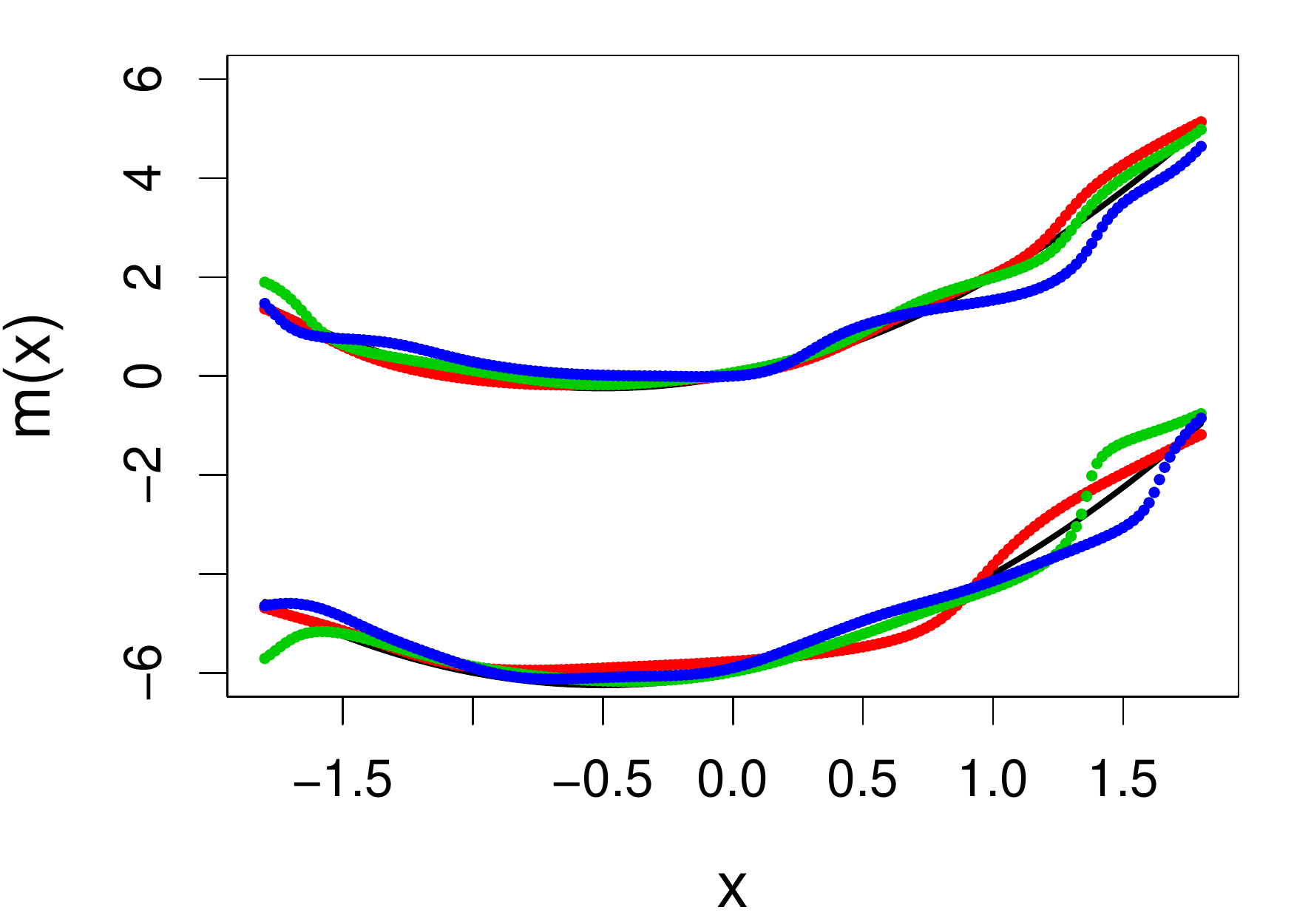} }
	\caption{Results under (C2) using the CV-SIMEX bandwidth selection. Panels (a), (c), and (e): boxplots of ISEs versus $\lambda$ for $\hat M_{\hbox {\tiny $N$}}(x)$, $\hat M_0(x)$, and $\hat M_1(x)$, respectively. Panels (b), (d), and (f): estimated mode curves, $\hat M_{\hbox {\tiny $N$}}(x)$, $\hat M_0(x)$, and $\hat M_1(x)$, respectively, when $\lambda=0.85$. In each panel with estimated mode curves associated with an estimator, the black lines depict the true mode curves, the red, green, and blue lines are three estimated mode curves from the same method that yield ISE being the first, second, and third quantiles among the 500 ISEs for that method from the simulation, respectively.}
	\label{Sim2LapLap500SIMEX-dens:box}
\end{figure}

\setcounter{equation}{0}
\setcounter{section}{0}
\renewcommand{\theequation}{I.\arabic{equation}}
\renewcommand{\thesection}{I.\arabic{section}}
\section*{Appendix I: A sketch of the arguments in Section 3.1 with uni-modality assumption relaxed}

Suppose $p(y|x)$ has $K$ modes, with the mode set $M(x)=\{y_{\hbox {\tiny $M,1$}}(x), \ldots, y_{\hbox {\tiny $M,K$}}(x)\}$, and the estimated mode set $\hat M(x)=\{\hat y_{\hbox {\tiny $M,1$}}(x), \ldots, \hat y_{\hbox {\tiny $M,K$}}(x)\}$. Then the pointwise error is $\Delta_n(x)=\max_{1\le k \le K} |\hat y_{\hbox {\tiny $M,k$}}(x)-y_{\hbox {\tiny $M,k$}}(x)|$. By the mean-value theorem, for each $k=1, \ldots, K$, one has 
$$\hat y_{\hbox {\tiny $M,k$}}(x)-y_{\hbox {\tiny $M,k$}}(x)=-\{g_{yy}(x, y_{\hbox {\tiny $M,k$}}\}^{-1}\hat g_y(x,y_{\hbox {\tiny $M,k$}})
+O(\|\hat g_{yy}-g_{yy}\|_\infty)\hat g_y(x,y_{\hbox {\tiny $M,k$}}),$$
as in (\ref{eq:diffym}). It follows that 
$$\Delta_n(x)=\max_{1\le k \le K}|\{g_{yy}(x, y_{\hbox {\tiny $M,k$}})\}^{-1}\hat g_y(x, y_{\hbox {\tiny $M,k$}})|+ O(\|\hat g_{yy} -g_{yy}  \|_\infty) \max_{1\le k \le K}|\hat g_y(x, y_{\hbox {\tiny $M,k$}})|,$$ 
and thus
\begin{eqnarray*}
& & \frac{\Delta_n(x)}{\max_{1 \le k \le K}|\{g_{yy}(x, y_{\hbox {\tiny $M,k$}})\}^{-1}\hat g_y(x, y_{\hbox {\tiny $M,k$}})|} \\
& = & 
1+ O(\|\hat g_{yy} -g_{yy}  \|_\infty)\frac{\max_{1\le k \le  K}|\hat g_y(x, y_{\hbox {\tiny $M,k$}})|}{\max_{1 \le k \le K}|\{g_{yy}(x, y_{\hbox {\tiny $M,k$}})\}^{-1}\hat g_y(x, y_{\hbox {\tiny $M,k$}})|} \\
& = &  1+ O(\|\hat g_{yy} -g_{yy}  \|_\infty), 
\end{eqnarray*}
where the last equality results from assumption (CP2). 

Hence, under the same conditions that supporting (\ref{eq:Delta2}), $\Delta_n(x)$ can be approximated by $\max_{1\le k \le K}|\{g_{yy}(x, y_{\hbox {\tiny $M,k$}})\}^{-1}\hat g_y(x, y_{\hbox {\tiny $M,k$}})|$, and thus the convergence rate of $\Delta_n(x)$ is the same as that of $\max_{1 \le k \le K}|\hat g_y(x, y_{\hbox {\tiny $M,k$}})|$. From this point on, all arguments in Section~\ref{s:asymptotics} regarding $\hat g_y(x, y_{\hbox {\tiny $M$}})$, which is $\hat p_y(x, y_{\hbox {\tiny $M$}})$ in Section~\ref{s:rate0} and is $\hat p_y(y_{\hbox {\tiny $M$}}|x)$ in Section~\ref{s:rate1}, carry over to $\hat g_y(x, y_{\hbox {\tiny $M,k$}})$ for each $k=1, \ldots, K$. And as long as $K$ is finite, the convergence rate of $\max_{1 \le k \le K}|\hat g_y(x, y_{\hbox {\tiny $M,k$}})|$ is the same as that of $|\hat g_y(x, y_{\hbox {\tiny $M,k$}})|$ for a $k\in \{1, \ldots, K\}$.

\end{document}